\documentclass[twocolumn]{aastex63}
\turnoffedit
\usepackage[caption=false]{subfig}
\usepackage{totcount}

\newtotcounter{citnum} 
\def\oldbibitem{} \let\oldbibitem=\bibitem
\def\bibitem{\stepcounter{citnum}\oldbibitem}

\usepackage{lipsum}

\usepackage{amsmath}
\usepackage[flushleft]{threeparttable} 
\usepackage{array}
\newcommand{\numnew}{11}
\newcommand{\nlobes}{67}
\newcommand{\example}{HOPS 68}
\newcommand{\tex}{$T_{\rm ex}$}

\newcommand{\kms}{km~s$^{-1}$}

\newcommand{\co}[1][]{\ensuremath{^{#1}}CO}

\shorttitle{Outflows in Orion A}
\shortauthors{Feddersen {\em et al.}}

\begin{document}

\title{The CARMA-NRO Orion Survey: Protostellar Outflows, Energetics, and Filamentary Alignment}

\author{Jesse R. Feddersen}
\affiliation{Department of Astronomy, Yale University, P.O. Box 208101, New Haven, CT 06520-8101, USA}
\author{H\'ector G. Arce}
\affiliation{Department of Astronomy, Yale University, P.O. Box 208101, New Haven, CT 06520-8101, USA}
\author{Shuo Kong}
\affiliation{Department of Astronomy, Yale University, P.O. Box 208101, New Haven, CT 06520-8101, USA}
\author{S\"umeyye Suri}
\affiliation{I. Physikalisches Institut, Universit\"at zu K\"oln, Z\"ulpicher Str. 77, 50937 K\"oln, Germany}
\author{\'Alvaro S\'anchez-Monge}
\affiliation{I. Physikalisches Institut, Universit\"at zu K\"oln, Z\"ulpicher Str. 77, 50937 K\"oln, Germany}
\author{Volker Ossenkopf-Okada}
\affiliation{I. Physikalisches Institut, Universit\"at zu K\"oln, Z\"ulpicher Str. 77, 50937 K\"oln, Germany}
\author{Michael M. Dunham}
\affiliation{Department of Physics, State University of New York at Fredonia, 280 Central Avenue, Fredonia, NY 14063, USA}
\author{Fumitaka Nakamura}
\affiliation{National Astronomical Observatory of Japan, 2-21-1 Osawa, Mitaka, Tokyo 181-8588, Japan}
\author{Yoshito Shimajiri}
\affiliation{Laboratoire AIM, CEA/DSM-CNRS-Universit\'e Paris Diderot, IRFU/Service d’Astrophysique, CEA Saclay, F-91191 Gif-sur-Yvette, France}
\author{John Bally}
\affiliation{Department of Astrophysical and Planetary Sciences, University of Colorado, Boulder, CO, USA}
\email{jesse.feddersen@yale.edu}

\begin{abstract}
We identify 45 protostellar outflows in CO maps of the Orion A giant molecular cloud from the CARMA-NRO Orion survey. Our sample includes \numnew{} newly detected outflows. We measure the mass and energetics of the outflows, including material at low-velocities by correcting for cloud contributions. The total momentum and kinetic energy injection rates of outflows is comparable to the turbulent dissipation rate of the cloud. We also compare the outflow position angles to the orientation of C$^{18}$O filaments. We find that the full sample of outflows is consistent with being randomly oriented with respect to the filaments. A subsample of the most reliable measurements shows a moderately perpendicular outflow-filament alignment which may reflect accretion of mass across filaments and onto the protostellar cores.
\end{abstract}

\keywords{ISM: clouds --- ISM: individual objects (Orion A) ---  stars: formation} 

\section{Introduction}
Stars form inside the densest parts of giant molecular clouds (GMCs) \citep{McKee07}. Bipolar, high-velocity flows of molecular gas, called outflows, are launched from forming protostars. These structures are so ubiquitous they can be observed even when the protostar itself is unseen \citep{Kong19}. While it is unclear how exactly outflows are launched, they likely arise from the interaction of the accretion disk with magnetic fields near the protostar \citep[e.g.][]{Konigl00,Shu00,Frank14}.

Outflows have long been considered as a mechanism for sustaining turbulence and slowing star formation in GMCs \citep{Nakamura07,Carroll09,Federrath15}. Surveys of outflows have repeatedly shown that they have enough aggregate momentum and kinetic energy to significantly offset the dissipation of turbulence, especially on cluster scales \citep{Arce10,Nakamura11,Plunkett13,Plunkett15,Li15}. More uncertain is the efficiency with which outflows inject momentum at larger scales in order to maintain the observed turbulence in GMCs \citep{Brunt09,Padoan09,Carroll10}. For a comprehensive review of outflows, see \citet{Arce07}, \citet{Frank14}, and \citet{Bally16}.

In many GMCs, most of the star formation takes place in relatively narrow, dense filaments \citep{Arzoumanian11,Suri19}. If the angular momentum of a growing protostar is inherited from the mass accretion onto a filament, the protostellar spin and the filament direction will be correlated \citep{Bodenheimer95,Andre14,Li19}. If, however, the link between mass accretion at filament scales and protostellar scales is disrupted, e.g. by turbulence or interaction with protostellar companions, then the protostellar spin may not be correlated with the filament orientation \citep{Offner16,Lee17}.

To distinguish between these scenarios, several studies have recently considered the relative orientation of outflows and filaments. Assuming that outflow direction traces the angular momentum of the protostar, a correlation between the outflow and filament direction could mean a connection between sub-AU scales, where the outflow is launched, to filaments at much larger scales. \citet{Davis09} measured the position angle of H$_2$ outflows in the Orion A GMC and found no correlation with the large-scale integral-shaped-filament (ISF). \citet{Stephens17} compared CO outflows with filaments extracted from \emph{Herschel} dust maps in the Perseus molecular cloud and likewise found a random outflow-filament alignment. However, \citet{Kong19} recently found a preferentially perpendicular outflow-filament alignment in the IRDC G28.37+0.07

The Orion A GMC is an ideal environment for studying protostellar outflows and their connection to filaments. At a distance of 400~pc, \footnote{We adopt a distance of 400~pc throughout this paper to be consistent with the recent parallax measurements from GAIA Data Release 2 \citep{Kounkel18,Grossschedl18,Kuhn19}.  \citet{Tanabe19}, like most previous studies of Orion A, assume a slightly greater distance of 414~pc \citep{Menten07}. Increasing the distance by this amount would increase mass, momentum, and kinetic energy reported here by 7\%. The mass loss rate, momentum injection rate and kinetic energy injection rate would increase by 4\%. Our conclusions are not sensitive to this uncertainty in distance.} it is one of the closest clouds forming both low- and high-mass stars. Following earlier outflow studies, \citet{Tanabe19} recently carried out a systematic search of CO outflows in Orion A, which complements our study. We use the same CO data as \citet{Tanabe19}, supplemented with interferometric observations which provide greater resolving power (see Section~\ref{sec:observations}). Additionally, we use different methods to identify outflows and derive their physical properties. In particular, we correct for velocity-dependent opacity and low-velocity outflow emission. We also use the C$^{18}$O filament catalog from \citet{Suri19} for an entirely new analysis of the outflow-filament connection.

In this paper, we present a study of protostellar outflows in the CARMA-NRO Orion survey. In Section~\ref{sec:observations}, we describe the CARMA-NRO Orion CO data and protostar catalogs. In Section~\ref{sec:identification}, we describe how we search for outflows in the CO maps and present the outflow catalog. In Section~\ref{sec:physics}, we calculate the physical properties of the outflows and discuss their impact on the cloud. In Section~\ref{sec:filaments}, we discuss the relative orientation of outflows and C$^{18}$O filaments and compare to models for random, parallel, and perpendicular outflow-filament alignment. Finally, in Section~\ref{sec:conclusions}, we summarize our conclusions and discuss future directions. The entire outflow catalog is presented in the Appendix.

\section{Observations}\label{sec:observations}

\subsection{CO Maps}
The CARMA-NRO Orion survey combines interferometric observations from the Combined Array for Research in Millimeter-wave Astronomy (CARMA) with single-dish observations from the 45 m telescope at the Nobeyama Radio Observatory (NRO). The combination of interferometric and single-dish observations results in an unprecedented spatial dynamic range of 0.01 - 10 pc in the Orion A molecular cloud.

We use the \co[12](1-0), \co[13](1-0), and C$^{18}$O(1-0) data first presented in \citet{Kong18}. The \co[12]{} data have a resolution of $10\arcsec\times8\arcsec$ and velocity resolution of $0.25$ \kms{}. The original \co[13]{} data have a resolution of $8\arcsec\times6\arcsec$ and a velocity resolution of $0.22$ \kms{}, but we smooth the \co[13]{} data to match the resolution of \co[12]{}. The C$^{18}$O data have a resolution of $10\arcsec\times8\arcsec$ and a velocity resolution of $0.22$ \kms{}. We do not smooth the C$^{18}$O data because we use it primarily for filament identification (Section~\ref{sec:filaments}) and do not directly compare it to the other lines. The area covered by the CARMA-NRO Orion survey is shown in Figure~\ref{fig:overview}.

\subsection{Source Catalogs}
We primarily use the Herschel Orion Protostar Survey \citep[HOPS;][]{Furlan16} to assign driving sources to outflows. The HOPS catalog is the result of \emph{Herschel} PACS 70-160~$\micron{}$ observations of the protostars identified in the \emph{Spitzer} Orion survey by \citet{Megeath12}. \citet{Furlan16} fit the spectral energy distribution (SED) of these 330 protostars from 1.2-870~$\micron{}$ with models to derive a variety of protostar, disk, and envelope properties.

We also use the study of H$_2$ outflows by \citet{Davis09} to guide our search. \citet{Davis09} surveyed Orion A using narrow-band images of the ro-vibrational H$_2$ 2.122 \micron{} line. Their work builds on previous H$_2$ mapping by \citet{Stanke02}. We search for outflows around the 17 sources of H$_2$ flows that are not in the HOPS catalog. In addition to the driving sources of H$_2$ outflows, \citet{Davis09} catalog H$_2$ features with no obvious driving source. We also search for the CO counterpart to the 29 H$_2$ flows without an identified source within the CO data footprint. \citet{Davis09} measure the proper motions of 33 flows. We use the H$_2$ images and, when available, the proper motions to help assign especially difficult CO outflows to driving sources.

\subsection{Other Outflow Studies}
Orion A has been mapped extensively in CO over the past 30 years \citep[e.g.][]{Bally87,Wilson05,Shimajiri11,Buckle12,Ripple13,Berne14}. Several studies have searched for outflows in the cloud using these CO observations. In the northern part of the cloud, previous searches have identified 18 outflows along the OMC 2/3 ridge \citep{Chini97,Aso00,Williams03,Takahashi08,Shimajiri08,Shimajiri09}. In the L1641N cluster in the southern part of the cloud, \citet{Stanke07} and \citet{Nakamura12} found six outflows. In the NGC 1999 region further south, previous studies have found five outflows \citep{Morgan91,Moro-Martin99,Davis2000,Choi17}. 

\citet{Tanabe19} have carried out a systematic search for outflows across the Orion A cloud, using the same single-dish NRO observations which are used in our CARMA-NRO Orion survey. They identify a total of 44 outflows across the cloud, including 17 new detections. Eleven of these are in the OMC 4/5 region where no outflows had previously been found. Although we expect our catalog to largely overlap with \citet{Tanabe19}, they use an automated outflow search procedure which may include false positives. The improved resolution of the combined CARMA-NRO Orion data allow us to search for smaller outflows, help us disentangle outflow emission in clustered regions, and help with matching outflows to driving sources. We describe our procedure for identifying outflows below.

\section{Outflow Identification}
\label{sec:identification}
We search for CO outflows around each HOPS protostar in \citet{Furlan16} and H$_2$ jet driving source in \citet{Davis09}. For each source, we first fit a Gaussian to the average \co[12] spectrum in a circular region of radius 15\arcsec{} centered on the source. Next, we integrate the \co[12] emission at various velocity intervals on either side of the mean velocity. We visually inspect the contour maps of the blue-shifted and red-shifted emission to look for collimated structures centered on the source which are detected above $5\sigma$ in these integrated blue- and red-shifted intensity maps.

In addition to the \co[12]{} blue/red contour maps, we use NIR images from the VISTA survey \citep{Meingast16} and the H$_2$ outflows in \citet{Davis09} to guide our assignment of high-velocity CO to driving sources. These ancillary data are especially useful in areas of overlapping outflows, or where outflow lobes are not detected close to the driving source. When a series of H$_2$ bow shocks or NIR nebulosity which are clearly associated with a particular source overlap a region of high-velocity CO emission, we can be more confident in the assignment of this CO outflow to its source.

\citet{Tanabe19} automatically defines any blue or redshifted emission above $5\sigma$ near a protostar as an outflow. By restricting our catalog to structures that have the expected morphology, or are correlated with H$_2$ flows, we limit the risk of false positives. Several of the outflows in \citet{Tanabe19} are not included here for these reasons (see Table~\ref{tab:outflows}).

Some regions (e.g. OMC 2/3) contain several overlapping outflows. In these cases, we try to follow previous authors' assignment of the high-velocity emission, unless we strongly disagree with their assessment. In Section~\ref{sec:extraction}, we describe the outflow region extraction.

Once we have identified an outflow, we adjust the velocity range over which we integrate \co[12]{} to produce contour maps that most clearly separate the blue/red lobes from surrounding cloud emission. Table~\ref{tab:outflows} lists these visually determined velocities $v_{\rm blue}$ and $v_{\rm red}$. $v_{\rm blue}$ and $v_{\rm red}$ denote the lowest velocity (closest to the mean cloud velocity) where the outflow emission is clearly separated from the cloud. 

For each outflow, we determine a confidence rating of "Definite" or "Marginal". Definite outflows are clearly associated with their driving source, are clearly distinct from surrounding outflows and cloud emission, and are often clearly correlated with H$_2$ outflows. Most definite outflows have been identified previously. Marginal outflows either have an unclear source assignment, are not clearly separated from the cloud or overlapping outflow emission, or are simply very weakly detected. In some cases, we are more confident of either the blue or red outflow lobe, so we assign confidence ratings independently to each. We expect the subset of definite outflows will have more reliable physical properties and position angles (see Section~\ref{sec:physics}).

While many studies have identified energetic outflows in the OMC 1 region \citep[e.g.][]{Schmid-Burgk90,Zapata05,Teixeira16,Bally17}, we avoid this region in our outflow search. Neither \citet{Davis09} nor \citet{Furlan16} cover the central part of the Orion Nebula, due to saturation and confusion with the bright nebulosity. Aside from the lack of good source catalogs in this region, the CO velocity dispersion here can be as high as 100~km~s$^{-1}$, due to the BN/KL ``explosion'' \citep{Bally17}, making it hard to define blue/red outflow lobes. Thus, as in \citet{Tanabe19}, we focus our outflow search outside of OMC 1. 

\begin{figure*}
\centering
\includegraphics[width=0.8\textwidth]{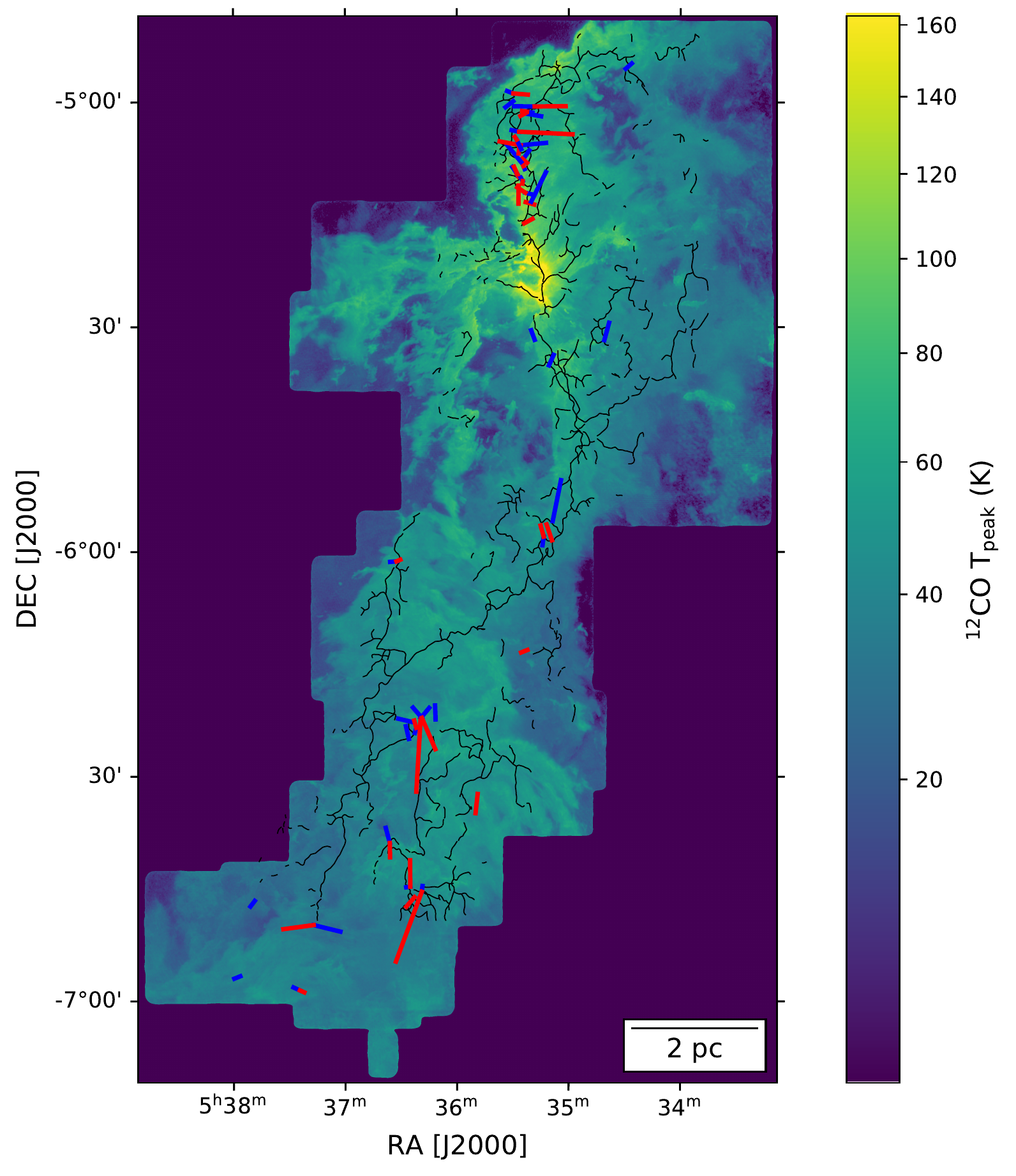}
\caption{Peak \co[12]{} antenna temperature in Orion A with outflows. The blue (red) lines indicate blue (red) outflow lobes. The length of the lines denote the maximum length of the outflow lobe, $R_{\rm max}$~(Section~\ref{sec:tdyn}). The orientation of the outflow lobes indicate the measured position angle (Section~\ref{sec:angles}). The black lines indicate C$^{18}$O filaments from \citet{Suri19} (Section~\ref{sec:filaments}). \label{fig:overview}}
\end{figure*}

In total, we identify 45 outflows with 67 individual lobes. Of these, 11 were not identified by \citet{Tanabe19}. While we expect all outflows to be bipolar, in several cases we only see one lobe. This could be due to interactions between the outflow and the turbulent environment or obscuration of one of the lobes by intervening dense gas \citep{Offner11}. Alternatively, the outflow source may be close to the cloud surface, so that only one side of the outflow entrains molecular gas \citep[e.g.][]{Chernin95} or the other outflow lobe may be impossible to separate from other nearby emission \citep{Arce01a}. We do not detect 10 of the outflows included in \citet{Tanabe19}. Most of these non-detections consist of the most dubious outflows in that study, which we suspect are spurious due to the automated identification method they use. Of the 67 outflow lobes, we classify 38 as ``definite'' and 29 as ``marginal''. Table~\ref{tab:outflows} lists the source, location, $v_{\rm blue}$ and $v_{\rm red}$, confidence score, and the corresponding outflow in \citet{Tanabe19} for our entire catalog. Figure~\ref{fig:overview} shows the outflow catalog on a map of the peak \co[12]{} temperature.

\begin{figure*}[htb!]
\centering
\includegraphics[width=0.6\textwidth]{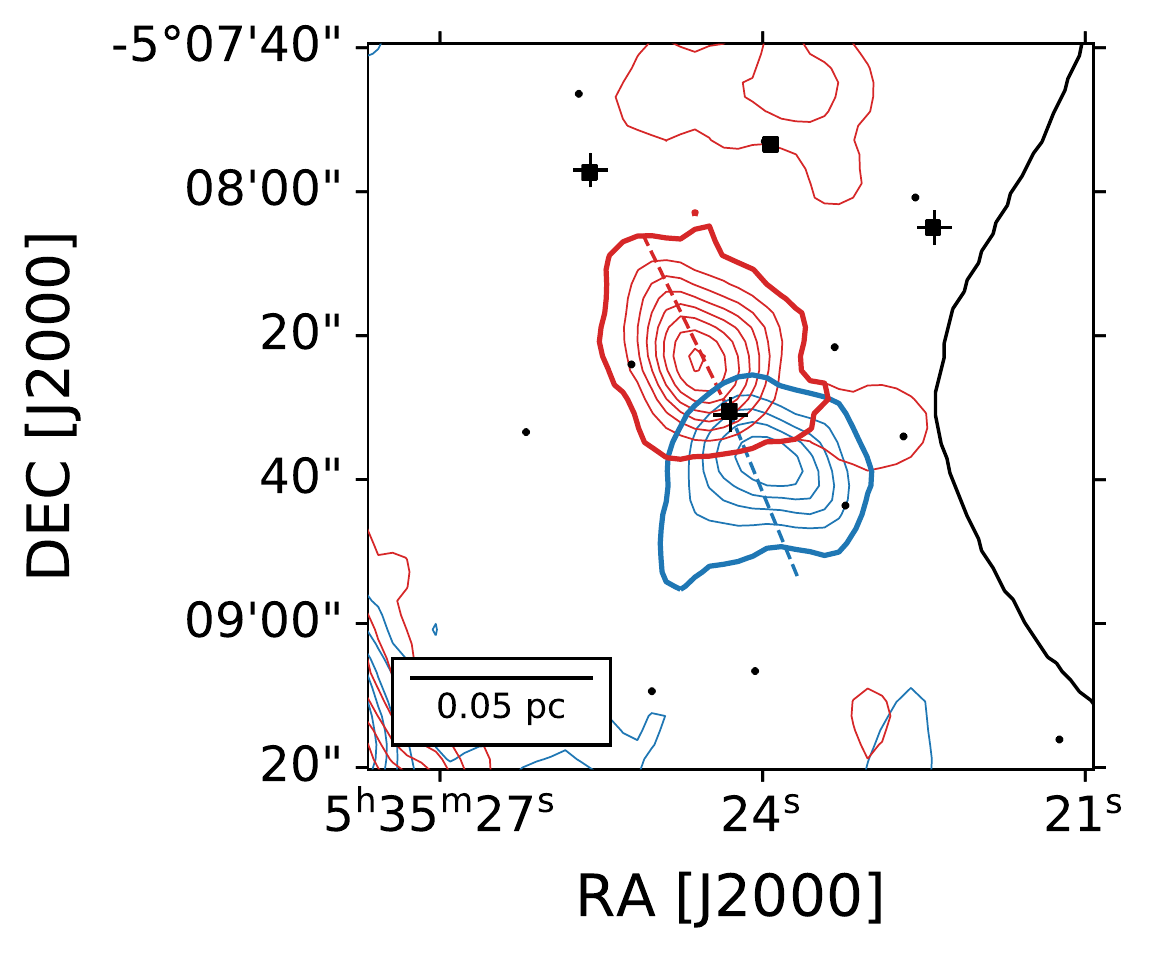}
\caption{Outflow around \example{}. The blue (red) contours show \co[12]{} integrated from -2 \kms{} to $v_{\rm blue}$ ($v_{\rm red}$ to 20 \kms{}), where $v_{\rm blue}$ and $v_{\rm red}$ are given in Table~\ref{tab:outflows}. Contour levels go from 5 to 50$\sigma$, in steps of 5$\sigma$ where $\sigma$ is the RMS error in the integrated map. The thick contours show the region we extract for each lobe.
The black solid line shows the closest C$^{18}$O filament from \citet{Suri19}. Black squares indicate HOPS protostars from \citet{Furlan16}, black crosses indicate H$_2$ outflow sources from \citet{Davis09}, and small black points indicate all \emph{Spitzer} YSOs from \citet{Megeath12}.\label{fig:stamp}}
\end{figure*}

Figure~\ref{fig:stamp} shows an outflow in OMC 2 driven by \example{}. This outflow is well-known in the literature (Outflow 12 in \citealp{Tanabe19} and FIR~2 in \citealp{Takahashi08}). We use this outflow in Figures~\ref{fig:c18o_fit}-\ref{fig:filament} to demonstrate our methods for calculating outflow properties. We present the entire outflow catalog in Appendix~\ref{sec:appendix}.


\section{Physical Properties of Outflows}\label{sec:physics}
\subsection{Calculating Physical Properties}
To calculate the physical properties of outflows, we adapt the methods described by \citet{Arce01}, \citet{Dunham14}, and \citet{ZhangY16}.
\subsubsection{Extracting Outflow Emission}\label{sec:extraction}
To measure the outflow mass, we must first extract each outflow from the surrounding cloud. This is particularly difficult in Orion, where many outflows are clustered and overlapping. We use a two-step approach to extract each outflow lobe.

First, we integrate the $^{12}$CO cube over the visually estimated velocity range of the outflow lobe (discussed in Section~\ref{sec:identification}). We integrate from $v_{\rm blue}$ ($v_{\rm red}$) given in Table~\ref{tab:outflows} to the velocity extremum of the cube, -2~\kms{} (20~\kms{}). We select pixels above $5\sigma$ in this integrated map of high-velocity emission. Next, we draw a region around each outflow lobe by hand to remove other overlapping outflows or other cloud structures in the area. 

To summarize, those pixels above $5\sigma$ in integrated high-velocity $^{12}$CO and which are visually inside the outflow are included in determining the physical properties of the outflow. Figure~\ref{fig:stamp} shows the regions extracted for the \example{} outflow. The regions extracted for the other outflows are shown in Appendix~\ref{sec:appendix}. 

\subsubsection{Systemic Velocity}\label{sec:vsys}
\begin{figure}
\plotone{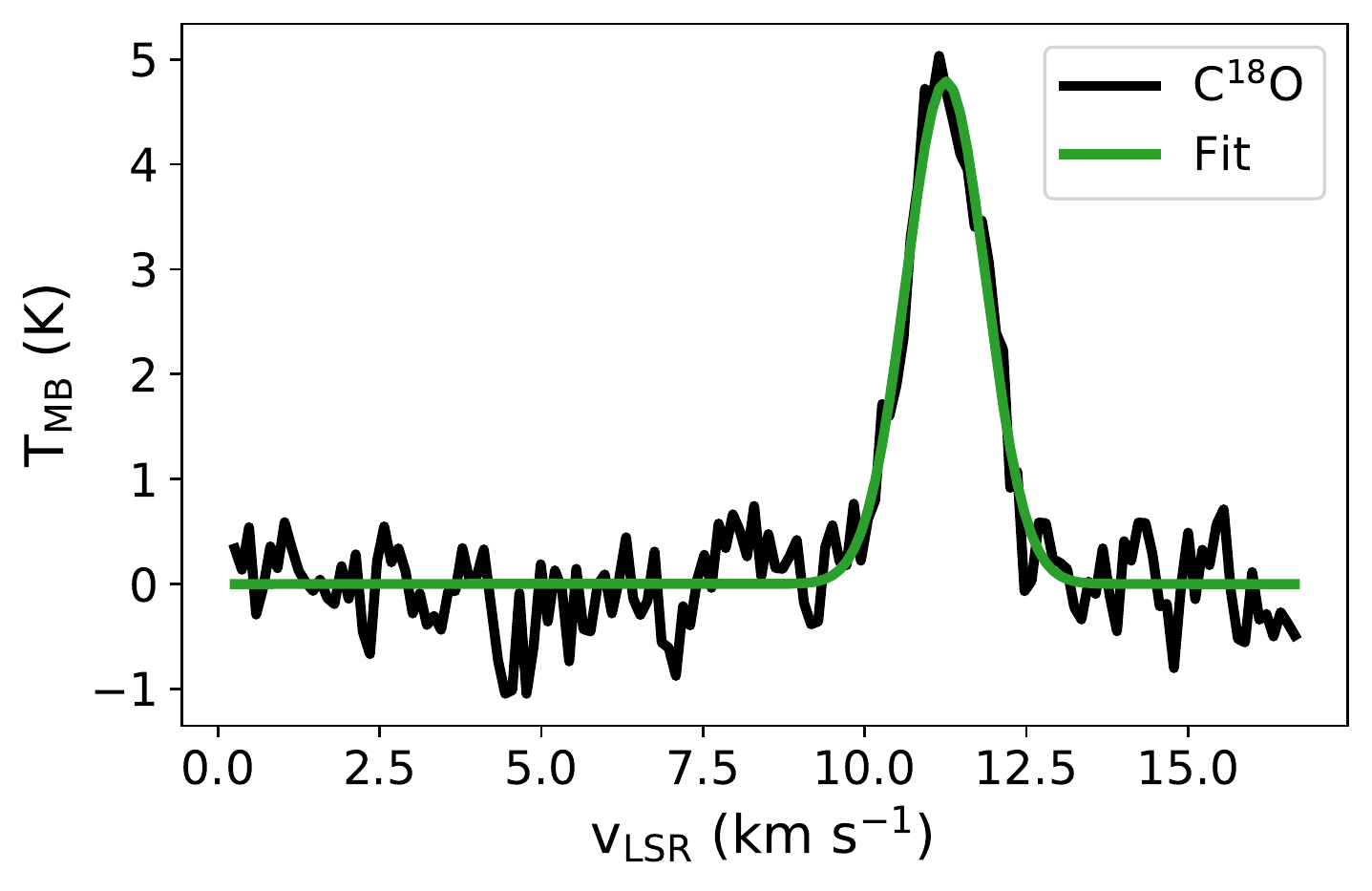}
\caption{C$^{18}$O spectrum toward \example{}. The black line shows the average C$^{18}$O spectrum within a radius of 15\arcsec{} around \example{}. The green line is a Gaussian fit to the spectrum. We define the mean of this fit to be the systemic velocity $v_{\rm sys}$ (see Section~\ref{sec:vsys}). \label{fig:c18o_fit}}
\end{figure}

To calculate outflow energetics, we need to know the systemic velocity $v_{\rm sys}$ of the outflow source. We use the CARMA-NRO Orion C$^{18}$O data \citep{Kong18} for this purpose. We fit a Gaussian model to the average C$^{18}$O spectrum within a 15\arcsec{} radius around the outflow source and define the mean of this Gaussian to be $v_{\rm sys}$. In most cases, there is only one significant velocity component in the C$^{18}$O spectrum. In the few outflows where multiple velocity components are detected, we use the velocity of the component with the highest peak intensity. Figure~\ref{fig:c18o_fit} shows the C$^{18}$O spectrum and fit for the \example{} outflow.

\subsubsection{Excitation Temperature}\label{sec:tex}
We estimate the excitation temperature of $^{12}$CO, \tex, using the equation from \citet{Rohlfs96}, which assumes $^{12}$CO is optically thick in the line center:

\begin{equation}\label{eq:Tex}
T_{\rm ex} = \frac{5.53}{ln(1 + [5.53/(T_{\rm peak} + 0.82)])}.
\end{equation}

\begin{figure}
\plotone{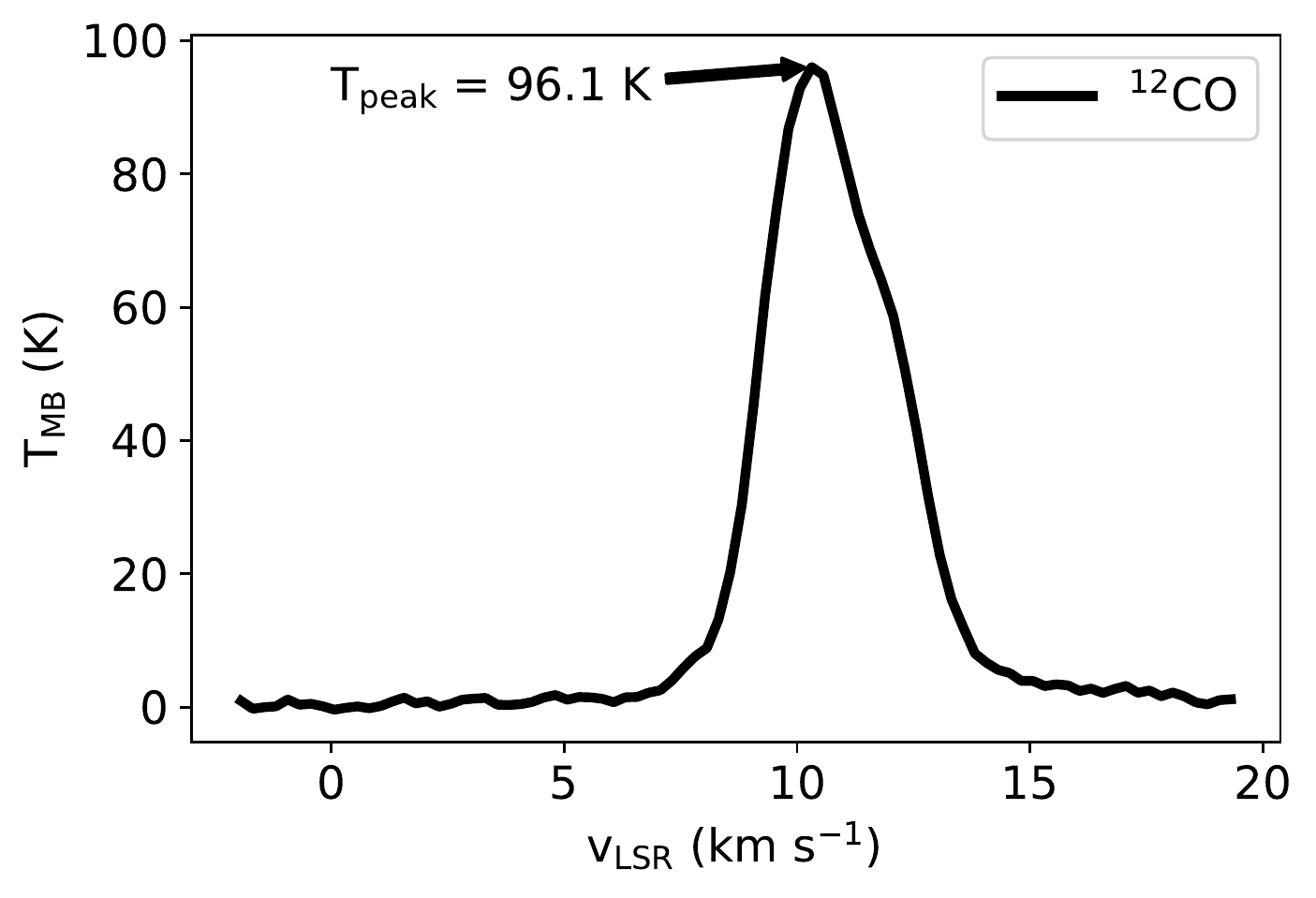}
\caption{\co[12]{} spectrum of the \example{} outflow. The black line shows the average \co[12]{} spectrum within the outflow mask including both lobes shown in Figure~\ref{fig:stamp}. We use the peak of this spectrum, $T_{\rm peak}$, to calculate \tex{} (see Section~\ref{sec:tex}). \label{fig:tex}}
\end{figure}

We define $T_{\rm peak}$ for each outflow to be the peak temperature of the average $^{12}$CO spectrum within the outflow area defined in Section~\ref{sec:extraction}, including both lobes if present. Figure~\ref{fig:tex} shows the average $^{12}$CO spectrum with $T_{\rm peak}$ indicated for the \example{} outflow. The average T$_{\rm ex}$ of the outflow sample is 64~K.

While the peak $^{12}$CO temperature traces the bulk of the gas in the region of the outflow source, the actual outflowing gas likely has a different excitation temperature. \citet{Yang18} find an average temperature of around 100~K for the entrained outflow gas traced by the low-J CO transitions. Above about 20~K, the mass dependence on T$_{\rm ex}$ is close to linear. To convert a mass from one excitation temperature to another, the following formula may be used, obtained by a linear least-squares fit to the mass-T$_{\rm ex}$ relation above 20~K:

\begin{equation}\label{eq:massTex}
\frac{M_{\rm new}}{M_{\rm old}} = \frac{0.128 + 0.0175~\rm T_{ex, new}}{0.128 + 0.0175~\rm T_{ex, old}}.
\end{equation}

If the outflows in our sample have a constant T$_{\rm ex}$ of 100~K, the outflow mass (and all properties derived from mass) will increase by an average factor of 1.5.

\subsubsection{$^{12}$CO Opacity Correction}\label{sec:opacity}
In Orion A, \co[12]{} is usually optically thick \citep{Kong18}. Therefore, if we do not correct for opacity we may miss a substantial amount of outflow mass. By comparing to more optically thin tracers, studies have long shown that outflows tend to be optically thick in \co[12]{}, at least close to cloud velocities \citep[e.g.,][]{Goldsmith84,Arce01}. \citet{Dunham14} find that correcting for the optical depth of outflows increases their mass by a factor of 3 on average. Despite this, outflow studies that lack an optically thin tracer often assume that all \co[12]{} outflow emission is optically thin \citep[e.g. in Orion,][]{Morgan91,Takahashi08}. 

\citet{Tanabe19} adopt an average \co[12]{} optical depth of 5 for their entire outflow catalog. They apply this constant correction factor to every velocity channel. But \citet{Dunham14} show that the optical depth varies with velocity: optical depth decreases away from the mean cloud velocity. We follow \citet{Dunham14} and \citet{ZhangY16} and use the ratio of $^{12}$CO / $^{13}$CO to derive a velocity-dependent opacity correction for each outflow.

We assume both \co[12]{} and \co[13]{} are in LTE with the same excitation temperature and \co[13]{} is optically thin, a reasonable assumption given the results reported in \citet{Kong18}. Then the ratio between the two isotopes is

\begin{equation}\label{eq:Tratio}
\frac{T_{\rm \co[12]}}{T_{\rm \co[13]}} = \frac{[\rm{\co[12]}]}{[\rm{\co[13]}]} \frac{1 - e^{-\tau_{12}}}{\tau_{12}},
\end{equation}
where we assume the abundance ratio [\co[12]]/[\co[13]] is 62 \citep{Langer93} and $\tau_{12}$ is the optical depth of \co[12]{}. In each velocity channel, we calculate the average ratio between \co[12]{} and \co[13]{} in pixels in the outflow region with both lines detected at $5\sigma$ or higher. 

\begin{figure}
\plotone{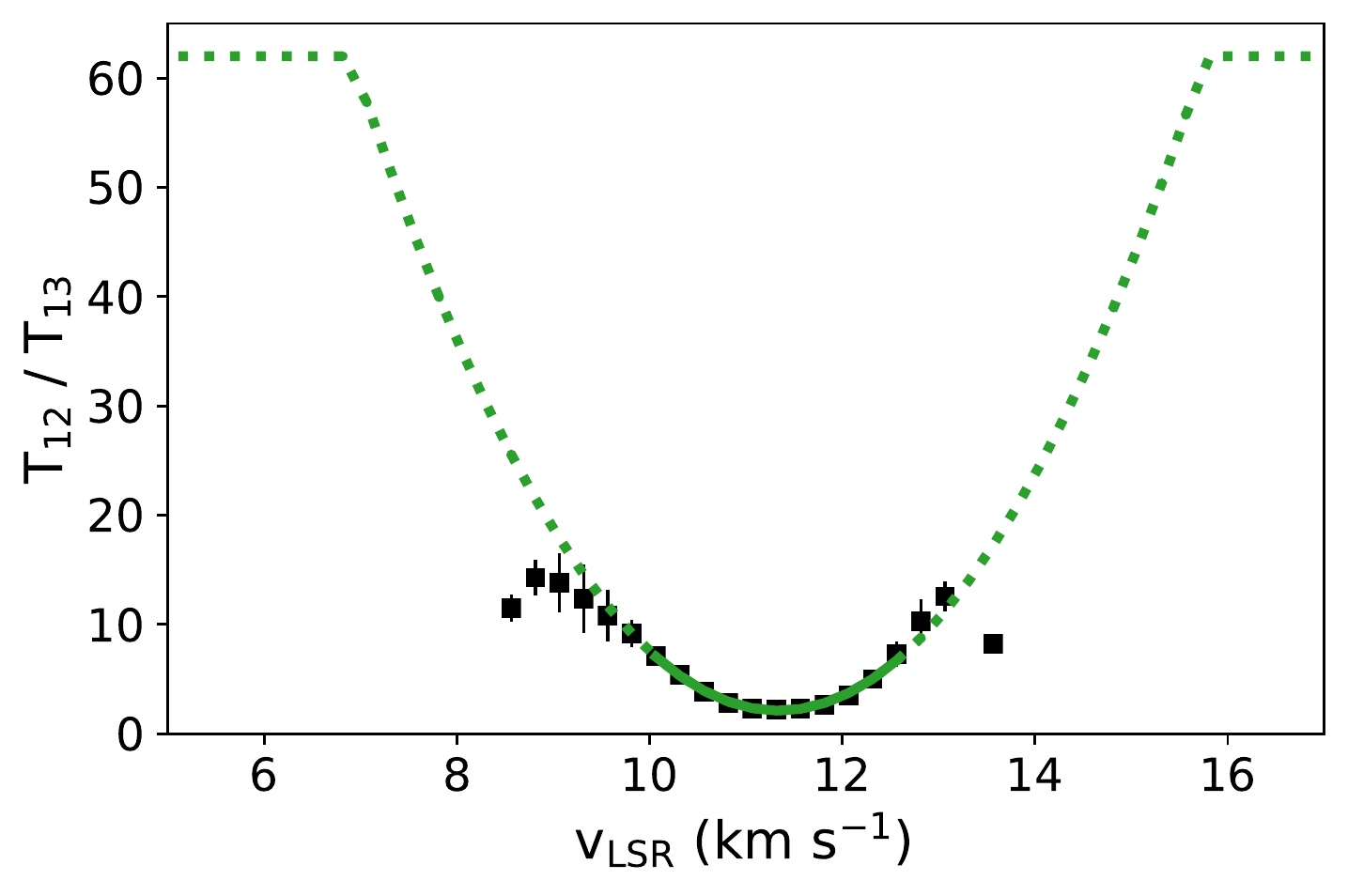}
\caption{\co[12]{}/\co[13]{} ratio of the \example{} outflow. The black points show the average ratio between \co[12]{} and \co[13]{} in each velocity channel using only pixels where both lines are detected at or above $5\sigma$. The error bars indicate the standard deviation of the ratio. The green parabola is the weighted least-squares fit to the ratio. The solid line shows the fitting range used, while the dotted line is an extrapolation of this fit. The parabola is capped at the assumed isotopic ratio of 62 (see Section~\ref{sec:opacity}). Note the uncertainty on the ratio at the most extreme velocities is likely underestimated because the only pixels at these velocities lie in a small region with a size similar to the beam. Thus, the standard deviation in these pixels is lower due to the correlation between pixels.\label{fig:opacity}}
\end{figure}

\co[13]{} is usually too weak to be detected more than 2-3 \kms{} away from the line core. Thus we extrapolate the measured ratio spectrum by fitting it with a 2nd-order polynomial, weighting each velocity channel by the standard deviation of the ratio in that channel. For most of the outflows, we use a fitting range of 1.5~\kms{} on either side of the minimum ratio. This fitting range is adjusted for the few outflows with multiple velocity components to ensure that the component corresponding to $v_{\rm sys}$ is fit. For each velocity, we use the value of this fit and Equation \ref{eq:Tratio} to calculate the correction factor $\tau_{12}$/(1 - $e^{-\tau_{12}}$) with which we multiply the observed \co[12]{}. Because the opacity correction factor cannot be less than unity for any value of $\tau_{12}$, the ratio spectrum fit is capped at the value of [\co[12]]/[\co[13]] and in this regime we consider \co[12]{} optically thin. In Figure~\ref{fig:opacity}, we show an example of the \co[12]{}/\co[13]{} ratio spectrum and fit for the \example{} outflow. 

\subsubsection{Outflow Mass}\label{sec:dmdv}
After correcting for opacity, we use the \co[12]{} emission to calculate the H$_2$ column density in each velocity channel. From Equation A6 in \citet{ZhangY16},

\begin{equation}\label{eq:dNdv}
\frac{dN}{dv} = \left(\frac{8\pi k \nu^2_{ul}}{h c^3 A_{ul} g_u}\right) Q_{\rm rot}(T_{\rm ex})~ e^{E_u / kT_{\rm ex}} \frac{T_R(v)}{f}
\end{equation}
where $\nu_{ul} = 115.271$ GHz is the frequency of the \co[12](1-0) transition, $A_{ul} = 7.203 \times 10^{-8}$~s$^{-1}$ is the Einstein A coefficient, $E_u/k = 5.53$~K is the energy of the upper level, $g_u = 3$ is the degeneracy of the upper level, $Q_{\rm rot}$ is the partition function (calculated to $J=100$), $T_{\rm ex}$ is the excitation temperature defined in Section~\ref{sec:tex}, $T_R(v)$ is the opacity-corrected brightness temperature of \co[12]{}, and $f$ is the abundance ratio of 
\co[12]/H$_{2}$. We assume an abundance ratio of $f = 1 \times 10 ^{-4}$ \citep{Frerking82}.

\begin{figure}
\plotone{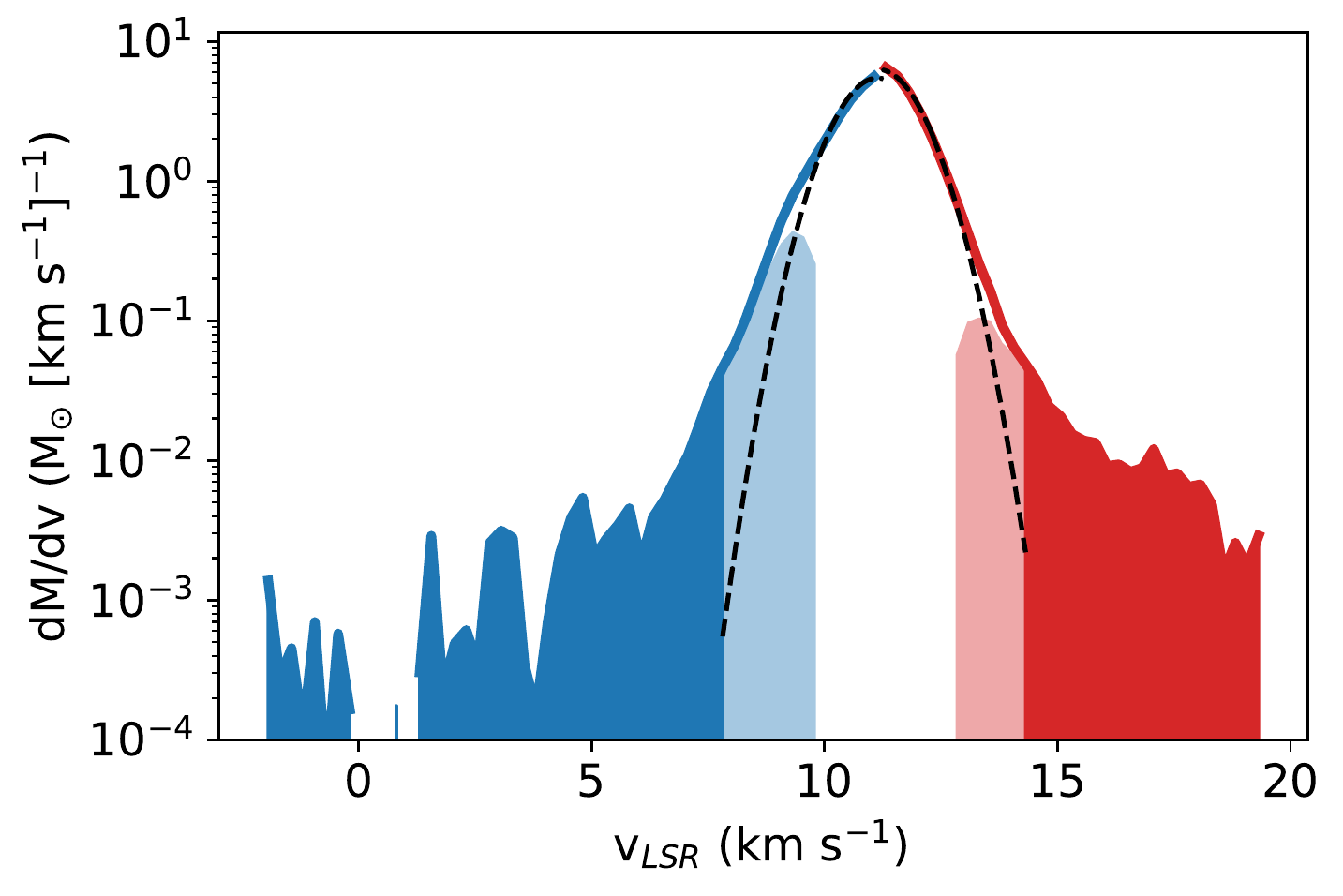}
\caption{Mass spectrum of the \example{} outflow. The blue (red) lines show the opacity-corrected mass spectrum in the blue (red) lobe region. The black dashed lines are Gaussian fits to these mass spectra. The dark blue (dark red) shaded regions indicate the region integrated to get the high-velocity mass, which is reported as the lower limit in Table~\ref{tab:physics}. The light blue (light red) shaded regions show the low-velocity mass after subtracting the cloud mass spectrum fit (see Section~\ref{sec:dmdv}).\label{fig:dmdv}}
\end{figure}

We then calculate the mass in each voxel:
\begin{equation}\label{eq:mass}
M = \mu_{\rm H_2} m_{\rm H} A_{\rm pixel} N_{\rm H_2},
\end{equation}
where $\mu_{\rm H_2} = 2.8$ is the mean molecular weight of H$_2$ \citep{Kauffmann08}, $m_{\rm H} = 1.674 \times 10^{-24}$~g is the mass of the hydrogen atom, and $A_{\rm pixel} = 1.6 \times 10^{-5}$~pc$^2$ is the spatial area subtended by each pixel at the distance of the cloud. In blue (red) outflow lobes we sum the total mass in each velocity channel blueward (redward) of $v_{\rm sys}$ and arrive at the outflow mass spectrum $dM/dv$. We only consider pixels above $3\sigma$ in a given channel and within their respective outflow lobe region in this mass calculation. Figure~\ref{fig:dmdv} shows an example mass spectrum for the \example{} outflow.

The total mass of each outflow lobe is obtained by integrating the mass spectrum over the relevant velocity range. For a lower limit on the mass, we consider only the high-velocity component: all velocity channels farther from $v_{\rm sys}$ than the minimum visually determined outflow velocity ($v_{\rm blue}$/$v_{\rm red}$). These velocities were chosen to include as much outflow emission as possible while avoiding contamination by the main cloud. However, if we only consider this high-velocity outflow material, we may miss a significant fraction of the total mass.

\subsubsection{Low-Velocity Outflow Emission}\label{sec:lowvelocity}
Outflows are conspicuous because of their high-velocity emission. But outflows also exist at lower velocities. \citet{Dunham14} note that escape velocities from protostars can be as low as 0.1 \kms. As these velocities are much lower than typical cloud CO line-widths, this low-velocity outflow material is often difficult to disentangle from the CO emission arising from the turbulent host molecular cloud.  

The low-velocity contribution to the total outflow mass can be quite significant. \citet{Dunham14} found that adding the inferred low-velocity emission increased the mass of outflows by a factor of 7.7 on average, with some outflows increasing by an order of magnitude or more. Outflow momentum and energy increased by factors of 3 to 5, on average. \citet{Offner11} found that using only high-velocity outflow emission underestimated the total outflow mass by a factor of 5 in synthetic observations of simulated outflows. Clearly, low-velocity emission should be accounted for when assessing the absolute impact of outflows on the cloud. We describe our method, adapted from \citet{Dunham14} for recovering this low-velocity outflow mass below. 

For each outflow lobe, we fit the opacity-corrected mass spectrum with a Gaussian. For ``low'' velocities between $v_{\rm sys}$ and the visually determined minimum outflow velocity ($v_{\rm blue}$ or $v_{\rm red}$), we subtract this Gaussian fit from the mass spectrum and define any excess mass as low-velocity outflow mass. To reduce the amount of extraneous cloud mass introduced with this method, we exclude all velocity channels within 1 \kms{} of $v_{\rm sys}$. Figure~\ref{fig:dmdv} demonstrates this procedure for the \example{} outflow. 

Generally, the low-velocity outflow mass is significantly greater than the mass at high velocities. Because this method assumes the cloud mass spectrum is fitted well by a Gaussian, we expect that the low-velocity mass will often be contaminated by ambient cloud material. Thus, for each outflow lobe, we consider the high-velocity outflow mass to be a lower limit and the high-velocity plus low-velocity mass to be an upper limit on the total outflow lobe mass. We report these mass ranges for each outflow lobe in Table~\ref{tab:physics}.

\subsubsection{Momentum and Kinetic Energy}

We define the momentum per velocity channel to be $dP/dv = (dM/dv) v_{\rm out}$, where $dM/dv$ is the mass spectrum discussed in Section~\ref{sec:dmdv} and $v_{\rm out}$ is the velocity relative to $v_{\rm sys}$. Similarly, the kinetic energy per velocity channel is $dE/dv = (1/2) (dM/dv)  v_{\rm out}^2$. We sum the momentum and kinetic energy separately for low velocities, with the ambient cloud corrected mass spectrum, and high velocities. In Table~\ref{tab:physics}, we report the momentum and kinetic energy of each outflow lobe.

\subsubsection{Dynamical Time}\label{sec:tdyn}
\begin{figure}
\plotone{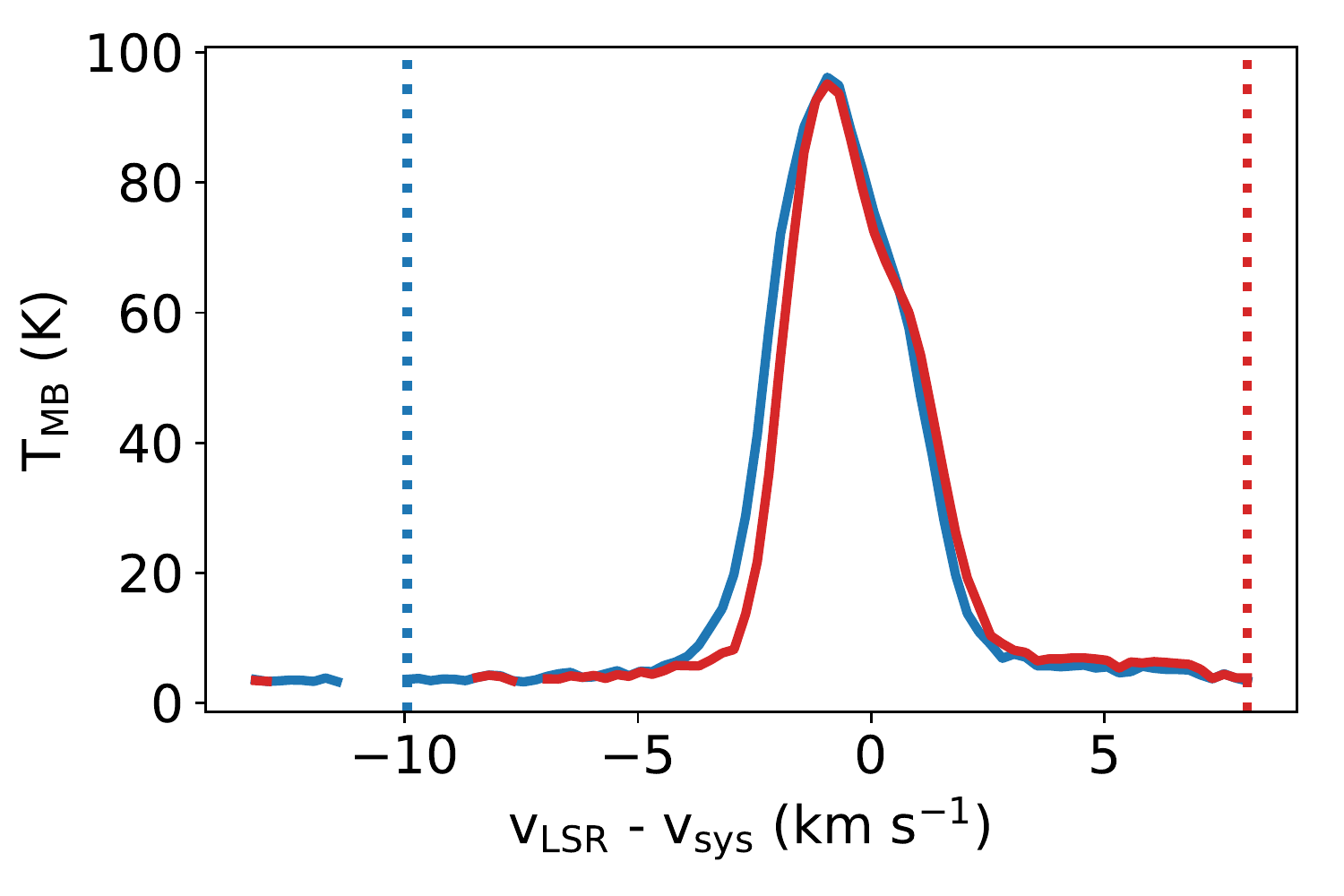}
\caption{Maximum velocity of the \example{} outflow. The blue (red) lines show the mean \co[12]{} spectrum in the blue (red) lobe regions. The blue (red) dotted line marks the maximum blue (red) outflow velocity, defined as the first channel where \co[12]{} is not detected at $3\sigma$ (see Section~\ref{sec:tdyn}).\label{fig:vmax}}
\end{figure}

We use the same method as \citet{Curtis10} to estimate the dynamical time $t_{\rm dyn}$ of each outflow lobe. Assuming the outflow has been expanding uniformly at the same velocity since it was launched, $t_{\rm dyn} = R / v_{\rm max}$, where $R$ is the length of the outflow and $v_{\rm max}$ is the maximum outflow velocity. We define $R$ to be the projected distance from the outflow source to the farthest part of the outflow lobe and $v_{\rm max}$ as the minimum velocity relative to $v_{\rm sys}$ where \co[12]{} is not detected at $3\sigma$. Some outflows are detectable all the way to the limits of the \co[12]{} spectral coverage (-2~\kms{} in the blue, 20~\kms{} in the red, relative to the LSR). In these cases, $v_{\rm max}$ is a lower limit, as are the mass and mass-derived properties. Figure~\ref{fig:vmax} shows our determination of $v_{\max}$ for the \example{} outflow. We report $R$, $v_{\max}$, and $t_{\rm dyn}$, for each outflow lobe in Table~\ref{tab:physics}.

We also calculate the mass loss rate, momentum injection rate, and energy injection rate by dividing the outflow mass, momentum, and kinetic energy by $t_{\rm dyn}$. In Section~\ref{sec:impact}, we use these quantities to compare the impact of outflows to turbulence in the cloud.

\subsubsection{Outflow Position Angle and Opening Angles\label{sec:angles}}
Most studies estimate the outflow position angle (PA) by eye \citep[e.g.][]{Morgan91,Takahashi08,Plunkett13,Stephens17,Kong19,Tanabe19}. We adopt a more reproducible and objective method to measure outflow position and opening angles (OA) modeled after the simulated outflow analysis carried out by \citet{Offner11}.

\begin{figure}
\plotone{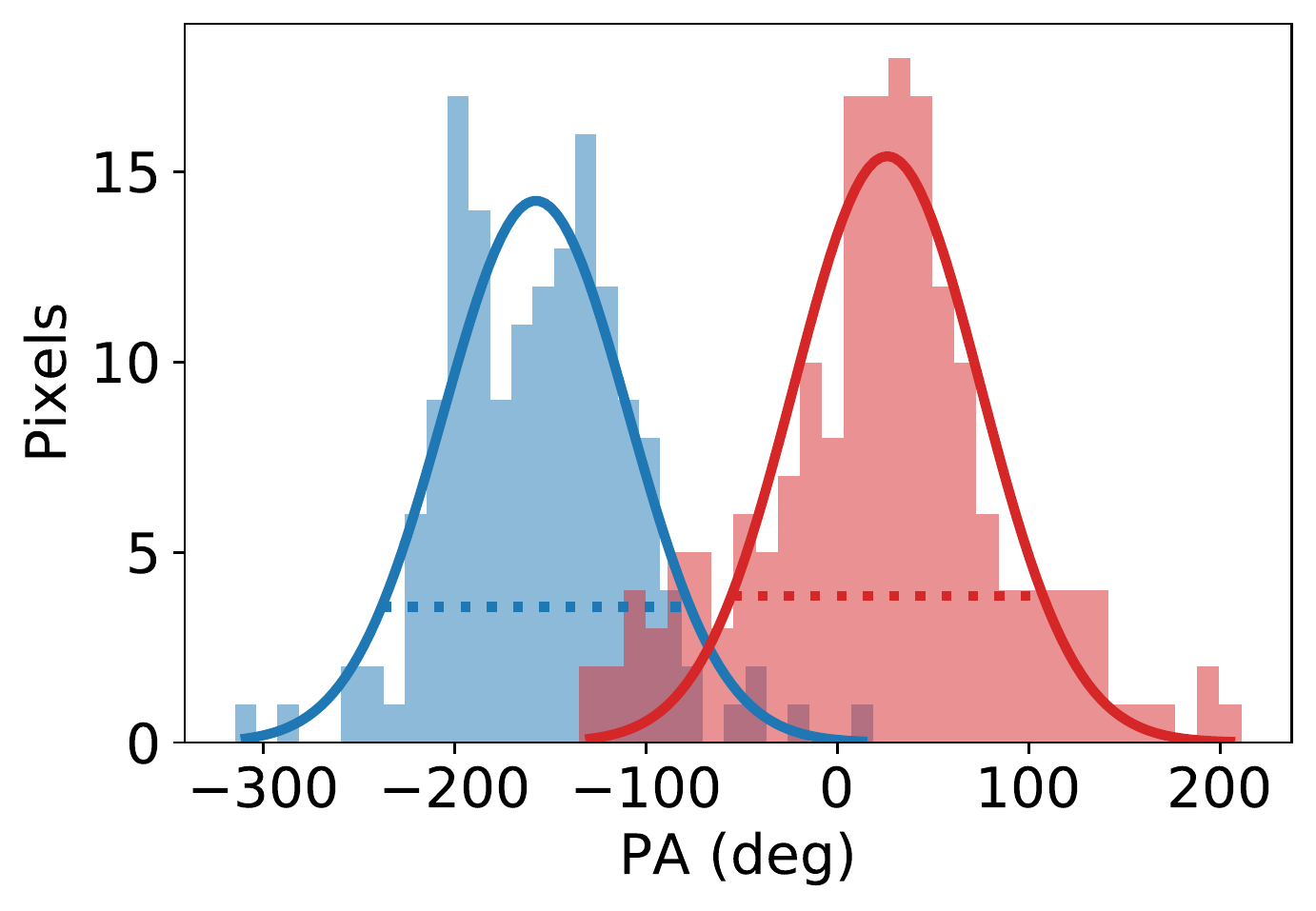}
\caption{Position angle and opening angle of \example{} outflow. The blue (red) histograms show the position angle distribution of all pixels in the blue (red) lobe region. The curves are Gaussian fits to these distributions. We adopt the mean of each lobe as the position angle and the full-width at quarter maximum (dotted lines) as the opening angle (see Section~\ref{sec:angles}).\label{fig:angles}}
\end{figure}

For each outflow lobe, we make an initial guess of the PA, measured counterclockwise (east) from the north celestial pole by convention. This initial guess is the angle from the outflow source to the peak of the integrated \co[12]{} over the velocity range of the outflow lobe. Then, we calculate the angle of every pixel in the outflow lobe relative to this initial guess. We fit the distribution of these angles with a Gaussian and define the mean of the Gaussian to be the PA of that outflow lobe. We define the OA of the outflow to be the full-width at quarter maximum (FWQM) of the Gaussian, following the definition by \cite{Offner11}. We find this automated method does a suitable job producing a similar PA and OA to a visual determination. Furthermore, when comparing the outflow PA with filament orientation (Section~\ref{sec:filaments}), we avoid the risk of an artificial correlation produced by unintentional measurement bias. Figure~\ref{fig:angles} shows the distribution of pixel position angles and Gaussian fit for the \example{} outflow. The PA and OA of each outflow lobe, along with their uncertainties from the Gaussian fit, are listed in Table~\ref{tab:angles}.

\subsection{Impact of Outflows on the Cloud}\label{sec:impact}
Protostellar outflows may be important in the maintenance of turbulence in clouds, at least at cluster scales \citep{Nakamura07}. The efficiency of turbulent driving by outflows is highly uncertain, depending on the transfer of momentum from outflowing gas to the cloud. To be even a plausible source of turbulence with perfect outflow-cloud coupling, the aggregate impact of outflows must be of a similar magnitude to the observed turbulent dissipation.

The total outflow momentum, kinetic energy, and their injection rates are given in the last row of Table~\ref{tab:physics}. We report lower and upper limits on these aggregate values by summing each outflow's high-velocity component and high-velocity + low-velocity components, respectively (see Section~\ref{sec:lowvelocity}). 

Te compare the aggregate outflow impact with the momentum and kinetic energy dissipation rates of the turbulent cloud, we follow the methods described in Sections~5.1.2 and 5.1.3 of \citet{Feddersen18}. Because we avoid the OMC 1 region in our outflow search, we remove the region with declination between $-5^\circ31{}^\prime44{}^{\prime\prime}$ and $-5^\circ15{}^\prime33{}^{\prime\prime}$, which corresponds to the gap in the HOPS protostar catalog. We select pixels where \co[13] is detected at 3$\sigma$ to calculate the total molecular mass, velocity dispersion, and kinetic energy. To calculate the dissipation time (Equation~5 in \citealp{Feddersen18}), we assume a cloud diameter of 12 pc (the geometric mean of the cloud's projected width and length) and calculate a mean one-dimensional velocity dispersion of 1.7 km~s$^{-1}$. To calculate the momentum dissipation rate (Equation 6 in \citealp{Feddersen18}), we assume a cloud radius of 6 pc and use the calculated total mass and median velocity dispersion. Excluding the OMC 1 region, we find a total momentum dissipation rate of \textbf{$\dot P_{\rm turb} = 1.3\times10^{-2}M_\odot~\rm{km}~\rm{s}^{-1}~\rm{yr}^{-1}$} and a total kinetic energy dissipation rate of \textbf{$\dot E_{\rm turb} = 1.4\times10^{34}~\rm{erg}~\rm{s}^{-1}$.}



We compare these turbulent dissipation rates to the aggregate outflow injection rates in Table~\ref{tab:physics}, $\dot P$ and $\dot E$. If we only account for the high-velocity outflow components, $\dot P$ is $14\%$ of $\dot P_{\rm turb}$ and $\dot E$ is $18\%$ of $\dot E_{\rm turb}$. If we add the low-velocity outflow emission, $\dot P$ is $51\%$ of $\dot P_{\rm turb}$ and $\dot E$ is $47\%$ of $\dot E_{\rm turb}$.

The outflow physical properties tabulated in Table~\ref{tab:physics} are not corrected for the inclination of outflows to the line-of-sight. Essentially, this means we assume that the radial velocity of outflow emission with respect to the source velocity is equivalent to the actual velocity of the outflowing gas. The closer an outflow is to the plane of the sky, the larger the discrepancy between observed and actual velocities. The outflow length $R_{\rm max}$, which we use to calculate $t_{\rm dyn}$ and all the derived injection rates, should also be corrected for the outflow inclination.

Because we do not know the actual inclination of each outflow, we estimate the effect of inclination on the aggregate outflow impact by assuming the same average inclination for every outflow. The average inclination angle, assuming any orientation is equally likely, is $57.3\degr$ (where $0\degr$ is a pole-on outflow).


\citet{Dunham14} summarize the inclination dependence of each outflow property in their Table 8. If all of the outflows are inclined $57.3\degr$, the total $\dot P$ ($\dot E$) will increase by a factor of 2.9 (5.3). Thus, after correcting for average inclination, $\dot P$ is  $41$-$148\%$ of $\dot P_{\rm turb}$ and $\dot E$ is $96$-$252\%$ of $\dot E_{\rm turb}$ (depending on whether low-velocity emission is included).

We note this inclination correction assumes that all outflow motions are along the axis of the flow, with no transverse motions. \citet{Dunham14} caution that the inclination correction factors may be significantly smaller if transverse motions are present, as demonstrated by simulations from \citet{Downes07}. The same simulations show that accounting for \emph{atomic} gas results in additional momentum and energy of about the same magnitude as the inclination corrections discussed above. While the absolute impact of outflows in Orion A remains highly uncertain, it is safe to say a significant fraction of the turbulent dissipation could be offset if outflows couple efficiently to the cloud. 

\section{Outflow-Filament Alignment}\label{sec:filaments}
Outflows are ejected along the angular momentum axis of the protostar. If mass accretion proceeds hierarchically, from larger scales of the cloud down to the protostellar scale, then the angular momentum axis of the protostar will trace the orientation of mass accretion flows \citep{Bodenheimer95}. Protostars tend to form along narrow filaments of dense gas \citep{Arzoumanian11}, and may accrete mass either along (parallel to) or across (perpendicular to) their host filaments, as seen for example in the simulations of \citet{Li18}. If one of these modes dominates protostellar mass accretion in filaments, we may expect to see a preferential direction of the angular momentum (and thus the outflow) with respect to the filament axis.

In simulations, \citet{Offner16} find that binaries formed via turbulent fragmentation of protostellar cores produce outflows with variable position angles, and they predict a random distribution of outflow orientations under this turbulent fragmentation model. \citet{Li18} find outflows form preferentially perpendicular to filaments in their simulations of a strongly magnetized cloud. 

The alignment between CO outflows and filaments has been studied in the Perseus molecular cloud by \citet{Stephens17} and in a massive infrared dark cloud (IRDC, \citealp{Kong19}). In Perseus, \citet{Stephens17} showed that a sample of 57 outflows are consistent with being randomly oriented with respect to the filament, neither parallel nor perpendicular. In the IRDC G28.37+0.07, \cite{Kong19} identified 64 outflows and showed that they are preferentially perpendicular to the filament axis. It remains to be seen whether this discrepancy arises from some meaningful difference between these clouds (e.g. evolutionary state or magnetic field strength) or by chance.

In Orion, \citet{Davis09} showed that H$_{2}$ outflows appear randomly oriented on the sky, showing no preferential alignment to the North-South integral-shaped filament. \citet{Tanabe19} studied the outflows in single-dish CO maps, finding no evidence for alignment between outflows and the large-scale filamentary structure in the cloud. However, the filamentary structure in Orion A is more complex than a single North-South integral-shaped filament. Therefore, we use the C$^{18}$O filaments identified by \citet{Suri19} in our analysis.

\subsection{C$^{18}$O Filaments}
\citet{Suri19} apply the Discrete Persistent Structures Extractor, DisPerSE \citep{Sousbie11} to extract filaments from the C$^{18}$O spectral cube. DisPerSE connects local maxima and saddle points in the intensity distribution, which are defined as filaments. \citet{Suri19} identify a total of 625 filaments across the Orion A cloud, each of which are defined by their PPV coordinates, allowing filaments that overlap spatially to be separated in velocity space. 

For each outflow source, we search for the closest filament. Because most of the outflow sources are located along lines of sight with a single significant C$^{18}$O velocity component, we ignore the filament velocity information and only consider projected distance on the sky. We use cubic spline interpolation to approximate the discrete filament coordinates with a smooth curve. For this spline interpolation, we used the \url{splrep} and \url{splev} functions from the \url{scipy.interpolate} package. 

\begin{figure}
\plotone{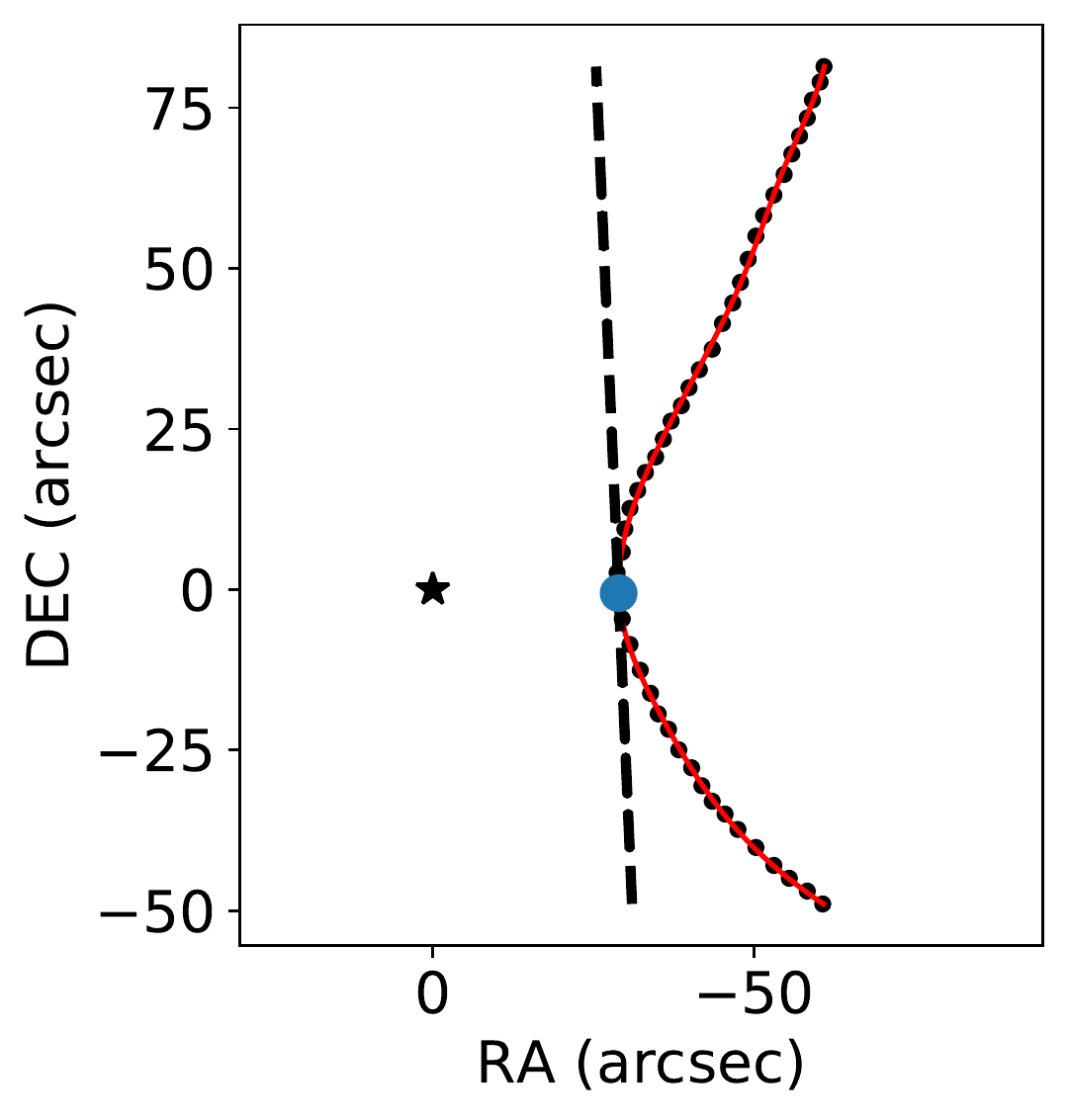}
\caption{C$^{18}$O filament near \example{}. The black points show the filament from the \citet{Suri19} catalog closest to \example{}. The red curve is a smooth cubic spline interpolation of the filament. The outflow source (\example{}) is marked with a black star. The blue circle shows the closest point on the filament to the source. To derive the filament position angle, we find the tangent at this nearest point, shown as a black dashed line (see Section~\ref{sec:filaments}).\label{fig:filament}}
\end{figure}

We take the slope of the tangent to the filament curve at the closest point to the outflow source to be the position angle of the filament, which is constrained to be between $-180$ and $180\degr$. Figure~\ref{fig:filament} shows an example of the spline interpolation and tangent fitting for the closest filament to the \example{} outflow.

\subsection{Projected Angle Between Outflows and Filaments}\label{sec:gamma}

\begin{figure}
\plotone{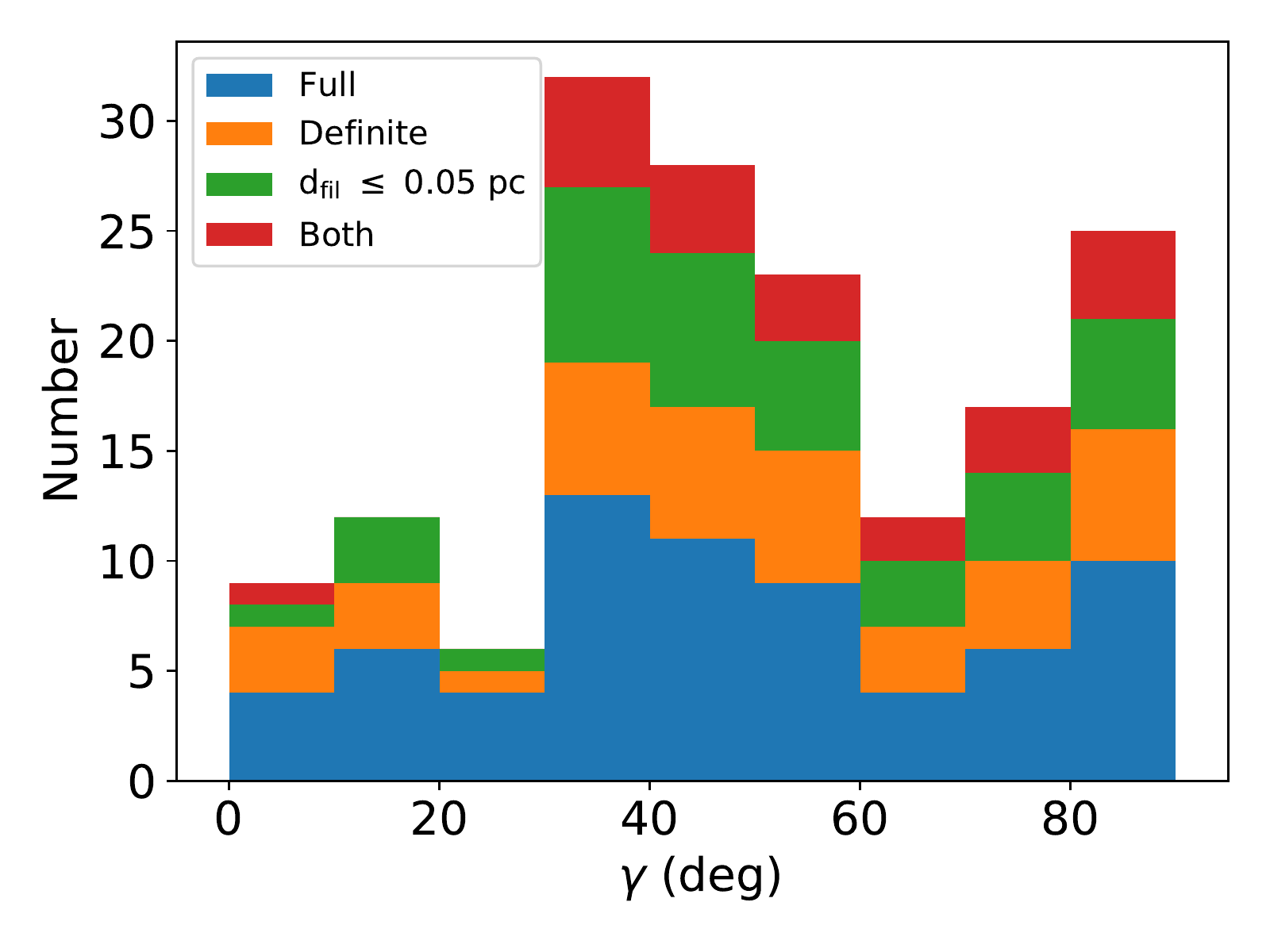}
\caption{Distribution of projected angle between outflows and filaments. The blue histogram shows the distribution of $\gamma$ in $10\degr$ bins for the full sample of \nlobes{} outflow lobes. The orange bars show the distribution of $\gamma$ for the subsample of 38 outflow lobes with a confidence grade of "Definite" (Table~\ref{tab:outflows}). The green bars show the subsample of 37 outflow lobes whose sources are within 0.05 pc of the nearest filament (Table~\ref{tab:angles}). The red bars show the subsample of 22 outflow lobes which satisfy both of these conditions. Note that the total height of the stacked bars does not equal the size of the full sample, since the subsamples are already contained within the full sample shown in blue. \label{fig:gamma_hist}}
\end{figure}

We follow \citet{Stephens17} and \citet{Kong19} in our definition of the angular separation between outflow and filament position angles, $\gamma$. For each outflow lobe, we define $\gamma$ to be 
\begin{equation}
    \gamma = \rm{MIN}\{|\rm{PA}_{\rm out} - \rm{PA}_{\rm fil}|, 180\degr - |\rm{PA}_{\rm out} - \rm{PA}_{\rm fil}|\},
\end{equation}
where PA$_{\rm out}$ and PA$_{\rm fil}$ are the position angles of the outflow lobe and the closest filament, respectively. The value of $\gamma$ for each outflow lobe is given in Table~\ref{tab:angles}. Figure~\ref{fig:gamma_hist} shows the distribution of $\gamma$. The full sample of $\gamma$ shows no obvious clustering at either 0 or 90$\degr$.

Many of the outflows in our catalog are not as clear as our example outflow driven by \example{}. Outflows categorized as "Marginal" in Table~\ref{tab:outflows} are likely to have less reliable position angles. In these cases, we either have difficulty disentangling the high-velocity emission near these sources and/or we are unsure of which protostar is driving the outflow. Both of these factors could greatly affect the measured position angle and, by extension, $\gamma$. \citet{Stephens17} argue that the incorrect assignment of driving sources led \citet{Anathpindika08} to erroneously conclude that outflows and filaments are perpendicular in the Perseus molecular cloud.

To test whether more reliable outflows have a different distribution of $\gamma$, we consider a subsample of outflow lobes which we consider "Definite" (Table~\ref{tab:outflows}). Figure~\ref{fig:gamma_hist} shows that these outflows are not distributed significantly differently from the full sample.

Aside from the uncertainties in calculating the outflow position angles discussed above, we consider the difficulty of assigning even a well-known outflow to a particular filament. Even though we select for the closest filament to each outflow source, we cannot exclude the possibility that other nearby filaments may be important to the environment of an outflow. In particular, the OMC-4 region contains spatially overlapping filaments which are distinct in velocity space \citep{Suri19} similar to the fibers identified by \citet{Hacar13}. The incorrect assignment of outflows to filaments may introduce noise to the distribution of $\gamma$ and mask an underlying correlation between outflow and filament orientations. To address this concern, we compare the full sample to the closest outflow-filament pairs. We adopt a threshold on the projected outflow-filament distance ($d_{\rm fil}$) based on the typical filament width.

\citet{Suri19} found an average filament FWHM of approximately 0.1~pc, similar to the ``characteristic'' filament width found using \emph{Herschel} dust continuum maps \citep[e.g.][]{Arzoumanian11,Koch15} However, \citet{Suri19} find a much larger spread in filament widths (about an order of magnitude around the mean). This is likely due to the fact that they allow the filament width to vary along its length, while most studies average the width over the entire filament. Motivated by the mean filament width, we choose a threshold of $d_{\rm fil} \leq 0.05$~pc, which corresponds to an outflow source within the FWHM of an average filament. Figure~\ref{fig:gamma_hist} shows the distribution of $\gamma$ for this subset of outflows.

Figure~\ref{fig:gamma_hist} shows that limiting the sample to the closest outflow-filament pairs with the highest confidence reduces the fraction of outflows at projected angles of 0-30\degr{} with respect to their filaments. We stress that these angles are projected on the plane of the sky. To determine whether these outflows are preferentially aligned with the filaments, we must consider the underlying distribution of deprojected $\gamma$: the outflow-filament alignment in 3D.

\subsection{3D Outflow-Filament Alignment}
An outflow-filament pair can appear at various relative orientations on the plane of the sky, depending on the line-of-sight. For example, an outflow observed parallel to a nearby filament may actually be perpendicular in space. Thus, we follow \citet{Stephens17} and \citet{Kong19} and run Monte Carlo simulations\footnote{See Appendix A of \citet{Stephens17} for a detailed description of the Monte Carlo simulations.} of random vector pairs to project different underlying distributions of $\gamma_{\rm 3D}$ onto the plane of the sky.

We first generate $10^7$ pairs of unit vectors uniformly distributed around the unit sphere. From this random uniform distribution, we make two subsets. In the ``parallel'' subset, we keep only those vector pairs separated by 0-20\degr. In the ``perpendicular'' subset, we keep the pairs separated by 70-90\degr. We also consider the full random uniform distribution of vector pairs, dubbed ``random''. Then, we project these vectors onto the plane of the sky by setting one coordinate to 0 and calculate the projected angle between them, $\gamma$. By comparing the distribution of observed $\gamma$ with the distribution of $\gamma$ in the simulated random, parallel, and perpendicular sets of vector pairs, we investigate which of these underlying scenarios is most likely given the observations.

\begin{figure*}
\epsscale{0.7}
\plotone{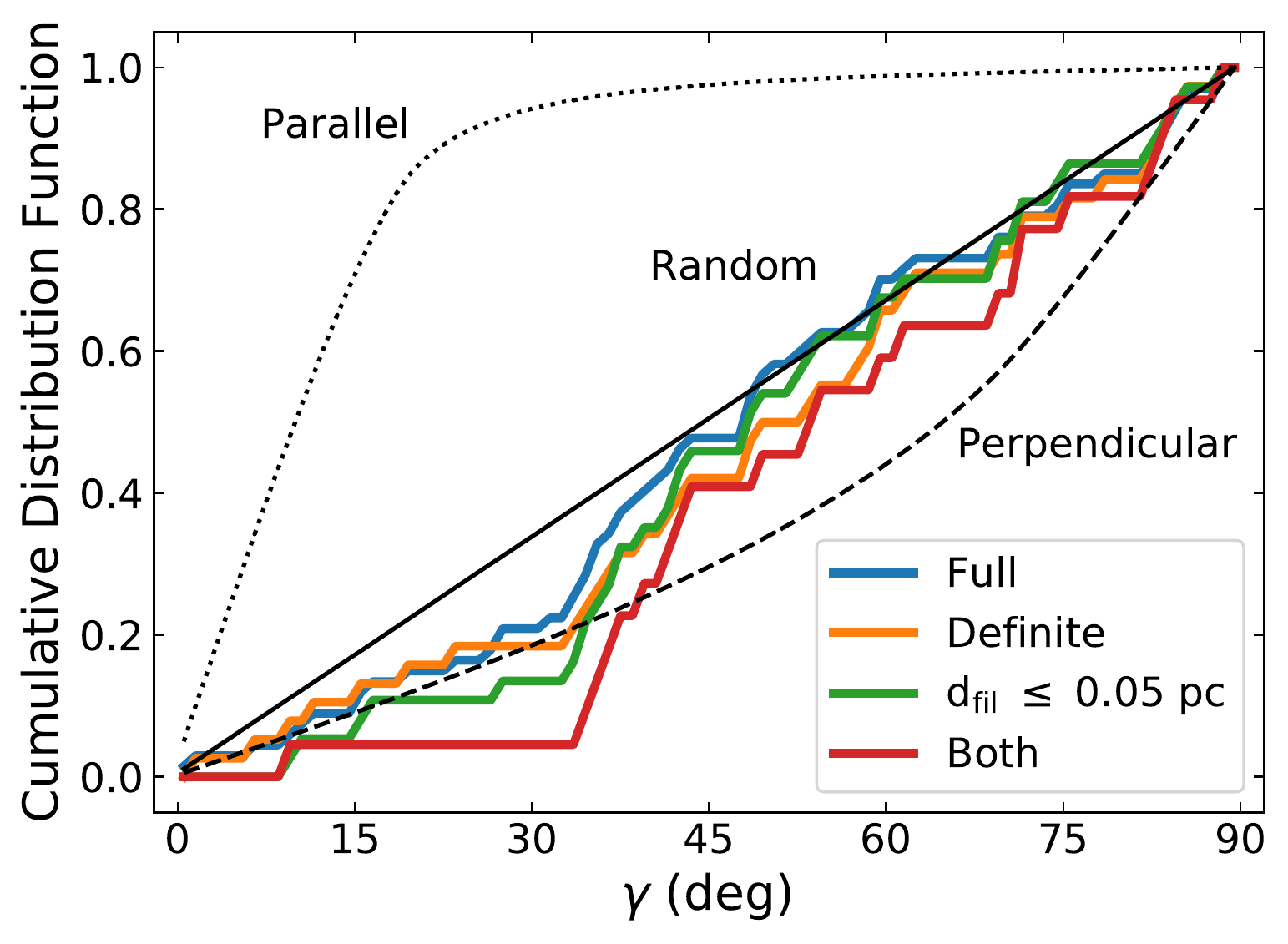}
\caption{Cumulative distribution of projected angle between outflows and filaments. The colored lines show the distribution of $\gamma$ for different subsets of our outflow catalog. The black lines show the expected distribution given three different model distributions. The parallel case contains only 3D angles between 0 and 20$\degr$, the random case contains all 3D angles between 0 and 90$\degr$, and the perpendicular case contains 3D angles between 70 and 90$\degr$. The results of Anderson-Darling tests between the observed and modeled distributions are in Table~\ref{tab:ad} (see Section~\ref{sec:gamma}).\label{fig:gamma_cdf}}
\end{figure*}

Figure~\ref{fig:gamma_cdf} shows the cumulative distribution of $\gamma$ for the Monte Carlo simulations and each of the outflow-filament samples discussed in Section~\ref{sec:gamma}. Compared to the random distribution of $\gamma$, the outflow samples have a deficit at $\gamma < 40\degr$. In particular, the "Both" subsample with the 22 definite outflow lobes closest to their filaments contains only one lobe with $\gamma < 33\degr$ compared to the $\sim8$ which would be expected if $\gamma$ were distributed randomly.

While Figure~\ref{fig:gamma_cdf} shows that the "Parallel" scenario is clearly inconsistent with the observed distribution of $\gamma$, the "Perpendicular" distribution is more difficult to rule out. We apply the Anderson-Darling (AD) test \citep{Stephens74} to determine if the perpendicular or random distributions can be rejected for each outflow subsample. The AD $p$-values listed in Table~\ref{tab:ad} represent the likelihood that the observed outflow-filament $\gamma$ distribution is drawn from either the random or perpendicular distribution. We do not report the results of AD tests for the parallel case; the $p$-values for the parallel distribution are all close to zero.

Based on the AD test, we reject the hypothesis that the full outflow sample is drawn from a perpendicular distribution with $>99.9\%$ confidence ($p < 0.001$), while the random distribution is not ruled out. For the definite subsample the results are similar, though we can only rule out the perpendicular distribution with $94\%$ confidence ($p = 0.06$). For the $d_{\rm fil} \leq 0.05~\rm{pc}$ subsample, we can rule out the perpendicular distribution with $98\%$ confidence ($p = 0.02$). Though these subsamples are unlikely to be drawn from a purely perpendicular distribution, their $p$-values are higher than the equivalent in \citet{Stephens17}. We do not test models with a mixture of perpendicular and parallel outflow-filament alignment, but such a bimodal distribution may be a better fit to our data than either random or perpendicular model alone.

If we limit the sample to the definite outflows with $d_{\rm fil} \leq 0.05~\rm{pc}$ ("Both"), and compare to the perpendicular model, we calculate an AD $p$-value of $> 0.25$. While we cannot rule out with better than $94\%$ confidence ($p = 0.06$) that this subsample of outflows is drawn from a random distribution, the perpendicular model more closely matches the angular distribution of these close outflow-filament pairs. We tentatively conclude that the clearest outflow-filament pairs in Orion A are perpendicularly aligned.

In Perseus, \citet{Stephens17} found a random distribution of outflow-filament angles. This random alignment could reflect processes at large or small scales. At large scales, the orientation of filaments may be irrelevant to the accretion of mass onto them, contradicting simulations of filament formation \citep[e.g.][]{Chen14,Clarke17}. At core scales, no matter the orientation of mass accretion, the protostellar angular momentum may be randomized by multiplicity or turbulence, as seen in simulations by \citet{Offner16} and \citet{Lee17}. Our finding of a moderately perpendicular alignment between outflows and filaments in Orion A may mean that filament formation or protostellar mass accretion is different in Perseus and Orion.

In the IRDC G28, \citet{Kong19} found a remarkably perpendicular outflow-filament alignment, rejecting a random distribution at high confidence (see Table~\ref{tab:ad}). Such a strong perpendicular alignment is predicted in simulations of IRDCs by \citet{Li18} and \citet{Li19} resulting from filament formation perpendicular to a strong magnetic field and continuous mass accretion along the magnetic field lines onto protostellar cores. As IRDCs are generally considered to be in the earliest phase of massive star formation \citep{Rathborne06}, the outflow-filament alignment may be related to evolutionary stage or magnetic field. A comprehensive study of outflow-filament alignment across many different star forming environments is necessary to answer this question.

While we find some evidence for perpendicular outflow-filament alignment in Orion A, previous studies have concluded that outflows and filaments in this cloud are randomly aligned. \citet{Davis09} show that the position angles of H$_2$ jets in Orion A are distributed uniformly and oriented randomly with respect to the $\sim$pc-scale integral shaped filament. \citet{Tanabe19} conclude the same using their CO outflow catalog. Our study is not directly comparable to these, as we use the C$^{18}$O filaments, which reveal the integral-shaped filament to be made up of many smaller filamentary structures.

The moderately perpendicular outflow-filament alignment may also be a result of observational biases. If we preferentially miss outflows that are parallel to their filaments, this could bias the outflow-filament alignment. An outflow launched parallel to its host filament will encounter more dense gas than a perpendicularly launched outflow. Thus, parallel outflows may be smaller or slower and thus harder to detect. This effect is seen in the simulations of \citet{Offner11}. If this bias is present, we would expect to detect more parallel outflows with higher resolution observations. The physical resolution of the Perseus observations used by \citet{Stephens17} is about 4$\times$ better than the CARMA-NRO Orion A survey used here. This could help explain the deficit of parallel outflows seen in Orion relative to Perseus. The \cite{Kong19} IRDC observations have a similar physical resolution to our data, so this observational bias cannot explain the difference between the outflow-filament alignment in the IRDC and Orion.

\section{Conclusions}\label{sec:conclusions}

We have identified 45 outflows in the CARMA-NRO Orion CO maps of Orion A. Eleven of these outflows are new detections. For the previously known outflows, we improve the earlier estimates of their mass by including a correction for low-velocity mass as well as a velocity-dependent opacity correction.
The outflows contain significant momentum and kinetic energy compared to estimates for the turbulent dissipation in Orion A. If outflows couple efficiently to the cloud, they can maintain cloud turbulence and slow star-formation. There is still considerable uncertainty in the outflow impact and a mechanism for transporting momentum from the outflow length scale to the larger cloud is needed (see \citealt{Offner18} for an option).

We compare the outflow position angles to the orientation of nearby filaments from the C$^{18}$O catalog of \citet{Suri19}. The full outflow catalog is consistent with random outflow-filament alignment. The most reliable outflows which are closest to their filaments show a moderately perpendicular outflow-filament alignment.


While we improve the specificity of the outflow-filament comparison compared to previous studies of Orion A by using the C$^{18}$O filament catalog, there is still uncertainty in pairing outflows and filaments. Future work should investigate the outflow-filament alignment over multiple length scales, and take into account the varying filament width.

The outflow-filament alignment may change as protostars evolve. A detailed comparison should be made between this alignment and protostellar properties such as bolometric temperature, multiplicity, and evolutionary stage. A combination of this sample with the studies in other clouds could provide enough statistical power to answer these more specific questions about outflow-filament alignment and the mass assembly of protostars.

\acknowledgements
We are grateful to the referee, Neal Evans, for his valuable suggestions and to Ian Stephens, Cheng-Han Hsieh, and the CARMA-NRO Orion collaboration for discussion that improved this study.

 CARMA operations were supported by the California Institute of Technology, the University of California-Berkeley, the University of Illinois at Urbana-Champaign, the University of Maryland College Park, and the University of Chicago. The Nobeyama 45 m telescope is operated by the Nobeyama Radio Observatory, a branch of the National Astronomical Observatory of Japan. J.R.F, H.G.A., and S.K. were funded by the National Science Foundation awards AST-1140063 and AST-1714710.

\software{Astropy \citep{Astropy-Collaboration13}, spectral-cube \citep{Robitaille16}}
\facilities{CARMA, No:45m, CDS}

\clearpage
\startlongtable
\begin{deluxetable*}{cccccc}
\tablecaption{Outflow Catalog.\label{tab:outflows}}
\tablehead{\colhead{Source\tablenotemark{a}} & \colhead{R.A.} & \colhead{Decl.} & \colhead{$v_{\rm blue}$/$v_{\rm red}$\tablenotemark{b}} & \colhead{Confidence\tablenotemark{c}} & \colhead{Tanabe}\\
\colhead{} & \colhead{(J2000)} & \colhead{(J2000)} & \colhead{(km s$^{-1}$)} & \colhead{} & \colhead{}}
\startdata
SMZ 11 & $5^\mathrm{h}35^\mathrm{m}23.30^\mathrm{s}$ & $-5^\circ07{}^\prime10.00{}^{\prime\prime}$ & 7.5/- & M/- & 9 \\
SMZ 17 & $5^\mathrm{h}35^\mathrm{m}27.00^\mathrm{s}$ & $-5^\circ09{}^\prime54.00{}^{\prime\prime}$ & 3/18 & D/D &  \\
SMZ 21 & $5^\mathrm{h}35^\mathrm{m}26.90^\mathrm{s}$ & $-5^\circ11{}^\prime07.00{}^{\prime\prime}$ & -/14 & -/M & 15 \\
SMZ 30 & $5^\mathrm{h}35^\mathrm{m}18.30^\mathrm{s}$ & $-5^\circ31{}^\prime42.00{}^{\prime\prime}$ & 4.5/- & M/- & 20 \\
SMZ 50 & $5^\mathrm{h}36^\mathrm{m}11.50^\mathrm{s}$ & $-6^\circ22{}^\prime22.00{}^{\prime\prime}$ & 4.5/- & M/- & 33 \\
HOPS 10 & $5^\mathrm{h}35^\mathrm{m}09.00^\mathrm{s}$ & $-5^\circ58{}^\prime27.48{}^{\prime\prime}$ & -/12.6 & -/D & 27 \\
HOPS 11 & $5^\mathrm{h}35^\mathrm{m}13.42^\mathrm{s}$ & $-5^\circ57{}^\prime57.96{}^{\prime\prime}$ & 4.7/12.8 & D/D & 26 \\
HOPS 12 & $5^\mathrm{h}35^\mathrm{m}08.59^\mathrm{s}$ & $-5^\circ55{}^\prime54.12{}^{\prime\prime}$ & 4.7/- & D/- & 25 \\
HOPS 44 & $5^\mathrm{h}35^\mathrm{m}10.58^\mathrm{s}$ & $-5^\circ35{}^\prime06.36{}^{\prime\prime}$ & 4.4/- & M/- &  \\
HOPS 50 & $5^\mathrm{h}34^\mathrm{m}40.90^\mathrm{s}$ & $-5^\circ31{}^\prime44.40{}^{\prime\prime}$ & 5.8/- & M/- & 21 \\
HOPS 56 & $5^\mathrm{h}35^\mathrm{m}19.46^\mathrm{s}$ & $-5^\circ15{}^\prime32.76{}^{\prime\prime}$ & -/14 & -/M & 19 \\
HOPS 58 & $5^\mathrm{h}35^\mathrm{m}18.50^\mathrm{s}$ & $-5^\circ13{}^\prime38.28{}^{\prime\prime}$ & -/14 & -/M &  \\
HOPS 59 & $5^\mathrm{h}35^\mathrm{m}20.14^\mathrm{s}$ & $-5^\circ13{}^\prime15.60{}^{\prime\prime}$ & 6.9/- & M/- & 17 \\
HOPS 60 & $5^\mathrm{h}35^\mathrm{m}23.33^\mathrm{s}$ & $-5^\circ12{}^\prime03.24{}^{\prime\prime}$ & 7.0/14.2 & D/D & 16 \\
HOPS 68 & $5^\mathrm{h}35^\mathrm{m}24.31^\mathrm{s}$ & $-5^\circ08{}^\prime30.48{}^{\prime\prime}$ & 7.8/14.5 & D/D & 12 \\
HOPS 70 & $5^\mathrm{h}35^\mathrm{m}22.42^\mathrm{s}$ & $-5^\circ08{}^\prime04.92{}^{\prime\prime}$ & -/14 & -/M & 11 \\
HOPS 71 & $5^\mathrm{h}35^\mathrm{m}25.61^\mathrm{s}$ & $-5^\circ07{}^\prime57.36{}^{\prime\prime}$ & 7.5/- & D/- & 10 \\
HOPS 75 & $5^\mathrm{h}35^\mathrm{m}26.66^\mathrm{s}$ & $-5^\circ06{}^\prime10.44{}^{\prime\prime}$ & -/14.3 & -/M & 8 \\
HOPS 78 & $5^\mathrm{h}35^\mathrm{m}25.82^\mathrm{s}$ & $-5^\circ05{}^\prime43.80{}^{\prime\prime}$ & 6/15 & D/D & 7 \\
HOPS 81 & $5^\mathrm{h}35^\mathrm{m}27.96^\mathrm{s}$ & $-5^\circ04{}^\prime58.08{}^{\prime\prime}$ & 7.5/13.5 & M/M &  \\
HOPS 84 & $5^\mathrm{h}35^\mathrm{m}26.57^\mathrm{s}$ & $-5^\circ03{}^\prime55.08{}^{\prime\prime}$ & 8.5/13.8 & D/D & 6 \\
HOPS 87 & $5^\mathrm{h}35^\mathrm{m}23.47^\mathrm{s}$ & $-5^\circ01{}^\prime28.56{}^{\prime\prime}$ & 8.6/13.8 & M/M & 5 \\
HOPS 88 & $5^\mathrm{h}35^\mathrm{m}22.44^\mathrm{s}$ & $-5^\circ01{}^\prime14.16{}^{\prime\prime}$ & 8.1/13.8 & D/D & 4 \\
HOPS 92 & $5^\mathrm{h}35^\mathrm{m}18.31^\mathrm{s}$ & $-5^\circ00{}^\prime33.12{}^{\prime\prime}$ & 7.6/13.7 & D/D & 3 \\
HOPS 96 & $5^\mathrm{h}35^\mathrm{m}29.71^\mathrm{s}$ & $-4^\circ58{}^\prime48.72{}^{\prime\prime}$ & 9.8/13.9 & D/D & 1 \\
HOPS 99 & $5^\mathrm{h}34^\mathrm{m}29.50^\mathrm{s}$ & $-4^\circ55{}^\prime30.72{}^{\prime\prime}$ & 8.3/- & M/- &  \\
HOPS 157 & $5^\mathrm{h}37^\mathrm{m}56.57^\mathrm{s}$ & $-6^\circ56{}^\prime39.12{}^{\prime\prime}$ & 3.5/- & M/- &  \\
HOPS 158 & $5^\mathrm{h}37^\mathrm{m}24.46^\mathrm{s}$ & $-6^\circ58{}^\prime32.88{}^{\prime\prime}$ & 5/9.5 & M/M &  \\
HOPS 160 & $5^\mathrm{h}37^\mathrm{m}51.05^\mathrm{s}$ & $-6^\circ47{}^\prime20.40{}^{\prime\prime}$ & 4.7/- & D/- &  \\
HOPS 166 & $5^\mathrm{h}36^\mathrm{m}25.13^\mathrm{s}$ & $-6^\circ44{}^\prime41.64{}^{\prime\prime}$ & 5.5/11.5 & M/D & 40 \\
HOPS 168 & $5^\mathrm{h}36^\mathrm{m}18.94^\mathrm{s}$ & $-6^\circ45{}^\prime22.68{}^{\prime\prime}$ & 4.9/12.1 & D/D & 41 \\
HOPS 169 & $5^\mathrm{h}36^\mathrm{m}36.12^\mathrm{s}$ & $-6^\circ38{}^\prime51.72{}^{\prime\prime}$ & 4.7/10 & D/D & 39 \\
HOPS 174 & $5^\mathrm{h}36^\mathrm{m}25.85^\mathrm{s}$ & $-6^\circ24{}^\prime58.68{}^{\prime\prime}$ & 4/- & M/- & 36 \\
HOPS 177 & $5^\mathrm{h}35^\mathrm{m}50.02^\mathrm{s}$ & $-6^\circ34{}^\prime53.40{}^{\prime\prime}$ & -/10.3 & -/D & 37 \\
HOPS 178 & $5^\mathrm{h}36^\mathrm{m}24.60^\mathrm{s}$ & $-6^\circ22{}^\prime41.16{}^{\prime\prime}$ & 4.3/- & D/- & 34 \\
HOPS 179 & $5^\mathrm{h}36^\mathrm{m}21.84^\mathrm{s}$ & $-6^\circ23{}^\prime29.76{}^{\prime\prime}$ & 4.5/11.9 & D/M & 35 \\
HOPS 181 & $5^\mathrm{h}36^\mathrm{m}19.51^\mathrm{s}$ & $-6^\circ22{}^\prime12.36{}^{\prime\prime}$ & 4/12 & D/D & 32 \\
HOPS 182 & $5^\mathrm{h}36^\mathrm{m}18.84^\mathrm{s}$ & $-6^\circ22{}^\prime10.20{}^{\prime\prime}$ & 4/11.7 & D/D & 31 \\
HOPS 192 & $5^\mathrm{h}36^\mathrm{m}32.45^\mathrm{s}$ & $-6^\circ01{}^\prime16.32{}^{\prime\prime}$ & 7.5/10 & M/M &  \\
HOPS 198 & $5^\mathrm{h}35^\mathrm{m}22.18^\mathrm{s}$ & $-6^\circ13{}^\prime06.24{}^{\prime\prime}$ & -/10 & -/M &  \\
HOPS 203 & $5^\mathrm{h}36^\mathrm{m}22.85^\mathrm{s}$ & $-6^\circ46{}^\prime06.24{}^{\prime\prime}$ & -/11.7 & -/D & 42 \\
HOPS 355 & $5^\mathrm{h}37^\mathrm{m}17.09^\mathrm{s}$ & $-6^\circ49{}^\prime49.44{}^{\prime\prime}$ & 4.3/10.5 & M/M &  \\
HOPS 368 & $5^\mathrm{h}35^\mathrm{m}24.72^\mathrm{s}$ & $-5^\circ10{}^\prime30.36{}^{\prime\prime}$ & 7/14 & D/D & 14 \\
HOPS 370 & $5^\mathrm{h}35^\mathrm{m}27.62^\mathrm{s}$ & $-5^\circ09{}^\prime33.48{}^{\prime\prime}$ & 5/17 & M/M & 13 \\
HOPS 383 & $5^\mathrm{h}35^\mathrm{m}29.81^\mathrm{s}$ & $-4^\circ59{}^\prime51.00{}^{\prime\prime}$ & 9.6/- & D/- &2
\enddata
\tablenotetext{a}{HOPS sources are protostars in the catalog from \citet{Furlan16}. SMZ sources are sources of H$_2$ outflows in \citet{Davis09} which are not in the HOPS catalog.}
\tablenotetext{b}{These are the visually determined velocities closest to the mean cloud velocity which most clearly separate the outflow lobes from the surrounding cloud. Entries marked with '-' indicate that one of the outflow lobes is not detected. See Section~\ref{sec:identification} for details.}
\tablenotetext{c}{Each entry has two values, referring to the blue/red lobes separately. Entries marked 'D' for Definite, are clearly outflows. Entries marked 'M' for Marginal are unclear. See Section~\ref{sec:identification} for details.}
\end{deluxetable*}

\clearpage
\startlongtable
\begin{deluxetable*}{ccccccccccc}
\tabletypesize{\footnotesize}
\centerwidetable
\tablecaption{Outflow Physical Properties\label{tab:physics}}
\tablehead{\colhead{Source} & \colhead{Lobe} & \colhead{$M$\tablenotemark{\footnotesize a}}& \colhead{$P$} & \colhead{$E$} & \colhead{$R_{\rm max}$} & \colhead{$v_{\rm max}$} & \colhead{$t_{\rm dyn}$} & \colhead{$\dot M$} & \colhead{$\dot P$} & \colhead{$\dot E$}\\
\colhead{} & \colhead{} & \colhead{($M_\odot$)} & \colhead{($M_\odot$ km s$^{-1}$)} & \colhead{($10^{43}$ erg)} & \colhead{(pc)} & \colhead{(km s$^{-1}$)} & \colhead{($10^4$ yr)} & \colhead{($10^{-6}$ $M_\odot$ yr$^{-1}$)} & \colhead{($10^{-6}$ $M_\odot$ km s$^{-1}$ yr$^{-1}$)} & \colhead{($10^{30}$ erg s$^{-1}$)}}

\startdata
SMZ 11 & B & 0.09 - 1.52 & 0.44 - 3.67 & 2.1 - 9.9 & 0.11 & 7.1 & 1.5 & 6.3 - 102.3 & 29.8 - 247.2 & 45.4 - 211.6 \\
SMZ 17 & B & 0.00 - 0.16 & 0.01 - 0.48 & 0.1 - 1.8 & 0.03 & 9.0 & 0.3 & 0.2 - 48.5 & 2.1 - 145.1 & 6.0 - 172.2 \\
- & R & 0.01 - 0.22 & 0.05 - 0.67 & 0.4 - 2.4 & 0.05 & 8.3 & 0.6 & 1.1 - 39.5 & 8.5 - 117.7 & 20.1 - 136.0 \\
SMZ 21 & R & 0.10 - 1.25 & 0.33 - 2.46 & 1.1 - 5.3 & 0.28 & 5.1 & 5.4 & 1.8 - 23.3 & 6.2 - 46.0 & 6.8 - 31.6 \\
SMZ 30 & B & 0.61 - 0.93 & 1.07 - 1.46 & 1.9 - 2.4 & 0.16 & 2.6 & 5.9 & 10.3 - 15.7 & 18.2 - 24.7 & 10.2 - 12.8 \\
SMZ 50 & B & 0.08 - 0.21 & 0.22 - 0.47 & 0.6 - 1.1 & 0.22 & 4.8 & 4.4 & 1.8 - 4.8 & 4.9 - 10.6 & 4.4 - 7.9 \\
HOPS 10 & R & 0.07 - 1.46 & 0.41 - 4.58 & 2.5 - 15.5 & 0.26 & 11.2 & 2.3 & 3.2 - 64.6 & 18.4 - 201.9 & 35.2 - 216.8 \\
HOPS 11 & B & 0.12 - 0.17 & 0.50 - 0.59 & 2.1 - 2.3 & 0.14 & 9.7 & 1.4 & 9.1 - 12.8 & 36.3 - 43.3 & 48.3 - 52.9 \\
- & R & 0.06 - 1.14 & 0.39 - 3.57 & 2.8 - 12.6 & 0.18 & 11.6 & 1.5 & 3.7 - 75.9 & 25.8 - 236.6 & 58.8 - 264.6 \\
HOPS 12 & B & 1.97 - 2.92 & 8.19 - 9.54 & 35.1 - 37.4 & 0.65 & 10.0 & 6.3 & 31.1 - 46.0 & 129.5 - 150.7 & 174.9 - 186.6 \\
HOPS 44 & B & 1.32 - 2.65 & 5.07 - 8.99 & 19.5 - 31.2 & 0.18 & 9.8 & 1.8 & 74.9 - 149.9 & 286.6 - 508.7 & 349.8 - 559.4 \\
HOPS 50 & B & 0.13 - 0.13 & 0.22 - 0.22 & 0.5 - 0.5 & 0.28 & 5.4 & 5.1 & 2.5 - 2.5 & 4.4 - 4.4 & 2.8 - 2.8 \\
HOPS 56 & R & 0.20 - 1.46 & 0.83 - 3.56 & 3.5 - 9.7 & 0.15 & 7.7 & 1.8 & 10.9 - 79.2 & 44.8 - 192.7 & 59.6 - 166.9 \\
HOPS 58 & R & 0.08 - 1.07 & 0.33 - 2.62 & 1.3 - 6.8 & 0.14 & 6.1 & 2.2 & 3.7 - 48.1 & 14.7 - 117.1 & 18.6 - 97.2 \\
HOPS 59 & B & 0.09 - 2.12 & 0.41 - 3.95 & 2.0 - 8.7 & 0.51 & 8.6 & 5.8 & 1.5 - 36.4 & 7.1 - 67.8 & 11.1 - 47.2 \\
HOPS 60 & B & 0.01 - 0.35 & 0.07 - 0.80 & 0.3 - 2.1 & 0.14 & 6.7 & 2.0 & 0.7 - 17.0 & 3.3 - 39.3 & 5.1 - 32.1 \\
- & R & 0.01 - 0.44 & 0.06 - 0.74 & 0.2 - 1.4 & 0.15 & 7.8 & 1.9 & 0.8 - 23.6 & 3.2 - 39.7 & 4.3 - 24.3 \\
HOPS 68 & B & 0.03 - 0.51 & 0.15 - 1.17 & 0.9 - 3.2 & 0.05 & 9.9 & 0.5 & 5.9 - 105.4 & 31.0 - 241.9 & 59.7 - 207.5 \\
- & R & 0.05 - 0.18 & 0.24 - 0.52 & 1.2 - 1.8 & 0.05 & 8.1 & 0.7 & 7.6 - 26.6 & 35.9 - 78.4 & 57.3 - 88.4 \\
HOPS 70 & R & 0.12 - 0.37 & 0.42 - 0.99 & 1.5 - 2.8 & 0.10 & 5.6 & 1.7 & 7.0 - 20.9 & 24.1 - 56.6 & 26.5 - 50.9 \\
HOPS 71 & B & 0.17 - 3.69 & 0.78 - 7.64 & 3.5 - 18.4 & 0.34 & 6.8 & 4.9 & 3.6 - 75.7 & 15.9 - 157.0 & 22.9 - 119.4 \\
HOPS 75 & R & 0.03 - 0.25 & 0.10 - 0.49 & 0.3 - 1.1 & 0.07 & 4.6 & 1.6 & 1.9 - 15.9 & 6.2 - 31.1 & 6.3 - 21.1 \\
HOPS 78 & B & 0.32 - 7.86 & 2.15 - 21.91 & 15.0 - 71.5 & 0.40 & 13.3 & 2.9 & 10.8 - 268.9 & 73.4 - 749.4 & 162.3 - 774.9 \\
- & R & 0.18 - 3.44 & 0.79 - 7.05 & 3.7 - 16.8 & 0.33 & 8.0 & 4.1 & 4.3 - 84.2 & 19.4 - 172.1 & 28.6 - 130.1 \\
HOPS 81 & B & 0.07 - 1.00 & 0.36 - 2.34 & 1.9 - 6.5 & 0.16 & 10.1 & 1.6 & 4.5 - 62.7 & 22.2 - 145.9 & 37.3 - 128.0 \\
- & R & 0.02 - 0.09 & 0.04 - 0.15 & 0.1 - 0.3 & 0.05 & 3.2 & 1.6 & 1.2 - 5.5 & 2.8 - 9.5 & 2.2 - 5.5 \\
HOPS 84 & B & 0.02 - 0.12 & 0.05 - 0.19 & 0.1 - 0.3 & 0.12 & 4.4 & 2.7 & 0.6 - 4.3 & 1.8 - 7.0 & 1.6 - 4.0 \\
- & R & 0.80 - 2.62 & 2.81 - 7.23 & 9.9 - 20.8 & 0.83 & 8.7 & 9.4 & 8.6 - 27.9 & 29.9 - 76.9 & 33.4 - 70.1 \\
HOPS 87 & B & 1.07 - 2.09 & 3.44 - 5.63 & 11.6 - 16.3 & 0.25 & 7.2 & 3.5 & 30.7 - 60.3 & 99.5 - 162.7 & 106.3 - 149.4 \\
- & R & 0.02 - 0.27 & 0.06 - 0.45 & 0.2 - 0.8 & 0.08 & 4.6 & 1.7 & 1.2 - 15.7 & 3.3 - 25.7 & 3.0 - 14.6 \\
HOPS 88 & B & 0.17 - 1.02 & 0.70 - 2.45 & 3.0 - 6.8 & 0.09 & 9.1 & 0.9 & 18.8 - 111.3 & 76.0 - 268.3 & 105.3 - 236.7 \\
- & R & 0.05 - 0.10 & 0.15 - 0.26 & 0.5 - 0.7 & 0.08 & 6.9 & 1.1 & 4.3 - 9.7 & 13.7 - 24.0 & 14.4 - 20.8 \\
HOPS 92 & B & 0.37 - 2.44 & 1.51 - 5.70 & 6.4 - 15.4 & 0.32 & 8.8 & 3.5 & 10.5 - 69.1 & 42.7 - 161.6 & 57.4 - 138.1 \\
- & R & 0.12 - 0.17 & 0.41 - 0.47 & 1.4 - 1.5 & 0.48 & 6.2 & 7.6 & 1.6 - 2.3 & 5.4 - 6.2 & 5.8 - 6.1 \\
HOPS 96 & B & 0.04 - 0.09 & 0.10 - 0.16 & 0.3 - 0.4 & 0.10 & 4.0 & 2.6 & 1.5 - 3.4 & 4.1 - 6.2 & 3.8 - 4.5 \\
- & R & 0.32 - 1.54 & 0.68 - 2.43 & 1.4 - 4.0 & 0.23 & 4.7 & 4.7 & 6.9 - 32.9 & 14.6 - 52.1 & 9.8 - 27.1 \\
HOPS 99 & B & 0.03 - 0.06 & 0.06 - 0.09 & 0.1 - 0.1 & 0.13 & 2.6 & 4.7 & 0.7 - 1.3 & 1.3 - 2.0 & 0.8 - 1.0 \\
HOPS 157 & B & 0.03 - 0.03 & 0.07 - 0.07 & 0.2 - 0.2 & 0.10 & 3.4 & 2.9 & 1.0 - 1.0 & 2.5 - 2.5 & 2.0 - 2.0 \\
HOPS 158 & B & 0.08 - 0.17 & 0.20 - 0.37 & 0.5 - 0.8 & 0.11 & 4.1 & 2.6 & 2.9 - 6.4 & 7.6 - 13.9 & 6.5 - 10.1 \\
- & R & 0.02 - 0.03 & 0.07 - 0.08 & 0.2 - 0.2 & 0.07 & 3.9 & 1.9 & 1.3 - 1.8 & 3.6 - 4.2 & 3.6 - 3.8 \\
HOPS 160 & B & 0.06 - 0.06 & 0.10 - 0.10 & 0.2 - 0.2 & 0.11 & 3.2 & 3.3 & 1.9 - 1.9 & 2.9 - 2.9 & 1.5 - 1.5 \\
HOPS 166 & B & 0.02 - 0.47 & 0.09 - 1.03 & 0.3 - 2.4 & 0.06 & 5.8 & 1.1 & 2.3 - 43.6 & 8.5 - 95.9 & 10.1 - 70.7 \\
- & R & 0.93 - 2.65 & 3.46 - 7.26 & 13.6 - 22.2 & 0.41 & 10.7 & 3.8 & 24.7 - 70.5 & 92.2 - 193.6 & 115.9 - 188.1 \\
HOPS 168 & B & 0.01 - 0.05 & 0.03 - 0.10 & 0.1 - 0.3 & 0.09 & 6.1 & 1.4 & 0.4 - 3.3 & 2.1 - 7.2 & 3.2 - 6.3 \\
- & R & 0.96 - 6.97 & 4.30 - 15.22 & 20.9 - 42.7 & 1.16 & 10.4 & 10.9 & 8.8 - 63.8 & 39.3 - 138.8 & 60.7 - 123.8 \\
HOPS 169 & B & 0.09 - 0.23 & 0.46 - 0.66 & 2.6 - 2.9 & 0.24 & 8.9 & 2.7 & 3.4 - 8.5 & 17.3 - 24.8 & 30.9 - 34.5 \\
- & R & 0.13 - 1.11 & 0.74 - 2.33 & 4.8 - 7.6 & 0.22 & 12.3 & 1.8 & 7.4 - 62.0 & 41.5 - 130.7 & 85.3 - 135.0 \\
HOPS 174 & B & 0.07 - 0.15 & 0.20 - 0.37 & 0.6 - 1.0 & 0.20 & 4.4 & 4.4 & 1.6 - 3.3 & 4.6 - 8.3 & 4.3 - 6.9 \\
HOPS 177 & R & 0.72 - 0.99 & 1.42 - 1.74 & 3.0 - 3.4 & 0.30 & 7.3 & 4.0 & 17.9 - 24.6 & 35.1 - 43.2 & 23.7 - 26.8 \\
HOPS 178 & B & 0.18 - 0.52 & 0.58 - 1.33 & 1.9 - 3.6 & 0.21 & 8.9 & 2.3 & 7.9 - 22.6 & 25.4 - 58.1 & 26.3 - 49.6 \\
HOPS 179 & B & 0.06 - 0.06 & 0.15 - 0.15 & 0.4 - 0.4 & 0.08 & 4.6 & 1.8 & 3.2 - 3.2 & 8.4 - 8.4 & 7.2 - 7.2 \\
- & R & 0.04 - 0.84 & 0.22 - 2.92 & 1.4 - 11.1 & 0.12 & 12.4 & 0.9 & 3.9 - 90.6 & 24.2 - 315.8 & 49.5 - 378.0 \\
HOPS 181 & B & 0.48 - 0.91 & 1.94 - 3.17 & 8.3 - 11.7 & 0.21 & 9.1 & 2.2 & 21.4 - 40.9 & 87.2 - 142.2 & 117.9 - 167.2 \\
- & R & 0.25 - 9.97 & 1.46 - 24.75 & 9.1 - 72.0 & 1.14 & 12.1 & 9.2 & 2.7 - 108.0 & 15.8 - 268.4 & 31.2 - 247.6 \\
HOPS 182 & B & 0.29 - 0.52 & 1.25 - 1.87 & 5.7 - 7.4 & 0.21 & 9.1 & 2.3 & 12.8 - 22.7 & 55.1 - 82.5 & 79.3 - 103.6 \\
- & R & 0.10 - 1.26 & 0.62 - 4.22 & 3.8 - 15.4 & 0.53 & 11.1 & 4.6 & 2.2 - 27.3 & 13.3 - 90.8 & 26.2 - 105.3 \\
HOPS 192 & B & 0.72 - 0.72 & 0.91 - 0.91 & 1.3 - 1.3 & 0.10 & 4.1 & 2.4 & 30.1 - 30.1 & 38.0 - 38.0 & 17.6 - 17.6 \\
- & R & 0.02 - 0.08 & 0.05 - 0.13 & 0.1 - 0.2 & 0.06 & 3.7 & 1.7 & 1.4 - 4.6 & 2.9 - 7.5 & 2.1 - 4.2 \\
HOPS 198 & R & 0.00 - 0.19 & 0.02 - 0.47 & 0.1 - 1.3 & 0.10 & 6.0 & 1.7 & 0.3 - 11.2 & 1.4 - 27.9 & 2.2 - 25.1 \\
HOPS 203 & R & 0.42 - 0.42 & 0.92 - 0.92 & 2.1 - 2.1 & 0.19 & 5.7 & 3.3 & 12.6 - 12.6 & 27.6 - 27.6 & 20.3 - 20.3 \\
HOPS 355 & B & 0.37 - 0.44 & 1.31 - 1.38 & 4.7 - 4.8 & 0.43 & 8.9 & 4.7 & 7.9 - 9.2 & 27.6 - 29.2 & 31.6 - 32.2 \\
- & R & 0.03 - 0.20 & 0.12 - 0.51 & 0.5 - 1.4 & 0.48 & 7.8 & 6.0 & 0.5 - 3.4 & 2.0 - 8.5 & 2.7 - 7.4 \\
HOPS 368 & B & 0.02 - 0.50 & 0.12 - 1.28 & 0.6 - 3.6 & 0.06 & 6.7 & 0.8 & 3.0 - 59.9 & 13.9 - 153.0 & 21.1 - 135.1 \\
- & R & 0.02 - 0.05 & 0.09 - 0.15 & 0.3 - 0.5 & 0.04 & 6.3 & 0.6 & 3.6 - 8.1 & 13.7 - 24.1 & 17.0 - 24.8 \\
HOPS 370 & B & 0.05 - 0.33 & 0.38 - 1.31 & 3.3 - 6.7 & 0.13 & 13.1 & 1.0 & 4.6 - 33.2 & 38.2 - 131.9 & 105.3 - 214.5 \\
- & R & 0.02 - 0.56 & 0.11 - 1.26 & 0.7 - 3.6 & 0.12 & 8.1 & 1.5 & 1.1 - 37.9 & 7.3 - 85.7 & 15.5 - 77.7 \\
HOPS 383 & B & 0.27 - 0.34 & 0.69 - 0.82 & 1.9 - 2.1 & 0.15 & 5.4 & 2.8 & 9.9 - 12.4 & 25.0 - 29.7 & 21.3 - 23.9\\
\hline
Total\tablenotemark{\footnotesize b}&&15.4 - 76.9&55.6 - 193&232 - 573&&&&499 - 2618&1855 - 6824&2551 - 6666\\
\enddata
\tablenotetext{a}{Mass and all properties derived from mass are given as lower and upper limits. The lower limit refers to the high-velocity component of the outflow only. The upper limit refers to the sum of the high-velocity and low-velocity components. See Section~\ref{sec:dmdv} for details.}
\tablenotetext{b}{The total of the lower-limits and upper-limits for each column are given. The physical properties in this table are not corrected for inclination angle. See Section~\ref{sec:impact} for a discussion of the inclination correction.}
\end{deluxetable*}

\startlongtable
\clearpage
\begin{deluxetable*}{ccccc}
\tabletypesize{\small}
\tablecaption{Outflow Angles and Filament Comparison.\label{tab:angles}}
\tablehead{\colhead{Source} & \colhead{Position Angle\tablenotemark{\footnotesize a}} & \colhead{Opening Angle} & \colhead{$\gamma$\tablenotemark{\footnotesize b}} & \colhead{$d_{\rm fil}$\tablenotemark{\footnotesize c}}\\
\colhead{} & \colhead{(\degr)} &\colhead{(\degr)}& \colhead{(\degr)} & \colhead{(pc)}}
\startdata
SMZ 11 & $-35 \pm 3.1$/- & $157 \pm 10$/- & $27$/- & 0.064 \\
SMZ 17 & $-118 \pm 11$/$34 \pm 3.1$ & $193 \pm 38$/$79 \pm 10$ & 10/38 & 0.01 \\
SMZ 21 & -/$179 \pm 0.5$ & -/$35 \pm 2$ & -/$43$ & 0.016 \\
SMZ 30 & $20 \pm 20$/- & $120 \pm 1.2e+02$/- & $35$/- & 0.11 \\
SMZ 50 & $2 \pm 1.4$/- & $77 \pm 8.8$/- & $36$/- & 0.19 \\
HOPS 10 & -/$19 \pm 2$ & -/$58 \pm 6.8$ & -/$71$ & 0.003 \\
HOPS 11 & $172 \pm 1.2$/$16 \pm 1.7$ & $92 \pm 4.1$/$68 \pm 5.6$ & 62/86 & 0.051 \\
HOPS 12 & $-11 \pm 0.1$/- & $39 \pm 0.4$/- & $54$/- & 0.031 \\
HOPS 44 & $-24 \pm 2.4$/- & $120 \pm 8.9$/- & $70$/- & 0.013 \\
HOPS 50 & $-15 \pm 0.4$/- & $42 \pm 1.3$/- & $16$/- & 0.005 \\
HOPS 56 & -/$119 \pm 6$ & -/$191 \pm 23$ & -/$33$ & 0.047 \\
HOPS 58 & -/$74 \pm 0.6$ & -/$44 \pm 1.9$ & -/$88$ & 0.056 \\
HOPS 59 & $-26 \pm 0.2$/- & $15 \pm 0.7$/- & $38$/- & 0.052 \\
HOPS 60 & $-104 \pm 0.3$/$57 \pm 0.3$ & $22 \pm 1.1$/$24 \pm 0.9$ & 78/83 & 0.055 \\
HOPS 68 & $-158 \pm 3.8$/$26 \pm 3.7$ & $161 \pm 13$/$163 \pm 12$ & 20/24 & 0.056 \\
HOPS 70 & -/$48 \pm 2$ & -/$83 \pm 7.5$ & -/$74$ & 0.015 \\
HOPS 71 & $39 \pm 0.6$/- & $35 \pm 2.3$/- & $59$/- & 0.1 \\
HOPS 75 & -/$175 \pm 1.6$ & -/$95 \pm 5.5$ & -/$17$ & 0.003 \\
HOPS 78 & $-85 \pm 0.5$/$80 \pm 0.2$ & $41 \pm 1.5$/$37 \pm 0.8$ & 59/43 & 0.045 \\
HOPS 81 & $-154 \pm 0.6$/$34 \pm 3.1$ & $51 \pm 2$/$83 \pm 12$ & 40/32 & 0.065 \\
HOPS 84 & $78 \pm 0.7$/$-93 \pm 0.2$ & $38 \pm 2.4$/$18 \pm 0.6$ & 58/49 & 0.054 \\
HOPS 87 & $-100 \pm 0.3$/$118 \pm 1.2$ & $49 \pm 1.2$/$47 \pm 4.5$ & 27/11 & 0.008 \\
HOPS 88 & $-100 \pm 3.5$/$82 \pm 1.7$ & $210 \pm 12$/$99 \pm 6.1$ & 36/35 & 0.022 \\
HOPS 92 & $89 \pm 0.3$/$-89 \pm 0.2$ & $33 \pm 1.2$/$24 \pm 0.4$ & 43/41 & 0.019 \\
HOPS 96 & $69 \pm 0.8$/$-95 \pm 5$ & $46 \pm 2.8$/$142 \pm 19$ & 55/71 & 0.007 \\
HOPS 99 & $-49 \pm 1.3$/- & $64 \pm 4.8$/- & $48$/- & 0.003 \\
HOPS 157 & $112 \pm 1.4$/- & $66 \pm 4.8$/- & $84$/- & 1.5 \\
HOPS 158 & $66 \pm 1.4$/$-114 \pm 2.7$ & $84 \pm 4.7$/$113 \pm 9.1$ & 50/50 & 1.1 \\
HOPS 160 & $-38 \pm 1.1$/- & $68 \pm 3.6$/- & $2$/- & 0.41 \\
HOPS 166 & $99 \pm 4.9$/$0 \pm 1.1$ & $239 \pm 17$/$80 \pm 3.7$ & 59/40 & 0.035 \\
HOPS 168 & $-7 \pm 3.2$/$160 \pm 0.2$ & $101 \pm 11$/$33 \pm 0.6$ & 83/69 & 0.001 \\
HOPS 169 & $15 \pm 0.5$/$-179 \pm 0.9$ & $46 \pm 1.7$/$62 \pm 2.9$ & 49/35 & 0.004 \\
HOPS 174 & $13 \pm 0.4$/- & $35 \pm 1.3$/- & $85$/- & 0.004 \\
HOPS 177 & -/$-6 \pm 0.7$ & -/$54 \pm 2.4$ & -/$6$ & 0.091 \\
HOPS 178 & $77 \pm 2$/- & $94 \pm 6.7$/- & $76$/- & 0.04 \\
HOPS 179 & $164 \pm 1.1$/$16 \pm 0.8$ & $60 \pm 3.6$/$71 \pm 2.6$ & 60/28 & 0.088 \\
HOPS 181 & $-40 \pm 2$/$177 \pm 0.4$ & $131 \pm 7.8$/$24 \pm 1.2$ & 48/12 & 0.2 \\
HOPS 182 & $41 \pm 2.5$/$-157 \pm 0.1$ & $77 \pm 8.2$/$9 \pm 0.3$ & 33/16 & 0.22 \\
HOPS 192 & $95 \pm 2.8$/$-67 \pm 1.6$ & $104 \pm 9.3$/$105 \pm 5.3$ & 53/34 & 0.044 \\
HOPS 198 & -/$112 \pm 4.2$ & -/$64 \pm 14$ & -/$0$ & 0.11 \\
HOPS 203 & -/$139 \pm 4.5$ & -/$60 \pm 15$ & -/$85$ & 0.011 \\
HOPS 355 & $-104 \pm 0.4$/$98 \pm 0.1$ & $45 \pm 1.3$/$8 \pm 0.3$ & 75/84 & 0.096 \\
HOPS 368 & $30 \pm 12$/$175 \pm 4.5$ & $289 \pm 44$/$135 \pm 15$ & 61/84 & 0.032 \\
HOPS 370 & $36 \pm 1.3$/$24 \pm 0.5$ & $66 \pm 4.8$/$40 \pm 1.9$ & 37/48 & 0.042 \\
HOPS 383 & $128 \pm 0.5$/- & $49 \pm 1.7$/- & $88$/- & 0.018
\enddata
\tablenotetext{a}{Columns with two entries refer to the blue/red outflow lobes separately. Entries marked '-' refer to lobes that are not detected.}
\tablenotetext{b}{$\gamma$ is the projected angle between the outflow and filament. See Section~\ref{sec:filaments} for details.}
\tablenotetext{c}{$d_{\rm fil}$ is the minimum distance between the outflow source and filament.}
\end{deluxetable*}

\begin{deluxetable*}{ccc}
\tabletypesize{\small}
\centerwidetable
\tablecaption{Anderson-Darling $p$-values.\label{tab:ad}}
\tablehead{\colhead{Sample} & \colhead{Random} & \colhead{Perpendicular}}
\startdata
Full & 0.23 & $<0.001$ \\
Definite & $>0.25$ & 0.06 \\
$d_{\rm fil}$ $\leq$ 0.05 pc & 0.13 & 0.02 \\
Both & 0.06 & $>0.25$ \\
\citet{Stephens17}\tablenotemark{a}&0.20&0.0045\\
\citet{Kong19}\tablenotemark{b}&$6.5\times10^{-5}$&0.53\\
\enddata
\tablenotetext{a}{Perseus outflows; $\gamma_{\rm F}$ in their Table 3.}
\tablenotetext{b}{IRDC G28 outflows}
\end{deluxetable*}

\clearpage

\bibliographystyle{aasjournal}
\bibliography{all.bib}

\newcommand{\noop}[1]{}
\begin{thebibliography}{}
\expandafter\ifx\csname natexlab\endcsname\relax\def\natexlab#1{#1}\fi
\providecommand{\url}[1]{\href{#1}{#1}}
\providecommand{\dodoi}[1]{doi:~\href{http://doi.org/#1}{\nolinkurl{#1}}}
\providecommand{\doeprint}[1]{\href{http://ascl.net/#1}{\nolinkurl{http://ascl.net/#1}}}
\providecommand{\doarXiv}[1]{\href{https://arxiv.org/abs/#1}{\nolinkurl{https://arxiv.org/abs/#1}}}

\bibitem[{{Anathpindika} \& {Whitworth}(2008)}]{Anathpindika08}
{Anathpindika}, S., \& {Whitworth}, A.~P. 2008, \aap, 487, 605

\bibitem[{{Andr{\'e}} {et~al.}(2014){Andr{\'e}}, {Di Francesco},
  {Ward-Thompson}, {Inutsuka}, {Pudritz}, \& {Pineda}}]{Andre14}
{Andr{\'e}}, P., {Di Francesco}, J., {Ward-Thompson}, D., {et~al.} 2014, in
  Protostars and Planets VI, ed. H.~{Beuther}, R.~S. {Klessen}, C.~P.
  {Dullemond}, \& T.~{Henning}, 27

\bibitem[{{Arce} {et~al.}(2010){Arce}, {Borkin}, {Goodman}, {Pineda}, \&
  {Halle}}]{Arce10}
{Arce}, H.~G., {Borkin}, M.~A., {Goodman}, A.~A., {Pineda}, J.~E., \& {Halle},
  M.~W. 2010, \apj, 715, 1170

\bibitem[{{Arce} \& {Goodman}(2001{\natexlab{a}})}]{Arce01a}
{Arce}, H.~G., \& {Goodman}, A.~A. 2001{\natexlab{a}}, \apj, 554, 132

\bibitem[{{Arce} \& {Goodman}(2001{\natexlab{b}})}]{Arce01}
---. 2001{\natexlab{b}}, \apjl, 551, L171

\bibitem[{{Arce} \& {Sargent}(2006)}]{Arce06}
{Arce}, H.~G., \& {Sargent}, A.~I. 2006, \apj, 646, 1070

\bibitem[{{Arce} {et~al.}(2007){Arce}, {Shepherd}, {Gueth}, {Lee}, {Bachiller},
  {Rosen}, \& {Beuther}}]{Arce07}
{Arce}, H.~G., {Shepherd}, D., {Gueth}, F., {et~al.} 2007, Protostars and
  Planets V, 245

\bibitem[{{Arzoumanian} {et~al.}(2011){Arzoumanian}, {Andr{\'e}}, {Didelon},
  {K{\"o}nyves}, {Schneider}, {Men'shchikov}, {Sousbie}, {Zavagno}, {Bontemps},
  \& {di Francesco}}]{Arzoumanian11}
{Arzoumanian}, D., {Andr{\'e}}, P., {Didelon}, P., {et~al.} 2011, \aap, 529, L6

\bibitem[{{Aso} {et~al.}(2000){Aso}, {Tatematsu}, {Sekimoto}, {Nakano},
  {Umemoto}, {Koyama}, \& {Yamamoto}}]{Aso00}
{Aso}, Y., {Tatematsu}, K., {Sekimoto}, Y., {et~al.} 2000, \apjs, 131, 465

\bibitem[{{Astropy Collaboration} {et~al.}(2013){Astropy Collaboration},
  {Robitaille}, {Tollerud}, {Greenfield}, {Droettboom}, {Bray}, {Aldcroft},
  {Davis}, {Ginsburg}, {Price-Whelan}, {Kerzendorf}, {Conley}, {Crighton},
  {Barbary}, {Muna}, {Ferguson}, {Grollier}, {Parikh}, {Nair}, {Unther},
  {Deil}, {Woillez}, {Conseil}, {Kramer}, {Turner}, {Singer}, {Fox}, {Weaver},
  {Zabalza}, {Edwards}, {Azalee Bostroem}, {Burke}, {Casey}, {Crawford},
  {Dencheva}, {Ely}, {Jenness}, {Labrie}, {Lim}, {Pierfederici}, {Pontzen},
  {Ptak}, {Refsdal}, {Servillat}, \& {Streicher}}]{Astropy-Collaboration13}
{Astropy Collaboration}, {Robitaille}, T.~P., {Tollerud}, E.~J., {et~al.} 2013,
  \aap, 558, A33

\bibitem[{{Bally}(2016)}]{Bally16}
{Bally}, J. 2016, \araa, 54, 491

\bibitem[{{Bally} {et~al.}(2017){Bally}, {Ginsburg}, {Arce}, {Eisner},
  {Youngblood}, {Zapata}, \& {Zinnecker}}]{Bally17}
{Bally}, J., {Ginsburg}, A., {Arce}, H., {et~al.} 2017, \apj, 837, 60

\bibitem[{{Bally} {et~al.}(1987){Bally}, {Langer}, {Stark}, \&
  {Wilson}}]{Bally87}
{Bally}, J., {Langer}, W.~D., {Stark}, A.~A., \& {Wilson}, R.~W. 1987, \apjl,
  312, L45

\bibitem[{{Bern{\'e}} {et~al.}(2014){Bern{\'e}}, {Marcelino}, \&
  {Cernicharo}}]{Berne14}
{Bern{\'e}}, O., {Marcelino}, N., \& {Cernicharo}, J. 2014, \apj, 795, 13

\bibitem[{{Bodenheimer}(1995)}]{Bodenheimer95}
{Bodenheimer}, P. 1995, \araa, 33, 199

\bibitem[{{Brunt} {et~al.}(2009){Brunt}, {Heyer}, \& {Mac Low}}]{Brunt09}
{Brunt}, C.~M., {Heyer}, M.~H., \& {Mac Low}, M.~M. 2009, \aap, 504, 883

\bibitem[{{Buckle} {et~al.}(2012){Buckle}, {Davis}, {Francesco}, {Graves},
  {Nutter}, {Richer}, {Roberts}, {Ward-Thompson}, {White}, {Brunt}, {Butner},
  {Cavanagh}, {Chrysostomou}, {Curtis}, {Duarte-Cabral}, {Etxaluze}, {Fich},
  {Friberg}, {Friesen}, {Fuller}, {Greaves}, {Hatchell}, {Hogerheijde},
  {Johnstone}, {Matthews}, {Matthews}, {Rawlings}, {Sadavoy}, {Simpson},
  {Tothill}, {Tsamis}, {Viti}, {Wouterloot}, \& {Yates}}]{Buckle12}
{Buckle}, J.~V., {Davis}, C.~J., {Francesco}, J.~D., {et~al.} 2012, \mnras,
  422, 521

\bibitem[{{Carroll} {et~al.}(2010){Carroll}, {Frank}, \&
  {Blackman}}]{Carroll10}
{Carroll}, J.~J., {Frank}, A., \& {Blackman}, E.~G. 2010, \apj, 722, 145

\bibitem[{{Carroll} {et~al.}(2009){Carroll}, {Frank}, {Blackman}, {Cunningham},
  \& {Quillen}}]{Carroll09}
{Carroll}, J.~J., {Frank}, A., {Blackman}, E.~G., {Cunningham}, A.~J., \&
  {Quillen}, A.~C. 2009, \apj, 695, 1376

\bibitem[{{Chen} \& {Ostriker}(2014)}]{Chen14}
{Chen}, C.-Y., \& {Ostriker}, E.~C. 2014, \apj, 785, 69

\bibitem[{{Chernin} \& {Masson}(1995)}]{Chernin95}
{Chernin}, L.~M., \& {Masson}, C.~R. 1995, \apj, 455, 182

\bibitem[{{Chini} {et~al.}(1997){Chini}, {Reipurth}, {Ward-Thompson}, {Bally},
  {Nyman}, {Sievers}, \& {Billawala}}]{Chini97}
{Chini}, R., {Reipurth}, B., {Ward-Thompson}, D., {et~al.} 1997, \apjl, 474,
  L135

\bibitem[{{Choi} {et~al.}(2017){Choi}, {Kang}, {Lee}, {Tatematsu}, {Kang},
  {Sayers}, {Evans}, {Cho}, {Kwon}, {Park}, {Ohashi}, {Yoo}, \& {Lee}}]{Choi17}
{Choi}, M., {Kang}, M., {Lee}, J.-E., {et~al.} 2017, \apjs, 232, 24

\bibitem[{{Clarke} {et~al.}(2017){Clarke}, {Whitworth}, {Duarte-Cabral}, \&
  {Hubber}}]{Clarke17}
{Clarke}, S.~D., {Whitworth}, A.~P., {Duarte-Cabral}, A., \& {Hubber}, D.~A.
  2017, \mnras, 468, 2489

\bibitem[{{Curtis} {et~al.}(2010){Curtis}, {Richer}, {Swift}, \&
  {Williams}}]{Curtis10}
{Curtis}, E.~I., {Richer}, J.~S., {Swift}, J.~J., \& {Williams}, J.~P. 2010,
  \mnras, 408, 1516

\bibitem[{{Davis} {et~al.}(2000){Davis}, {Dent}, {Matthews}, {Coulson}, \&
  {McCaughrean}}]{Davis2000}
{Davis}, C.~J., {Dent}, W.~R.~F., {Matthews}, H.~E., {Coulson}, I.~M., \&
  {McCaughrean}, M.~J. 2000, \mnras, 318, 952

\bibitem[{{Davis} {et~al.}(2009){Davis}, {Froebrich}, {Stanke}, {Megeath},
  {Kumar}, {Adamson}, {Eisl{\"o}ffel}, {Gredel}, {Khanzadyan}, {Lucas},
  {Smith}, \& {Varricatt}}]{Davis09}
{Davis}, C.~J., {Froebrich}, D., {Stanke}, T., {et~al.} 2009, \aap, 496, 153

\bibitem[{{Downes} \& {Cabrit}(2007)}]{Downes07}
{Downes}, T.~P., \& {Cabrit}, S. 2007, \aap, 471, 873

\bibitem[{{Dunham} {et~al.}(2014){Dunham}, {Arce}, {Mardones}, {Lee},
  {Matthews}, {Stutz}, \& {Williams}}]{Dunham14}
{Dunham}, M.~M., {Arce}, H.~G., {Mardones}, D., {et~al.} 2014, \apj, 783, 29

\bibitem[{{Feddersen} {et~al.}(2018){Feddersen}, {Arce}, {Kong}, {Shimajiri},
  {Nakamura}, {Hara}, {Ishii}, {Sasaki}, \& {Kawabe}}]{Feddersen18}
{Feddersen}, J.~R., {Arce}, H.~G., {Kong}, S., {et~al.} 2018, \apj, 862, 121

\bibitem[{{Federrath}(2015)}]{Federrath15}
{Federrath}, C. 2015, \mnras, 450, 4035

\bibitem[{{Frank} {et~al.}(2014){Frank}, {Ray}, {Cabrit}, {Hartigan}, {Arce},
  {Bacciotti}, {Bally}, {Benisty}, {Eisl{\"o}ffel}, {G{\"u}del}, {Lebedev},
  {Nisini}, \& {Raga}}]{Frank14}
{Frank}, A., {Ray}, T.~P., {Cabrit}, S., {et~al.} 2014, Protostars and Planets
  VI, 451

\bibitem[{{Frerking} {et~al.}(1982){Frerking}, {Langer}, \&
  {Wilson}}]{Frerking82}
{Frerking}, M.~A., {Langer}, W.~D., \& {Wilson}, R.~W. 1982, \apj, 262, 590

\bibitem[{{Furlan} {et~al.}(2016){Furlan}, {Fischer}, {Ali}, {Stutz}, {Stanke},
  {Tobin}, {Megeath}, {Osorio}, {Hartmann}, {Calvet}, {Poteet}, {Booker},
  {Manoj}, {Watson}, \& {Allen}}]{Furlan16}
{Furlan}, E., {Fischer}, W.~J., {Ali}, B., {et~al.} 2016, \apjs, 224, 5

\bibitem[{{Goldsmith} {et~al.}(1984){Goldsmith}, {Snell}, {Hemeon-Heyer}, \&
  {Langer}}]{Goldsmith84}
{Goldsmith}, P.~F., {Snell}, R.~L., {Hemeon-Heyer}, M., \& {Langer}, W.~D.
  1984, \apj, 286, 599

\bibitem[{{Gro{\ss}schedl} {et~al.}(2018){Gro{\ss}schedl}, {Alves}, {Meingast},
  {Ackerl}, {Ascenso}, {Bouy}, {Burkert}, {Forbrich}, {F{\"u}rnkranz},
  {Goodman}, {Hacar}, {Herbst-Kiss}, {Lada}, {Larreina}, {Leschinski},
  {Lombardi}, {Moitinho}, {Mortimer}, \& {Zari}}]{Grossschedl18}
{Gro{\ss}schedl}, J.~E., {Alves}, J., {Meingast}, S., {et~al.} 2018, \aap, 619,
  A106

\bibitem[{{Hacar} {et~al.}(2013){Hacar}, {Tafalla}, {Kauffmann}, \&
  {Kov{\'a}cs}}]{Hacar13}
{Hacar}, A., {Tafalla}, M., {Kauffmann}, J., \& {Kov{\'a}cs}, A. 2013, \aap,
  554, A55

\bibitem[{{Kauffmann} {et~al.}(2008){Kauffmann}, {Bertoldi}, {Bourke}, {Evans},
  \& {Lee}}]{Kauffmann08}
{Kauffmann}, J., {Bertoldi}, F., {Bourke}, T.~L., {Evans}, N.~J., I., \& {Lee},
  C.~W. 2008, \aap, 487, 993

\bibitem[{{Koch} \& {Rosolowsky}(2015)}]{Koch15}
{Koch}, E.~W., \& {Rosolowsky}, E.~W. 2015, \mnras, 452, 3435

\bibitem[{{Kong} {et~al.}(2019){Kong}, {Arce}, {Jos{\'e} Maureira}, {Caselli},
  {Tan}, \& {Fontani}}]{Kong19}
{Kong}, S., {Arce}, H.~G., {Jos{\'e} Maureira}, M., {et~al.} 2019, \apj, 874,
  104

\bibitem[{{Kong} {et~al.}(2018){Kong}, {Arce}, {Feddersen}, {Carpenter},
  {Nakamura}, {Shimajiri}, {Isella}, {Ossenkopf-Okada}, {Sargent},
  {S{\'a}nchez-Monge}, {Suri}, {Kauffmann}, {Pillai}, {Pineda}, {Koda},
  {Bally}, {Lis}, {Padoan}, {Klessen}, {Mairs}, {Goodman}, {Goldsmith},
  {McGehee}, {Schilke}, {Teuben}, {Jos{\'e} Maureira}, {Hara}, {Ginsburg},
  {Burkhart}, {Smith}, {Schmiedeke}, {Pineda}, {Ishii}, {Sasaki}, {Kawabe},
  {Urasawa}, {Oyamada}, \& {Tanabe}}]{Kong18}
{Kong}, S., {Arce}, H.~G., {Feddersen}, J.~R., {et~al.} 2018, \apjs, 236, 25

\bibitem[{{Konigl} \& {Pudritz}(2000)}]{Konigl00}
{Konigl}, A., \& {Pudritz}, R.~E. 2000, in Protostars and Planets IV, ed.
  V.~{Mannings}, A.~P. {Boss}, \& S.~S. {Russell}, 759

\bibitem[{{Kounkel} {et~al.}(2018){Kounkel}, {Covey}, {Su{\'a}rez},
  {Rom{\'a}n-Z{\'u}{\~n}iga}, {Hernandez}, {Stassun}, {Jaehnig}, {Feigelson},
  {Pe{\~n}a Ram{\'{\i}}rez}, {Roman-Lopes}, {Da Rio}, {Stringfellow}, {Kim},
  {Borissova}, {Fern{\'a}ndez-Trincado}, {Burgasser},
  {Garc{\'{\i}}a-Hern{\'a}ndez}, {Zamora}, {Pan}, \& {Nitschelm}}]{Kounkel18}
{Kounkel}, M., {Covey}, K., {Su{\'a}rez}, G., {et~al.} 2018, \aj, 156, 84

\bibitem[{Kuhn {et~al.}(2019)Kuhn, Hillenbrand, Sills, Feigelson, \&
  Getman}]{Kuhn19}
Kuhn, M.~A., Hillenbrand, L.~A., Sills, A., Feigelson, E.~D., \& Getman, K.~V.
  2019, \apj, 870, 32

\bibitem[{{Langer} \& {Penzias}(1993)}]{Langer93}
{Langer}, W.~D., \& {Penzias}, A.~A. 1993, \apj, 408, 539

\bibitem[{{Lee} {et~al.}(2017){Lee}, {Hull}, \& {Offner}}]{Lee17}
{Lee}, J.~W.~Y., {Hull}, C.~L.~H., \& {Offner}, S.~S.~R. 2017, \apj, 834, 201

\bibitem[{{Li} {et~al.}(2015){Li}, {Li}, {Qian}, {Xu}, {Goldsmith},
  {Noriega-Crespo}, {Wu}, {Song}, \& {Nan}}]{Li15}
{Li}, H., {Li}, D., {Qian}, L., {et~al.} 2015, \apjs, 219, 20

\bibitem[{{Li} \& {Klein}(2019)}]{Li19}
{Li}, P.~S., \& {Klein}, R.~I. 2019, \mnras, 485, 4509

\bibitem[{{Li} {et~al.}(2018){Li}, {Klein}, \& {McKee}}]{Li18}
{Li}, P.~S., {Klein}, R.~I., \& {McKee}, C.~F. 2018, \mnras, 473, 4220

\bibitem[{{McKee} \& {Ostriker}(2007)}]{McKee07}
{McKee}, C.~F., \& {Ostriker}, E.~C. 2007, \araa, 45, 565

\bibitem[{{Megeath} {et~al.}(2012){Megeath}, {Gutermuth}, {Muzerolle},
  {Kryukova}, {Flaherty}, {Hora}, {Allen}, {Hartmann}, {Myers}, {Pipher},
  {Stauffer}, {Young}, \& {Fazio}}]{Megeath12}
{Megeath}, S.~T., {Gutermuth}, R., {Muzerolle}, J., {et~al.} 2012, \aj, 144,
  192

\bibitem[{{Meingast} {et~al.}(2016){Meingast}, {Alves}, {Mardones}, {Teixeira},
  {Lombardi}, {Gro{\ss}schedl}, {Ascenso}, {Bouy}, {Forbrich}, \&
  {Goodman}}]{Meingast16}
{Meingast}, S., {Alves}, J., {Mardones}, D., {et~al.} 2016, \aap, 587, A153

\bibitem[{{Menten} {et~al.}(2007){Menten}, {Reid}, {Forbrich}, \&
  {Brunthaler}}]{Menten07}
{Menten}, K.~M., {Reid}, M.~J., {Forbrich}, J., \& {Brunthaler}, A. 2007, \aap,
  474, 515

\bibitem[{{Morgan} {et~al.}(1991){Morgan}, {Schloerb}, {Snell}, \&
  {Bally}}]{Morgan91}
{Morgan}, J.~A., {Schloerb}, F.~P., {Snell}, R.~L., \& {Bally}, J. 1991, \apj,
  376, 618

\bibitem[{{Moro-Mart{\'{\i}}n} {et~al.}(1999){Moro-Mart{\'{\i}}n},
  {Cernicharo}, {Noriega-Crespo}, \& {Mart{\'{\i}}n-Pintado}}]{Moro-Martin99}
{Moro-Mart{\'{\i}}n}, A., {Cernicharo}, J., {Noriega-Crespo}, A., \&
  {Mart{\'{\i}}n-Pintado}, J. 1999, \apjl, 520, L111

\bibitem[{{Myers} \& {Ladd}(1993)}]{Myers93}
{Myers}, P.~C., \& {Ladd}, E.~F. 1993, \apjl, 413, L47

\bibitem[{{Nakamura} \& {Li}(2007)}]{Nakamura07}
{Nakamura}, F., \& {Li}, Z.-Y. 2007, \apj, 662, 395

\bibitem[{{Nakamura} {et~al.}(2011){Nakamura}, {Kamada}, {Kamazaki}, {Kawabe},
  {Kitamura}, {Shimajiri}, {Tsukagoshi}, {Tachihara}, {Akashi}, \&
  {Azegami}}]{Nakamura11}
{Nakamura}, F., {Kamada}, Y., {Kamazaki}, T., {et~al.} 2011, \apj, 726, 46

\bibitem[{{Nakamura} {et~al.}(2012){Nakamura}, {Miura}, {Kitamura},
  {Shimajiri}, {Kawabe}, {Akashi}, {Ikeda}, {Tsukagoshi}, {Momose}, {Nishi}, \&
  {Li}}]{Nakamura12}
{Nakamura}, F., {Miura}, T., {Kitamura}, Y., {et~al.} 2012, \apj, 746, 25

\bibitem[{{Offner} {et~al.}(2016){Offner}, {Dunham}, {Lee}, {Arce}, \&
  {Fielding}}]{Offner16}
{Offner}, S.~S.~R., {Dunham}, M.~M., {Lee}, K.~I., {Arce}, H.~G., \&
  {Fielding}, D.~B. 2016, \apjl, 827, L11

\bibitem[{{Offner} {et~al.}(2011){Offner}, {Lee}, {Goodman}, \&
  {Arce}}]{Offner11}
{Offner}, S. S.~R., {Lee}, E.~J., {Goodman}, A.~A., \& {Arce}, H. 2011, \apj,
  743, 91

\bibitem[{Offner \& Liu(2018)}]{Offner18}
Offner, S. S.~R., \& Liu, Y. 2018, NatAs, 2, 896

\bibitem[{{Padoan} {et~al.}(2009){Padoan}, {Juvela}, {Kritsuk}, \&
  {Norman}}]{Padoan09}
{Padoan}, P., {Juvela}, M., {Kritsuk}, A., \& {Norman}, M.~L. 2009, \apj, 707,
  L153

\bibitem[{Plunkett {et~al.}(2015)Plunkett, Arce, Corder, Dunham, Garay, \&
  Mardones}]{Plunkett15}
Plunkett, A.~L., Arce, H.~G., Corder, S.~A., {et~al.} 2015, \apj, 803, 22

\bibitem[{{Plunkett} {et~al.}(2013){Plunkett}, {Arce}, {Corder}, {Mardones},
  {Sargent}, \& {Schnee}}]{Plunkett13}
{Plunkett}, A.~L., {Arce}, H.~G., {Corder}, S.~A., {et~al.} 2013, \apj, 774, 22

\bibitem[{{Rathborne} {et~al.}(2006){Rathborne}, {Jackson}, \&
  {Simon}}]{Rathborne06}
{Rathborne}, J.~M., {Jackson}, J.~M., \& {Simon}, R. 2006, \apj, 641, 389

\bibitem[{{Ripple} {et~al.}(2013){Ripple}, {Heyer}, {Gutermuth}, {Snell}, \&
  {Brunt}}]{Ripple13}
{Ripple}, F., {Heyer}, M.~H., {Gutermuth}, R., {Snell}, R.~L., \& {Brunt},
  C.~M. 2013, \mnras, 431, 1296

\bibitem[{{Robitaille} {et~al.}(2016){Robitaille}, {Ginsburg}, {Beaumont},
  {Leroy}, \& {Rosolowsky}}]{Robitaille16}
{Robitaille}, T., {Ginsburg}, A., {Beaumont}, C., {Leroy}, A., \& {Rosolowsky},
  E. 2016, {spectral-cube: Read and analyze astrophysical spectral data cubes},
  Astrophysics Source Code Library.
\newblock \doeprint{1609.017}

\bibitem[{{Rohlfs} \& {Wilson}(1996)}]{Rohlfs96}
{Rohlfs}, K., \& {Wilson}, T.~L. 1996, {Tools of Radio Astronomy}
  (Springer-Verlag Berlin Heidelberg New York), 127

\bibitem[{{Schmid-Burgk} {et~al.}(1990){Schmid-Burgk}, {Guesten},
  {Mauersberger}, {Schulz}, \& {Wilson}}]{Schmid-Burgk90}
{Schmid-Burgk}, J., {Guesten}, R., {Mauersberger}, R., {Schulz}, A., \&
  {Wilson}, T.~L. 1990, \apjl, 362, L25

\bibitem[{{Shimajiri} {et~al.}(2008){Shimajiri}, {Takahashi}, {Takakuwa},
  {Saito}, \& {Kawabe}}]{Shimajiri08}
{Shimajiri}, Y., {Takahashi}, S., {Takakuwa}, S., {Saito}, M., \& {Kawabe}, R.
  2008, \apj, 683, 255

\bibitem[{{Shimajiri} {et~al.}(2009){Shimajiri}, {Takahashi}, {Takakuwa},
  {Saito}, \& {Kawabe}}]{Shimajiri09}
---. 2009, \pasj, 61, 1055

\bibitem[{{Shimajiri} {et~al.}(2011){Shimajiri}, {Kawabe}, {Takakuwa}, {Saito},
  {Tsukagoshi}, {Momose}, {Ikeda}, {Akiyama}, {Austermann}, {Ezawa}, {Fukue},
  {Hiramatsu}, {Hughes}, {Kitamura}, {Kohno}, {Kurono}, {Scott}, {Wilson},
  {Yoshida}, \& {Yun}}]{Shimajiri11}
{Shimajiri}, Y., {Kawabe}, R., {Takakuwa}, S., {et~al.} 2011, \pasj, 63, 105

\bibitem[{{Shu} {et~al.}(2000){Shu}, {Najita}, {Shang}, \& {Li}}]{Shu00}
{Shu}, F.~H., {Najita}, J.~R., {Shang}, H., \& {Li}, Z.~Y. 2000, in Protostars
  and Planets IV, ed. V.~{Mannings}, A.~P. {Boss}, \& S.~S. {Russell}, 789--814

\bibitem[{{Sousbie}(2011)}]{Sousbie11}
{Sousbie}, T. 2011, \mnras, 414, 350

\bibitem[{{Stanke} {et~al.}(2002){Stanke}, {McCaughrean}, \&
  {Zinnecker}}]{Stanke02}
{Stanke}, T., {McCaughrean}, M.~J., \& {Zinnecker}, H. 2002, \aap, 392, 239

\bibitem[{{Stanke} \& {Williams}(2007)}]{Stanke07}
{Stanke}, T., \& {Williams}, J.~P. 2007, \aj, 133, 1307

\bibitem[{{Stephens} {et~al.}(2017){Stephens}, {Dunham}, {Myers}, {Pokhrel},
  {Sadavoy}, {Vorobyov}, {Tobin}, {Pineda}, {Offner}, {Lee}, {Kristensen},
  {J{\o}rgensen}, {Goodman}, {Bourke}, {Arce}, \& {Plunkett}}]{Stephens17}
{Stephens}, I.~W., {Dunham}, M.~M., {Myers}, P.~C., {et~al.} 2017, \apj, 846,
  16

\bibitem[{Stephens(1974)}]{Stephens74}
Stephens, M.~A. 1974, Journal of the American Statistical Association, 69, 730

\bibitem[{{Suri} {et~al.}(2019){Suri}, {S{\'a}nchez-Monge}, {Schilke},
  {Clarke}, {Smith}, {Ossenkopf-Okada}, {Klessen}, {Padoan}, {Goldsmith},
  {Arce}, {Bally}, {Carpenter}, {Ginsburg}, {Johnstone}, {Kauffmann}, {Kong},
  {Lis}, {Mairs}, {Pillai}, {Pineda}, \& {Duarte-Cabral}}]{Suri19}
{Suri}, S., {S{\'a}nchez-Monge}, {\'A}., {Schilke}, P., {et~al.} 2019, \aap,
  623, A142

\bibitem[{{Takahashi} {et~al.}(2008){Takahashi}, {Saito}, {Ohashi}, {Kusakabe},
  {Takakuwa}, {Shimajiri}, {Tamura}, \& {Kawabe}}]{Takahashi08}
{Takahashi}, S., {Saito}, M., {Ohashi}, N., {et~al.} 2008, \apj, 688, 344

\bibitem[{Tanabe {et~al.}(2019)Tanabe, Nakamura, Tsukagoshi, Shimajiri, Ishii,
  Kawabe, Feddersen, Kong, Arce, Bally, Carpenter, \& Momose}]{Tanabe19}
Tanabe, Y., Nakamura, F., Tsukagoshi, T., {et~al.} 2019, Publications of the
  Astronomical Society of Japan, 71

\bibitem[{{Teixeira} {et~al.}(2016){Teixeira}, {Takahashi}, {Zapata}, \&
  {Ho}}]{Teixeira16}
{Teixeira}, P.~S., {Takahashi}, S., {Zapata}, L.~A., \& {Ho}, P.~T.~P. 2016,
  \aap, 587, A47

\bibitem[{{Williams} {et~al.}(2003){Williams}, {Plambeck}, \&
  {Heyer}}]{Williams03}
{Williams}, J.~P., {Plambeck}, R.~L., \& {Heyer}, M.~H. 2003, \apj, 591, 1025

\bibitem[{{Wilson} {et~al.}(2005){Wilson}, {Dame}, {Masheder}, \&
  {Thaddeus}}]{Wilson05}
{Wilson}, B.~A., {Dame}, T.~M., {Masheder}, M.~R.~W., \& {Thaddeus}, P. 2005,
  \aap, 430, 523

\bibitem[{Yang {et~al.}(2018)Yang, Green, II, Lee, J{\o}rgensen, Kristensen,
  Mottram, Herczeg, Karska, Dionatos, Bergin, Bouwman, van Dishoeck, van
  Kempen, Larson, \& Y{\i}ld{\i}z}]{Yang18}
Yang, Y.-L., Green, J.~D., II, N. J.~E., {et~al.} 2018, The Astrophysical
  Journal, 860, 174

\bibitem[{{Zapata} {et~al.}(2005){Zapata}, {Rodr{\'{\i}}guez}, {Ho}, {Zhang},
  {Qi}, \& {Kurtz}}]{Zapata05}
{Zapata}, L.~A., {Rodr{\'{\i}}guez}, L.~F., {Ho}, P.~T.~P., {et~al.} 2005,
  \apjl, 630, L85

\bibitem[{{Zhang} {et~al.}(2016){Zhang}, {Arce}, {Mardones}, {Cabrit},
  {Dunham}, {Garay}, {Noriega-Crespo}, {Offner}, {Raga}, \&
  {Corder}}]{ZhangY16}
{Zhang}, Y., {Arce}, H.~G., {Mardones}, D., {et~al.} 2016, \apj, 832, 158

\end{thebibliography}

\appendix
\section{Protostellar Properties}
\begin{figure*}
\begin{minipage}{.5\linewidth}
\centering
\subfloat{\includegraphics[scale=.6]{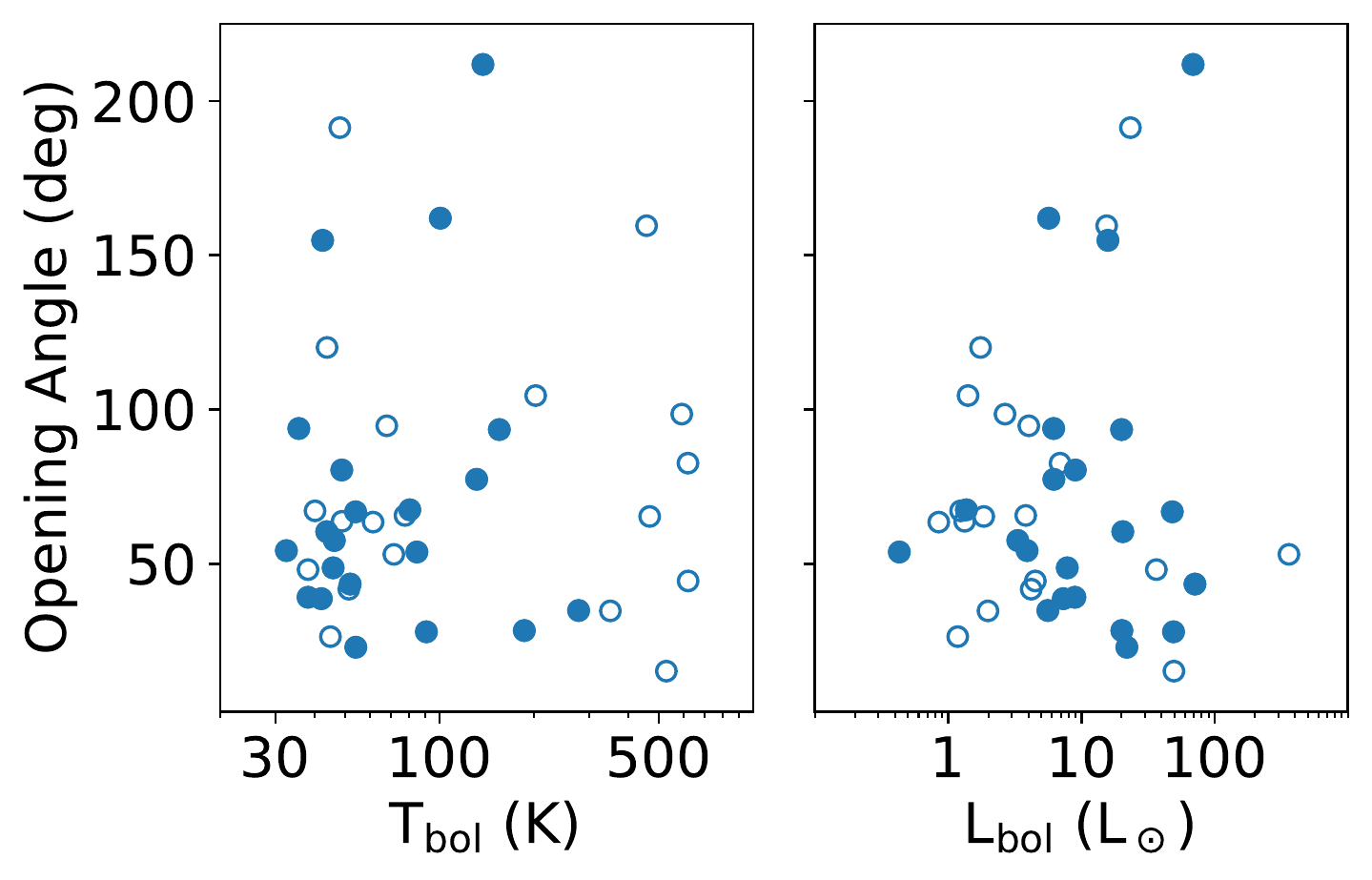}}
\end{minipage}%
\begin{minipage}{.5\linewidth}
\centering
\subfloat{\includegraphics[scale=.6]{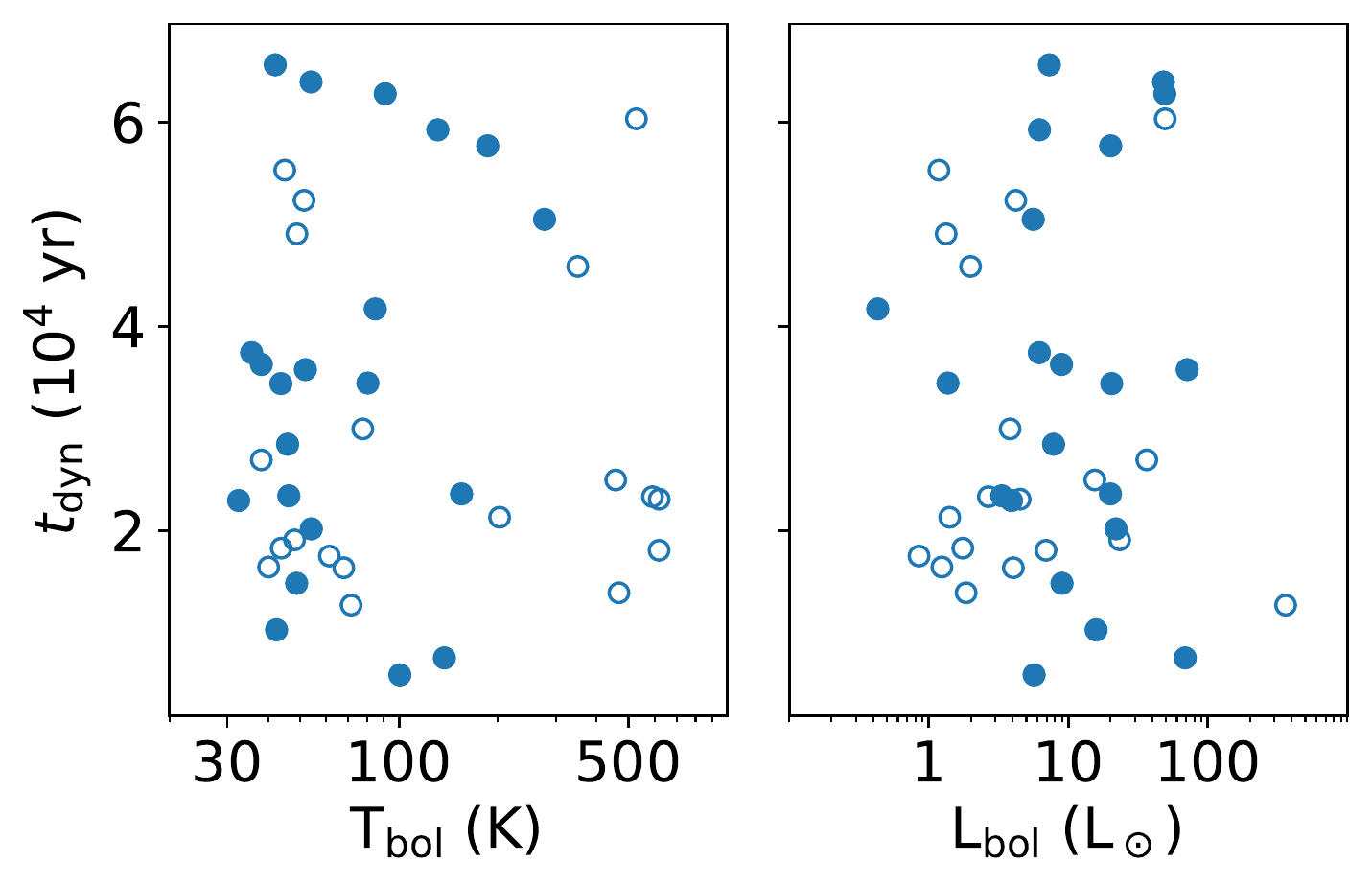}}
\end{minipage}\par\medskip
\begin{minipage}{.5\linewidth}
\centering
\subfloat{\includegraphics[scale=.6]{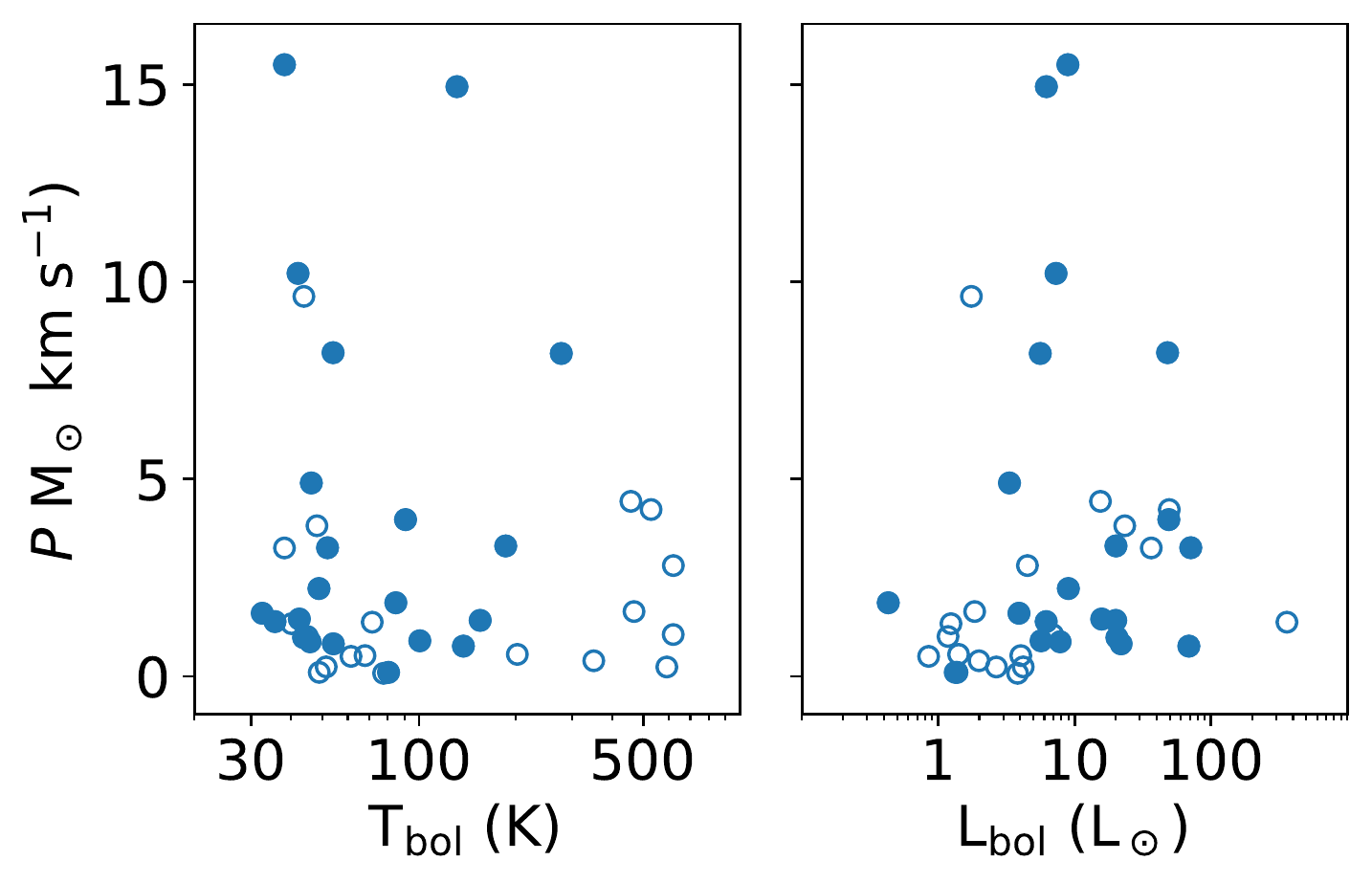}}
\end{minipage}%
\begin{minipage}{.5\linewidth}
\centering
\subfloat{\includegraphics[scale=.6]{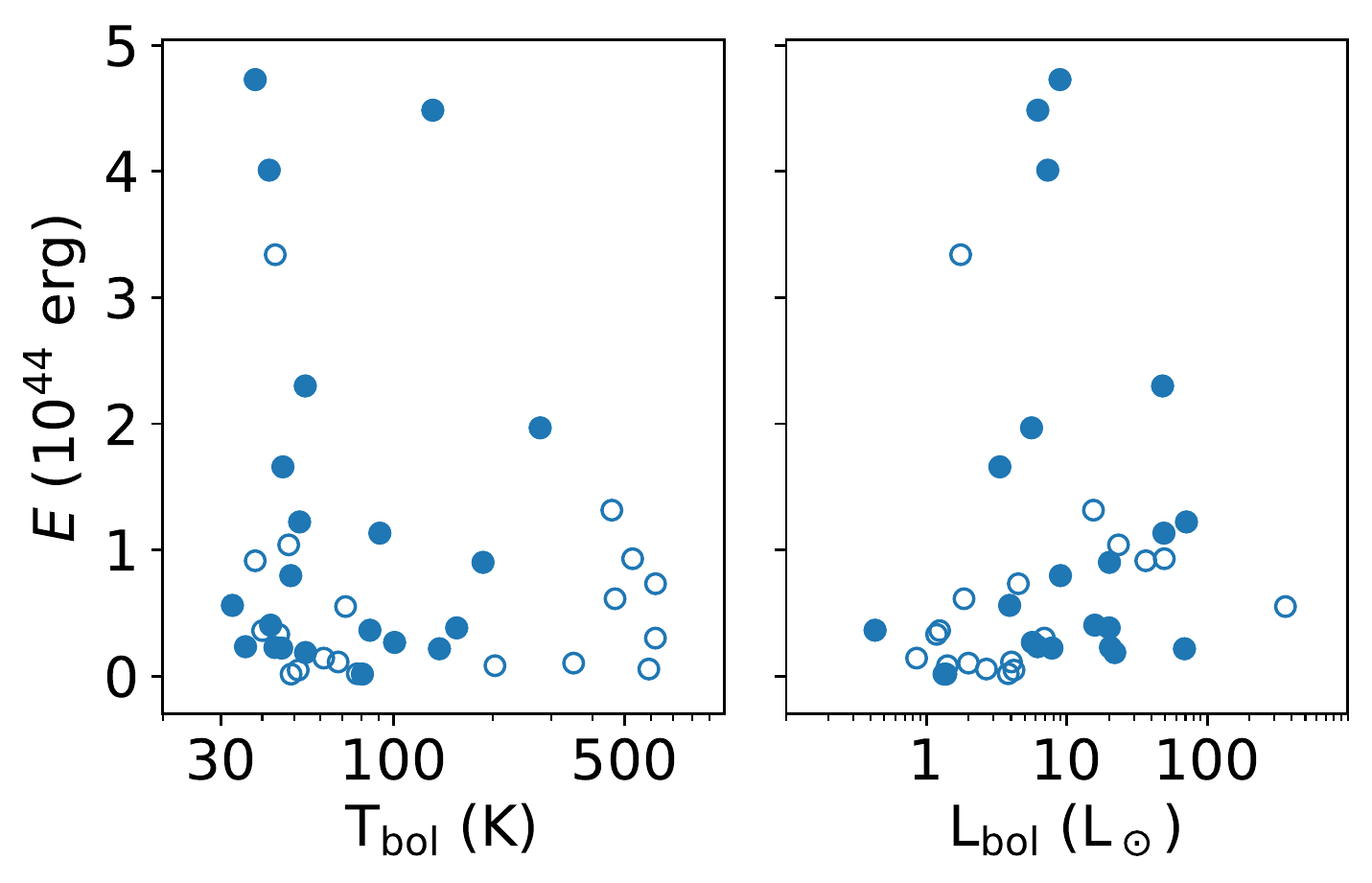}}
\end{minipage}\par\medskip
\begin{minipage}{.5\linewidth}
\centering
\subfloat{\includegraphics[scale=.6]{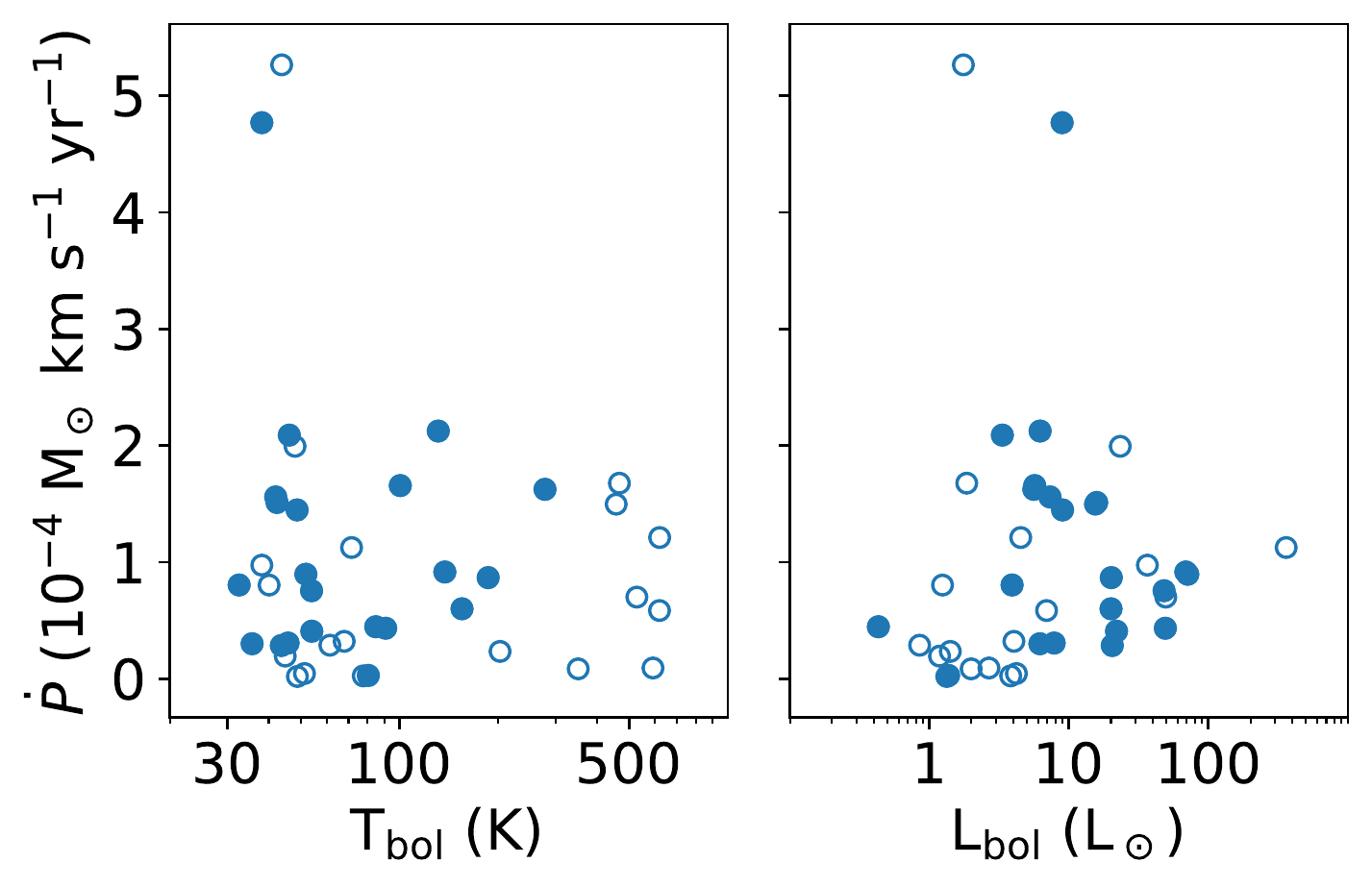}}
\end{minipage}%
\begin{minipage}{.5\linewidth}
\centering
\subfloat{\includegraphics[scale=.6]{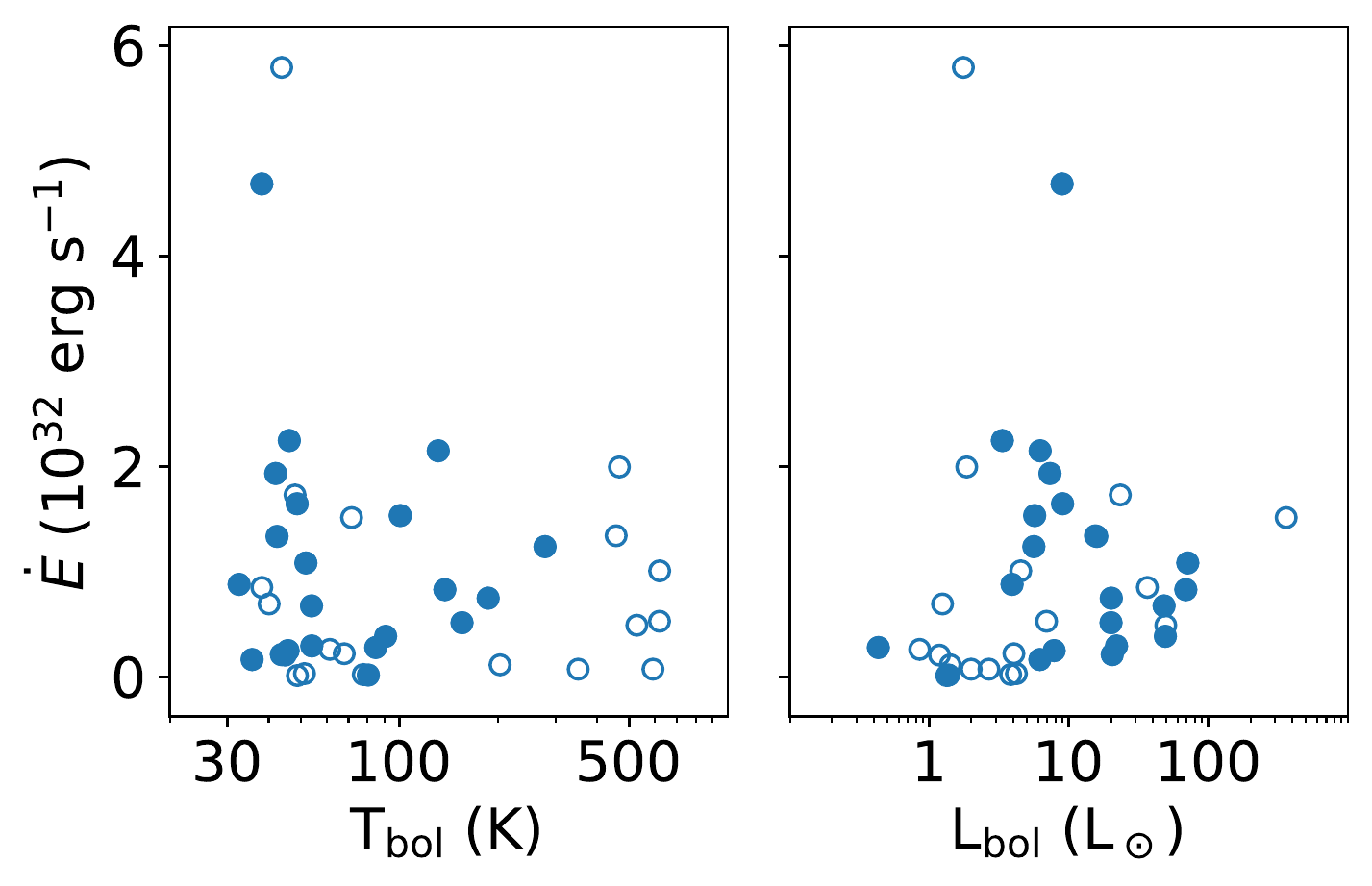}}
\end{minipage}\par\medskip
\caption{Protostellar $T_{\rm bol}$ and $L_{\rm bol}$ plotted against outflow properties for 40 outflows driven by HOPS sources. The outflow properties are the average of both lobes, when detected. Filled circles indicate outflows with a confidence level of ``definite'' for all detected lobes. Open circles indicate outflows with at least one ``marginal'' lobe. The outflow properties are given in Tables~\ref{tab:physics} and \ref{tab:angles}.}
\label{fig:appendix_scatter}
\end{figure*}

The HOPS catalog from \citet{Furlan16} compiles photometric measurements for protostars from 1.2-870 $\mu$m. From these SEDs, they calculate the bolometric luminosity $L_{\rm bol}$ and temperature $T_{\rm bol}$. $L_{\rm bol}$ is calculated by integrating the observed SED directly. $L_{\rm bol}$, if dominated by accretion luminosity, is a function of the accretion rate and protostellar mass \citep{Takahashi08}. Since accretion adds mass to the protostar, $L_{\rm bol}$ may be considered a proxy for the protostellar mass. $T_{\rm bol}$ is the temperature of a blackbody with the same mean frequency as the observed SED. $T_{\rm bol}$ increases as a protostar evolves and clears its envelope, making it a proxy for protostellar age \citep{Myers93}. While $L_{\rm bol}$ and $T_{\rm bol}$ are generally thought to trace the protostellar mass and age, respectively, other factors such as extinction and inclination may also affect these measurements. For the 40 outflows presented here with associated HOPS sources, we show the correlation between the protostellar $L_{\rm bol}$ and $T_{\rm bol}$ and outflow properties in Figure~\ref{fig:appendix_scatter}. \citet{Arce06} discovered a correlation between outflow opening angle and protostellar age, as traced by $T_{\rm bol}$. They found that more evolved (hotter $T_{\rm bol}$) protostars drive wider outflows. Figure~\ref{fig:appendix_scatter} shows that our measured outflow opening angles are not correlated with $T_{\rm bol}$ or $L_{\rm bol}$.

\citet{Takahashi08} investigated the relationship between momentum flux (identical to our $\dot P$) and $L_{\rm bol}$ in the outflows they identified in OMC-2/3. In their Figure 12, they show these outflows are consistent with a trend of increasing $\dot P$ with increasing $L_{\rm bol}$ spanning six orders of magnitude in $L_{\rm bol}$. As shown in Figure~\ref{fig:appendix_scatter}, we find no evidence for a correlation between $\dot P$ and $L_{\rm bol}$ within our outflow sample. However, our measurements are consistent with the scatter seen in Figure 12 of \citet{Takahashi08}. Figure~\ref{fig:appendix_scatter} also shows the lack of correlation between the protostellar $T_{\rm bol}$ or $L_{\rm bol}$ with the outflow momentum $P$, kinetic energy $E$, energy injection rate $\dot E$, and dynamical time $t_{\rm dyn}$.

\begin{figure*}
\begin{minipage}{\linewidth}
\centering
\subfloat{\includegraphics[scale=.7]{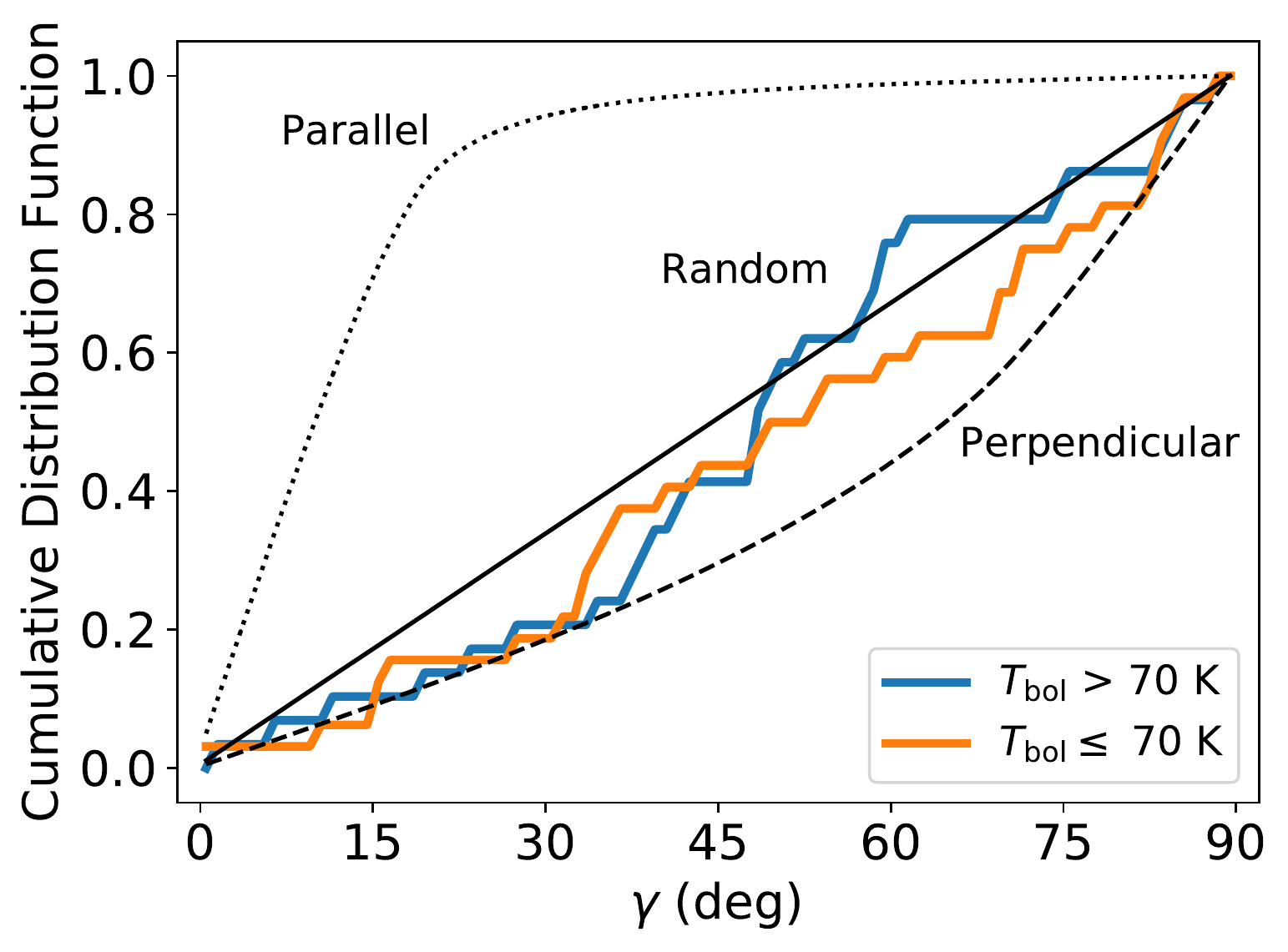}}
\end{minipage}\par\medskip
\begin{minipage}{\linewidth}
\centering
\subfloat{\includegraphics[scale=.7]{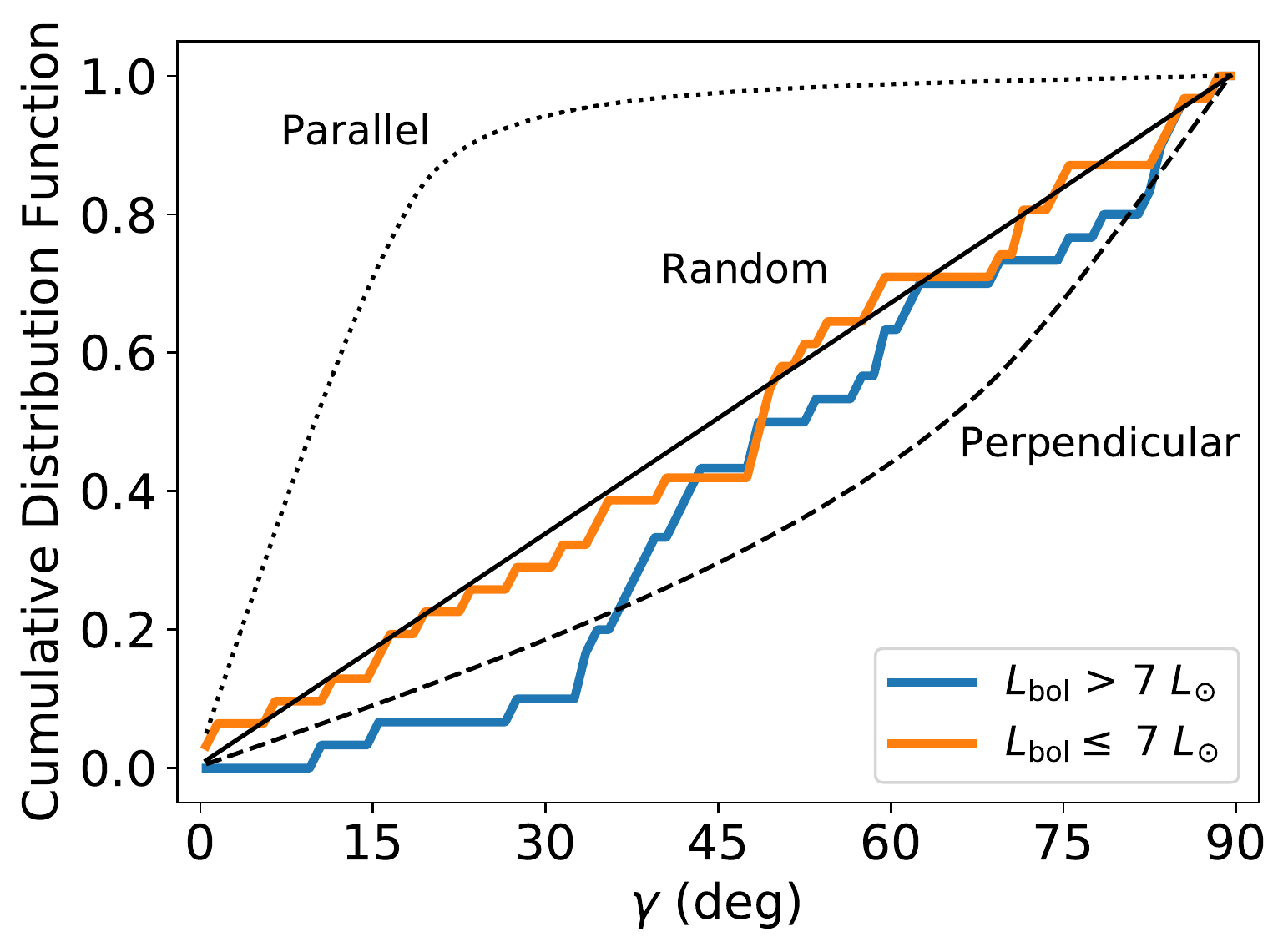}}
\end{minipage}\par\medskip
\caption{Outflow-filament alignment in low- and high-$T_{\rm bol}$ and $L_{\rm bol}$ samples. Symbols are the same as Figure~\ref{fig:gamma_cdf}.}
\label{fig:appendix_gamma_cdf}
\end{figure*}

We also look for any difference in the outflow-filament alignment as a function of protostellar $T_{\rm bol}$ and $L_{\rm bol}$. In Perseus, older, less embedded, hotter $T_{\rm bol}$ protostars show a slightly more perpendicular outflow-filament alignment than younger protostars \citep{Stephens17}. In Figure~\ref{fig:appendix_gamma_cdf}, we show the outflow-filament alignment in Orion A among HOPS protostars, split into low- and high-$T_{\rm bol}$ and $L_{\rm bol}$ samples. There is no significant difference between the low- and high-$T_{\rm bol}$ outflow-filament alignment. The high-$L_{\rm bol}$ outflows are slightly more consistent with a perpendicular outflow-filament alignment compared to the low-$L_{\rm bol}$ sample.

\clearpage
\section{All Outflows in CARMA-NRO Orion}\label{sec:appendix}

\begin{figure*}[h]
\centering
\includegraphics[width=\textwidth]{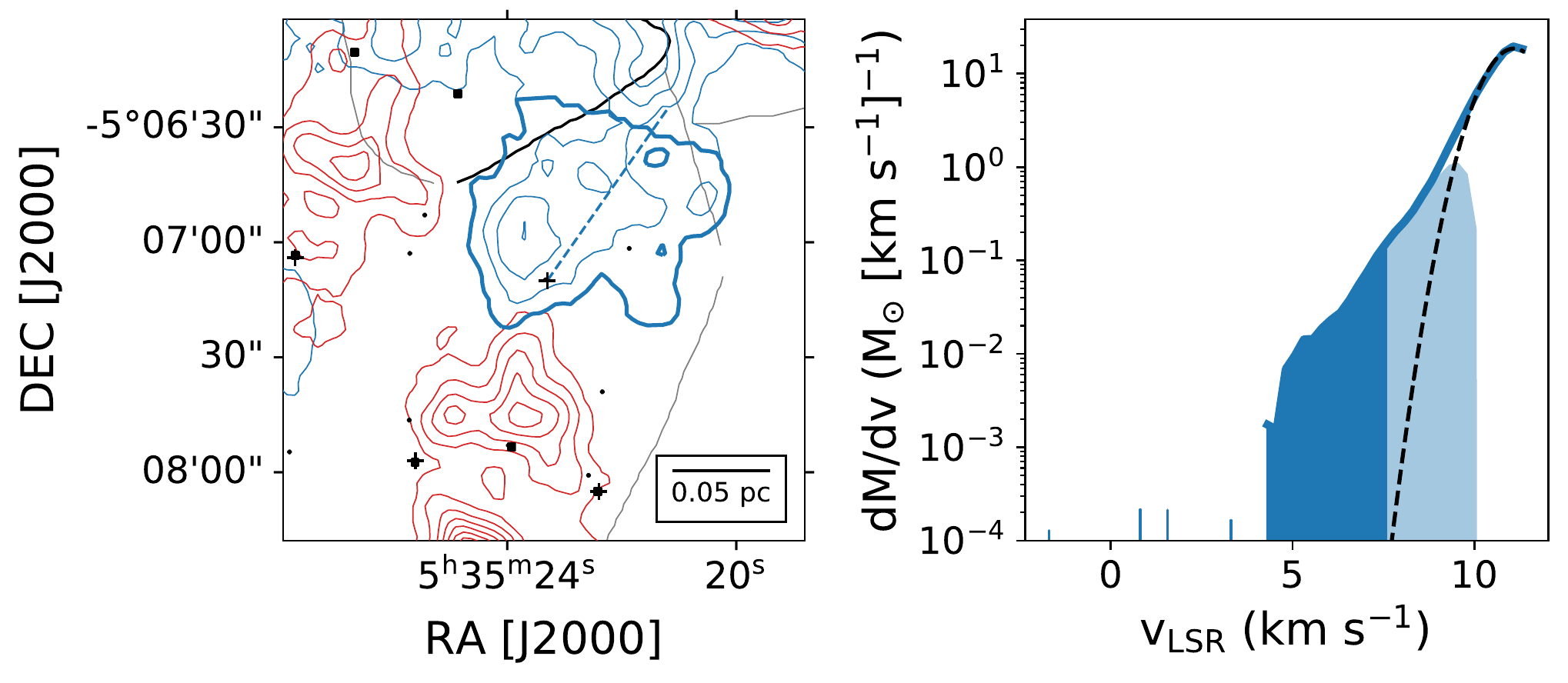}

    \caption{SMZ 11 outflow. The left panel shows the outflow, position angle, nearby sources, and filaments.
    The velocity range of integration is given by v$_{\rm blue}$/v$_{\rm red}$ in Table~\ref{tab:outflows}
    and the contours go from $5$ to $50\sigma$ in steps of 5$\sigma$, where $\sigma$ is the RMS error in the integrated map. Symbols are the same as Figure~\ref{fig:stamp}.
    The right panel shows the mass spectrum with fit, where $\sigma$ is the RMS error in the integrated map. Symbols are the same as Figure~\ref{fig:dmdv}.}
    \end{figure*}
\begin{figure*}[h]
\centering
\includegraphics[width=\textwidth]{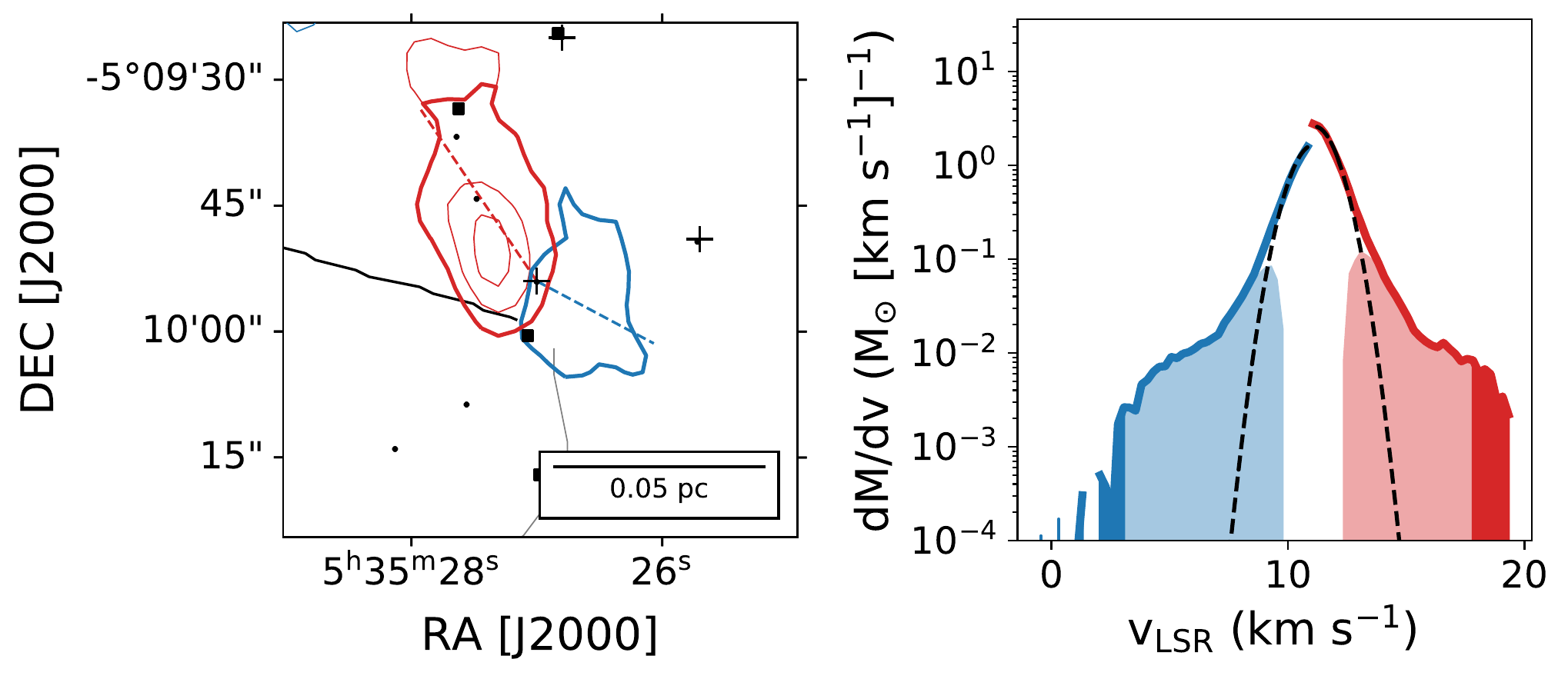}

    \caption{SMZ 17 outflow. The left panel shows the outflow, position angle, nearby sources, and filaments.
    The velocity range of integration is given by v$_{\rm blue}$/v$_{\rm red}$ in Table~\ref{tab:outflows}
    and the contours go from $5$ to $50\sigma$ in steps of 5$\sigma$, where $\sigma$ is the RMS error in the integrated map. Symbols are the same as Figure~\ref{fig:stamp}.
    The right panel shows the mass spectrum with fit, where $\sigma$ is the RMS error in the integrated map. Symbols are the same as Figure~\ref{fig:dmdv}.}
    \end{figure*}
\begin{figure*}[p]
\centering
\includegraphics[width=\textwidth]{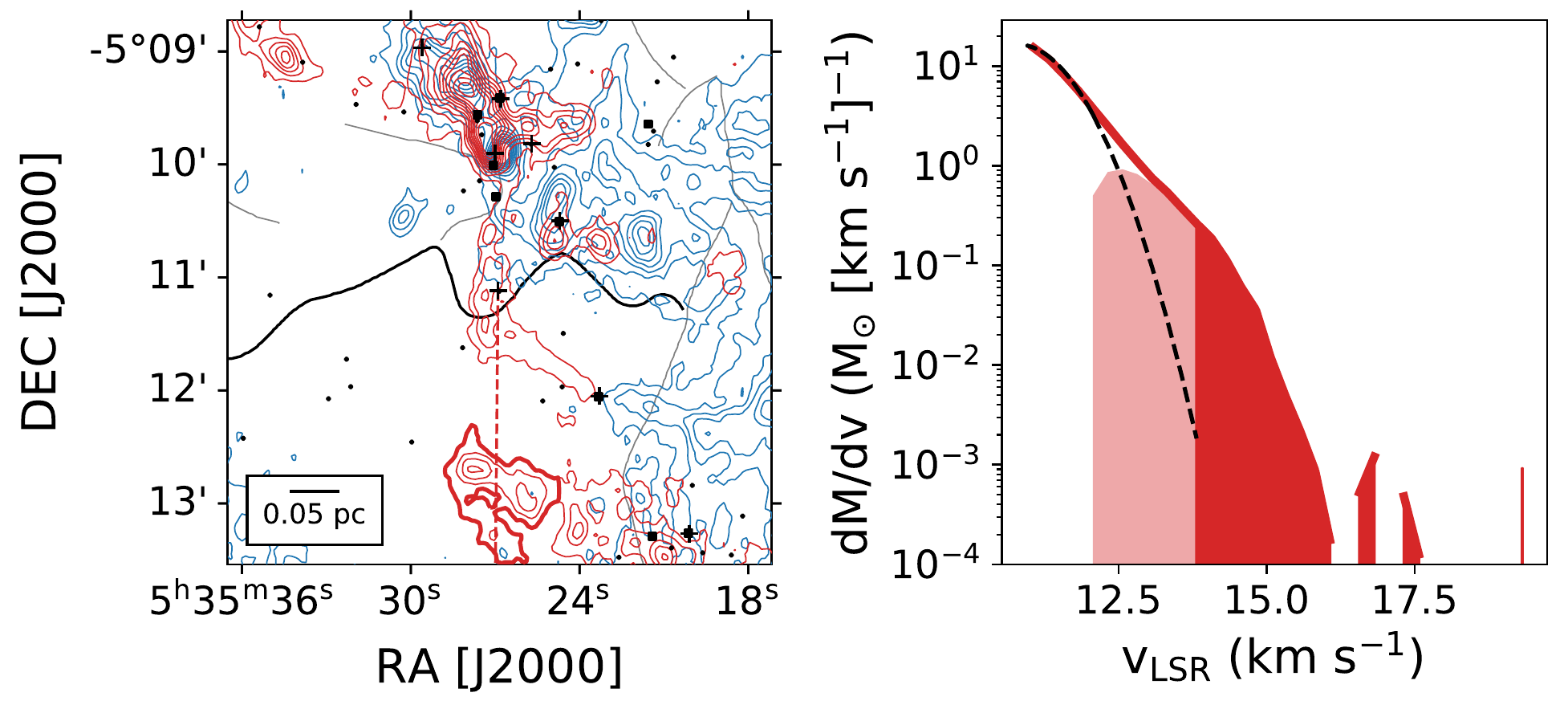}

    \caption{SMZ 21 outflow. The left panel shows the outflow, position angle, nearby sources, and filaments.
    The velocity range of integration is given by v$_{\rm blue}$/v$_{\rm red}$ in Table~\ref{tab:outflows}
    and the contours go from $5$ to $50\sigma$ in steps of 5$\sigma$, where $\sigma$ is the RMS error in the integrated map. Symbols are the same as Figure~\ref{fig:stamp}.
    The right panel shows the mass spectrum with fit, where $\sigma$ is the RMS error in the integrated map. Symbols are the same as Figure~\ref{fig:dmdv}.}
    \end{figure*}
\begin{figure*}[p]
\centering
\includegraphics[width=\textwidth]{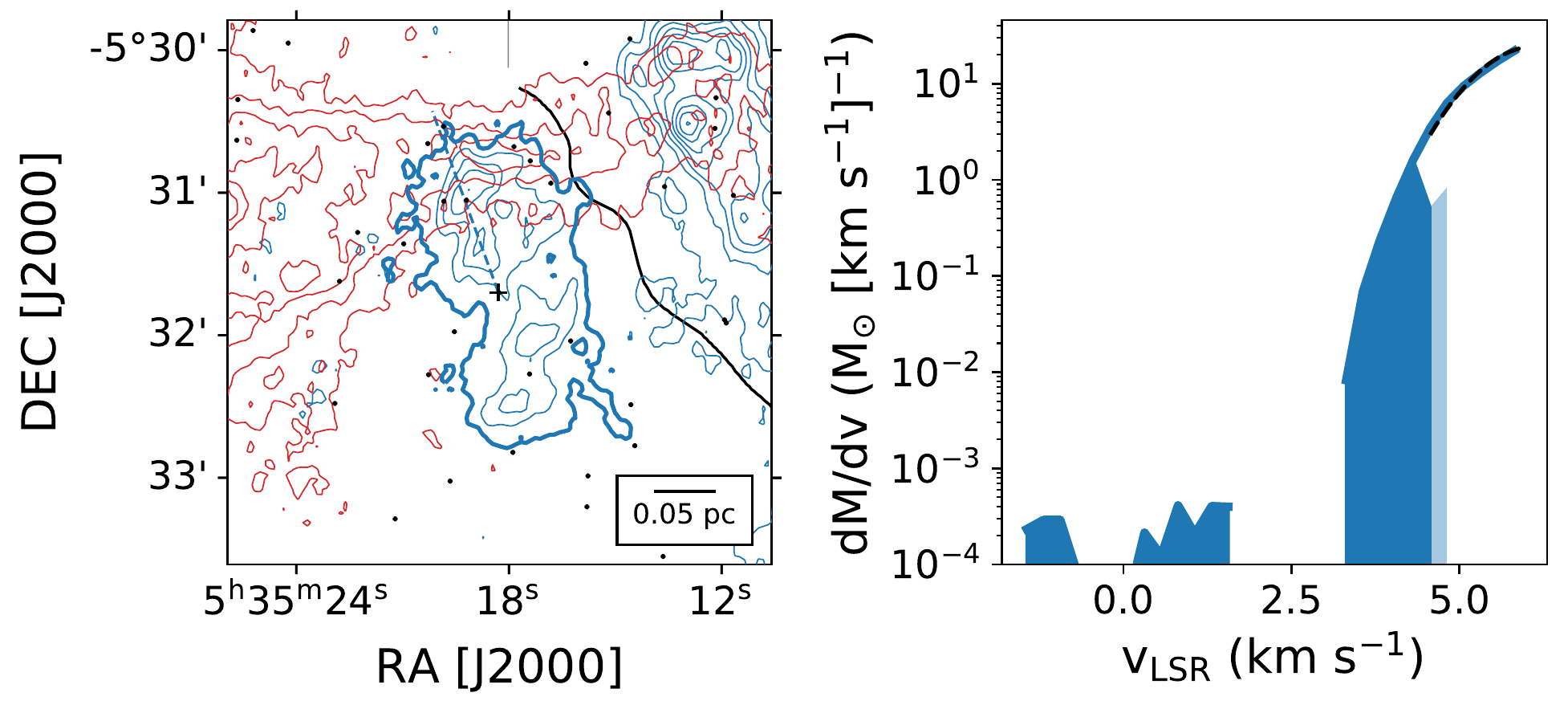}

    \caption{SMZ 30 outflow. The left panel shows the outflow, position angle, nearby sources, and filaments.
    The velocity range of integration is given by v$_{\rm blue}$/v$_{\rm red}$ in Table~\ref{tab:outflows}
    and the contours go from $5$ to $50\sigma$ in steps of 5$\sigma$, where $\sigma$ is the RMS error in the integrated map. Symbols are the same as Figure~\ref{fig:stamp}.
    The right panel shows the mass spectrum with fit, where $\sigma$ is the RMS error in the integrated map. Symbols are the same as Figure~\ref{fig:dmdv}.}
    \end{figure*}
\begin{figure*}[p]
\centering
\includegraphics[width=\textwidth]{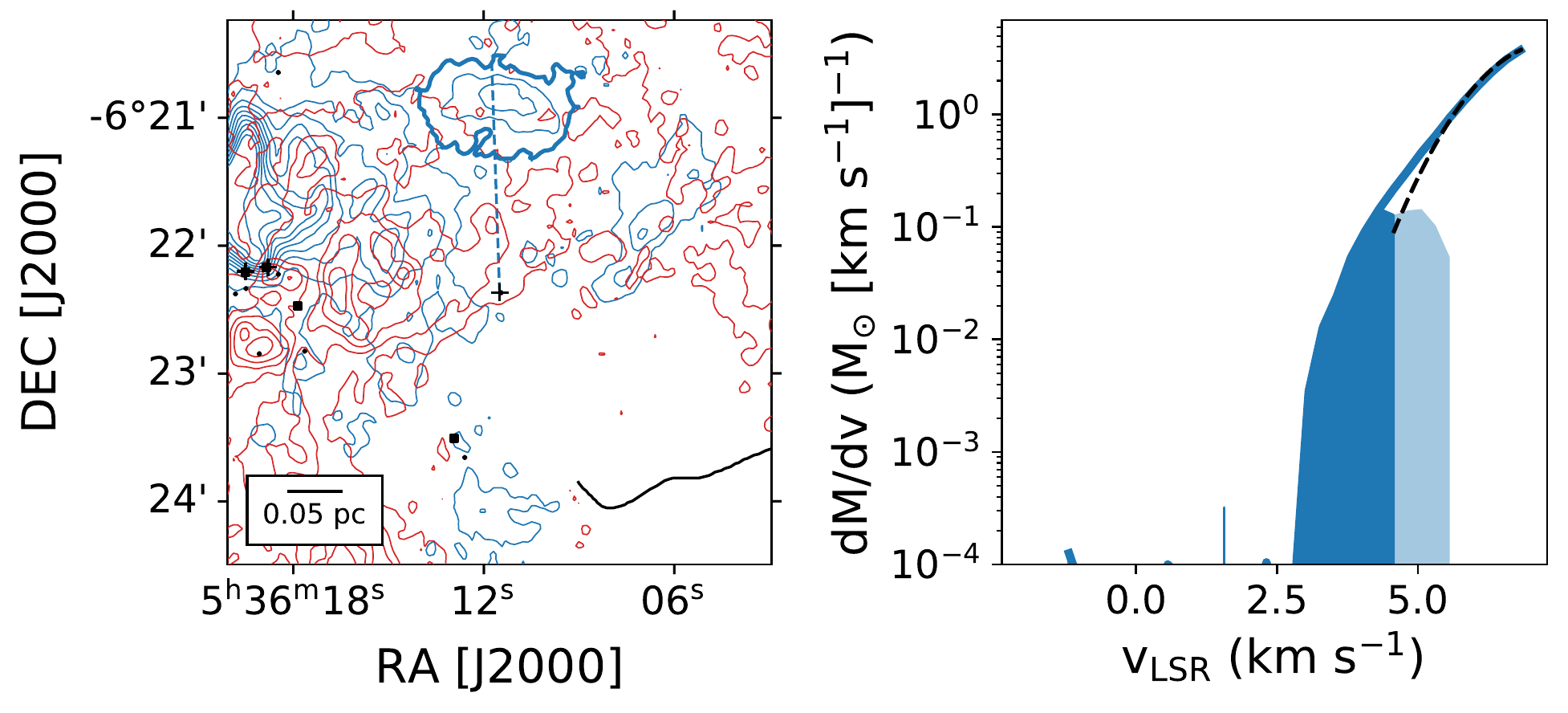}

    \caption{SMZ 50 outflow. The left panel shows the outflow, position angle, nearby sources, and filaments.
    The velocity range of integration is given by v$_{\rm blue}$/v$_{\rm red}$ in Table~\ref{tab:outflows}
    and the contours go from $5$ to $50\sigma$ in steps of 5$\sigma$, where $\sigma$ is the RMS error in the integrated map. Symbols are the same as Figure~\ref{fig:stamp}.
    The right panel shows the mass spectrum with fit, where $\sigma$ is the RMS error in the integrated map. Symbols are the same as Figure~\ref{fig:dmdv}.}
    \end{figure*}
\begin{figure*}[p]
\centering
\includegraphics[width=\textwidth]{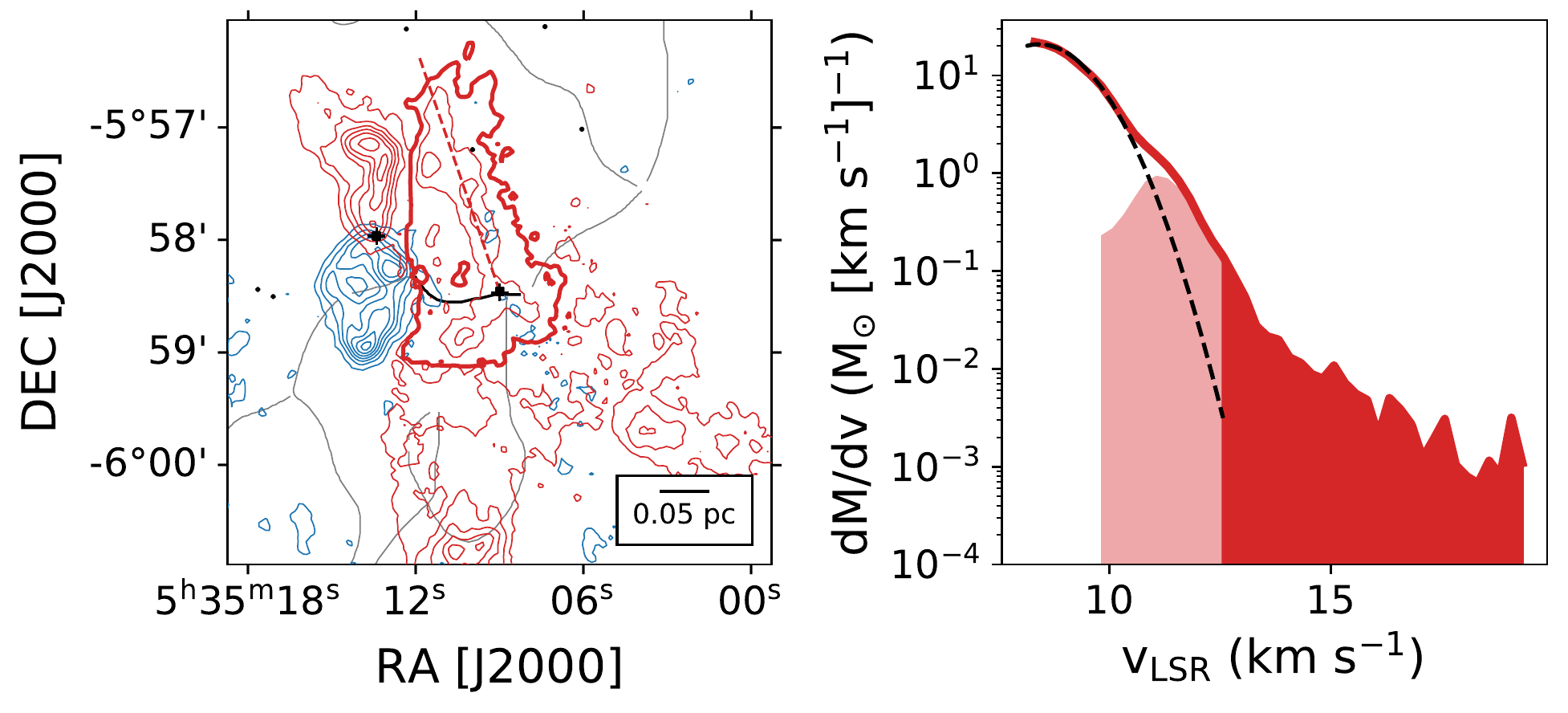}

    \caption{HOPS 10 outflow. The left panel shows the outflow, position angle, nearby sources, and filaments.
    The velocity range of integration is given by v$_{\rm blue}$/v$_{\rm red}$ in Table~\ref{tab:outflows}
    and the contours go from $5$ to $50\sigma$ in steps of 5$\sigma$, where $\sigma$ is the RMS error in the integrated map. Symbols are the same as Figure~\ref{fig:stamp}.
    The right panel shows the mass spectrum with fit, where $\sigma$ is the RMS error in the integrated map. Symbols are the same as Figure~\ref{fig:dmdv}.}
    \end{figure*}
\begin{figure*}[p]
\centering
\includegraphics[width=\textwidth]{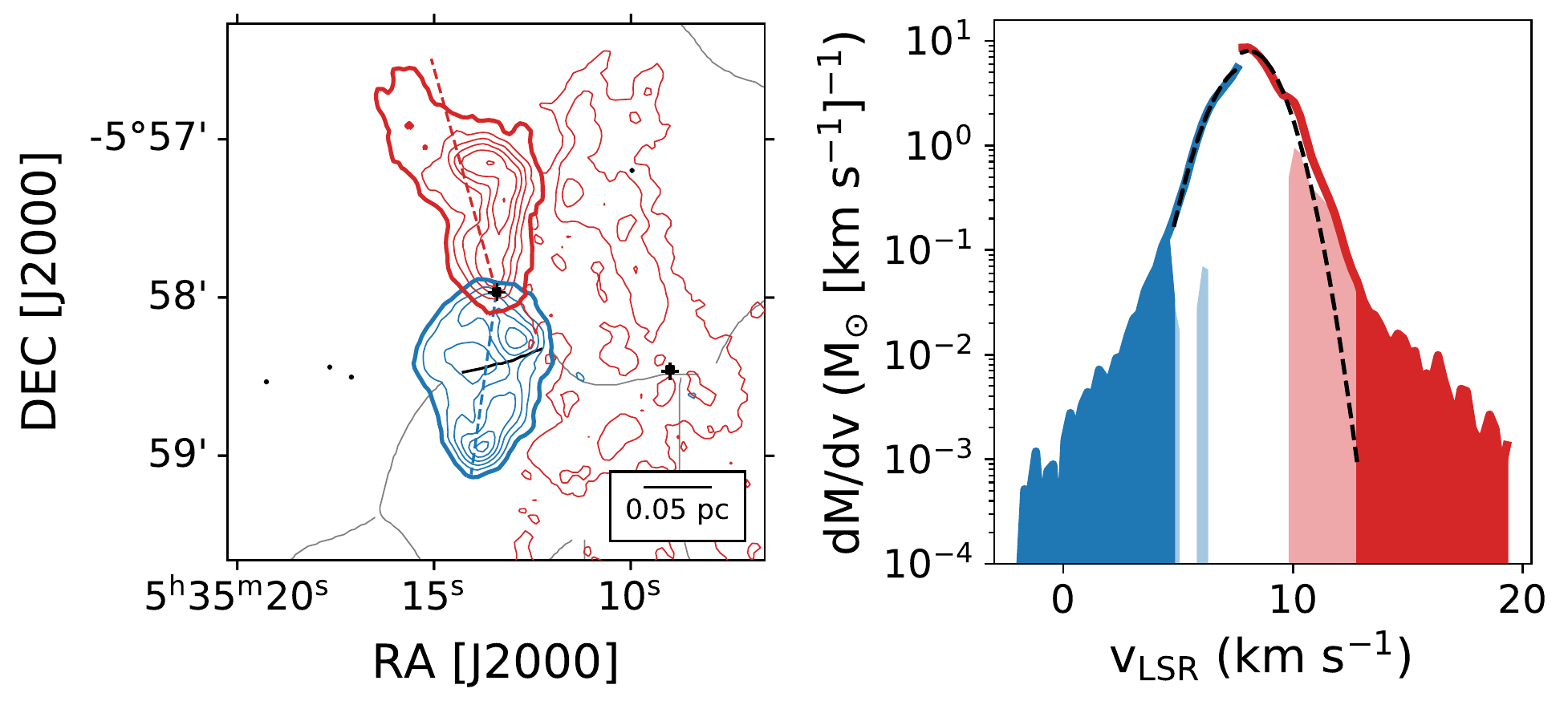}

    \caption{HOPS 11 outflow. The left panel shows the outflow, position angle, nearby sources, and filaments.
    The velocity range of integration is given by v$_{\rm blue}$/v$_{\rm red}$ in Table~\ref{tab:outflows}
    and the contours go from $5$ to $50\sigma$ in steps of 5$\sigma$, where $\sigma$ is the RMS error in the integrated map. Symbols are the same as Figure~\ref{fig:stamp}.
    The right panel shows the mass spectrum with fit, where $\sigma$ is the RMS error in the integrated map. Symbols are the same as Figure~\ref{fig:dmdv}.}
    \end{figure*}
\begin{figure*}[p]
\centering
\includegraphics[width=\textwidth]{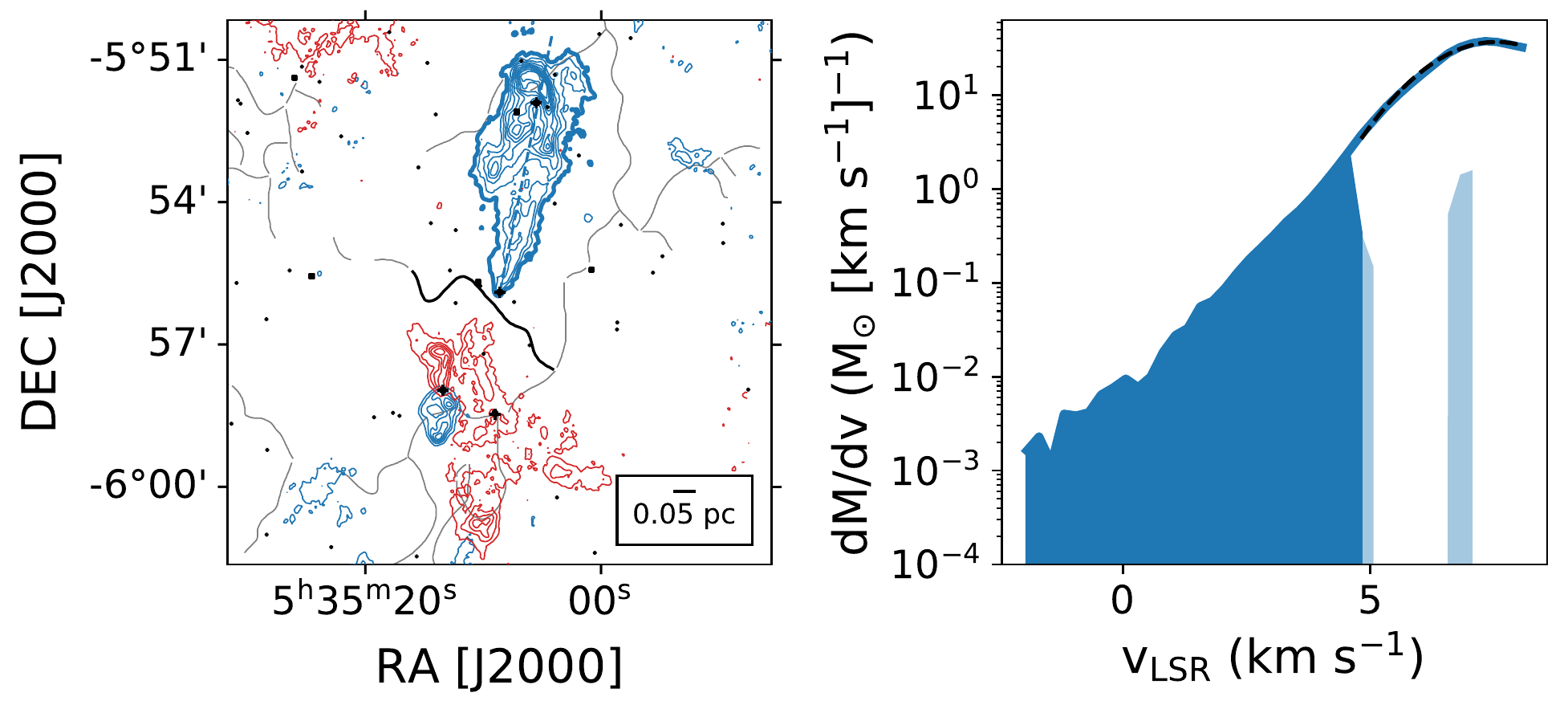}

    \caption{HOPS 12 outflow. The left panel shows the outflow, position angle, nearby sources, and filaments.
    The velocity range of integration is given by v$_{\rm blue}$/v$_{\rm red}$ in Table~\ref{tab:outflows}
    and the contours go from $5$ to $50\sigma$ in steps of 5$\sigma$, where $\sigma$ is the RMS error in the integrated map. Symbols are the same as Figure~\ref{fig:stamp}.
    The right panel shows the mass spectrum with fit, where $\sigma$ is the RMS error in the integrated map. Symbols are the same as Figure~\ref{fig:dmdv}.}
    \end{figure*}
\begin{figure*}[p]
\centering
\includegraphics[width=\textwidth]{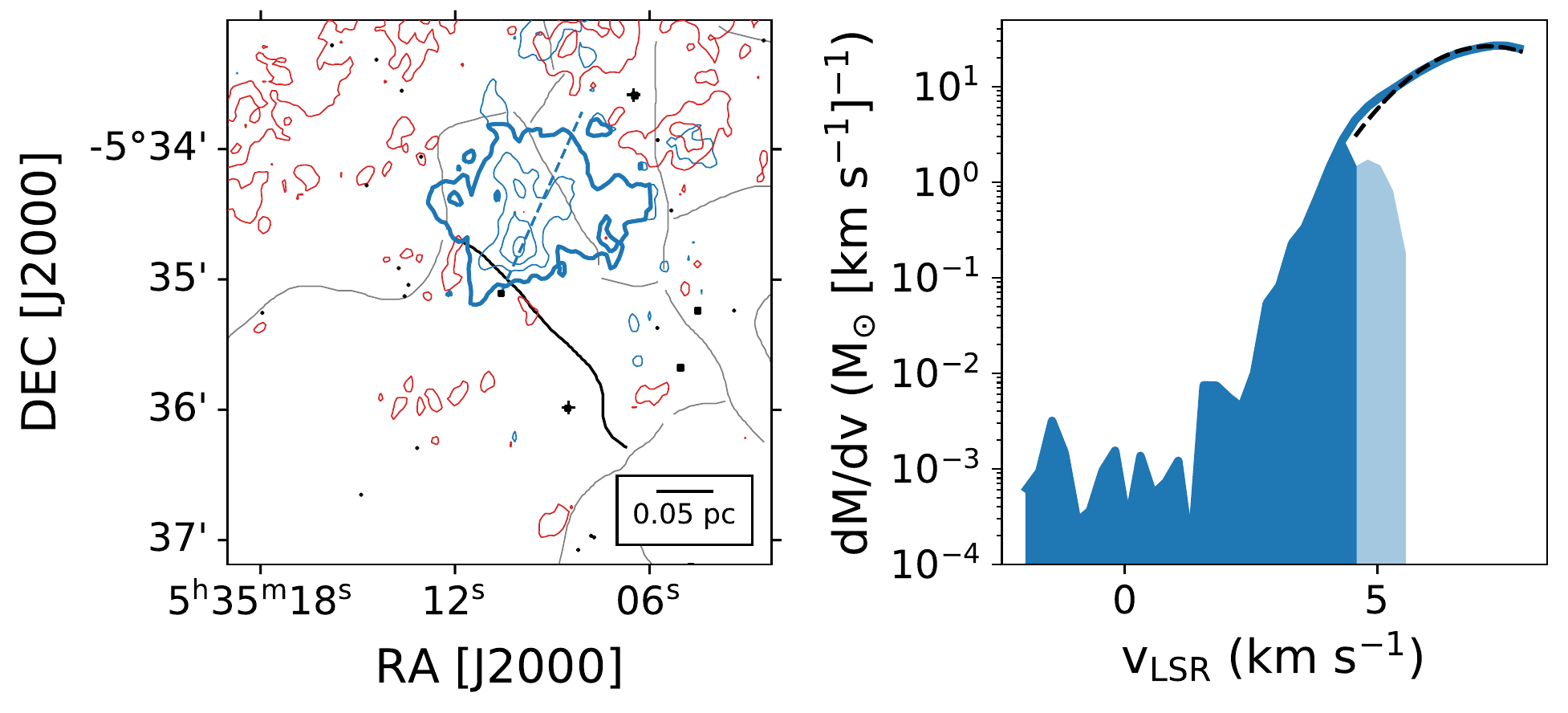}

    \caption{HOPS 44 outflow. The left panel shows the outflow, position angle, nearby sources, and filaments.
    The velocity range of integration is given by v$_{\rm blue}$/v$_{\rm red}$ in Table~\ref{tab:outflows}
    and the contours go from $5$ to $50\sigma$ in steps of 5$\sigma$, where $\sigma$ is the RMS error in the integrated map. Symbols are the same as Figure~\ref{fig:stamp}.
    The right panel shows the mass spectrum with fit, where $\sigma$ is the RMS error in the integrated map. Symbols are the same as Figure~\ref{fig:dmdv}.}
    \end{figure*}
\begin{figure*}[p]
\centering
\includegraphics[width=\textwidth]{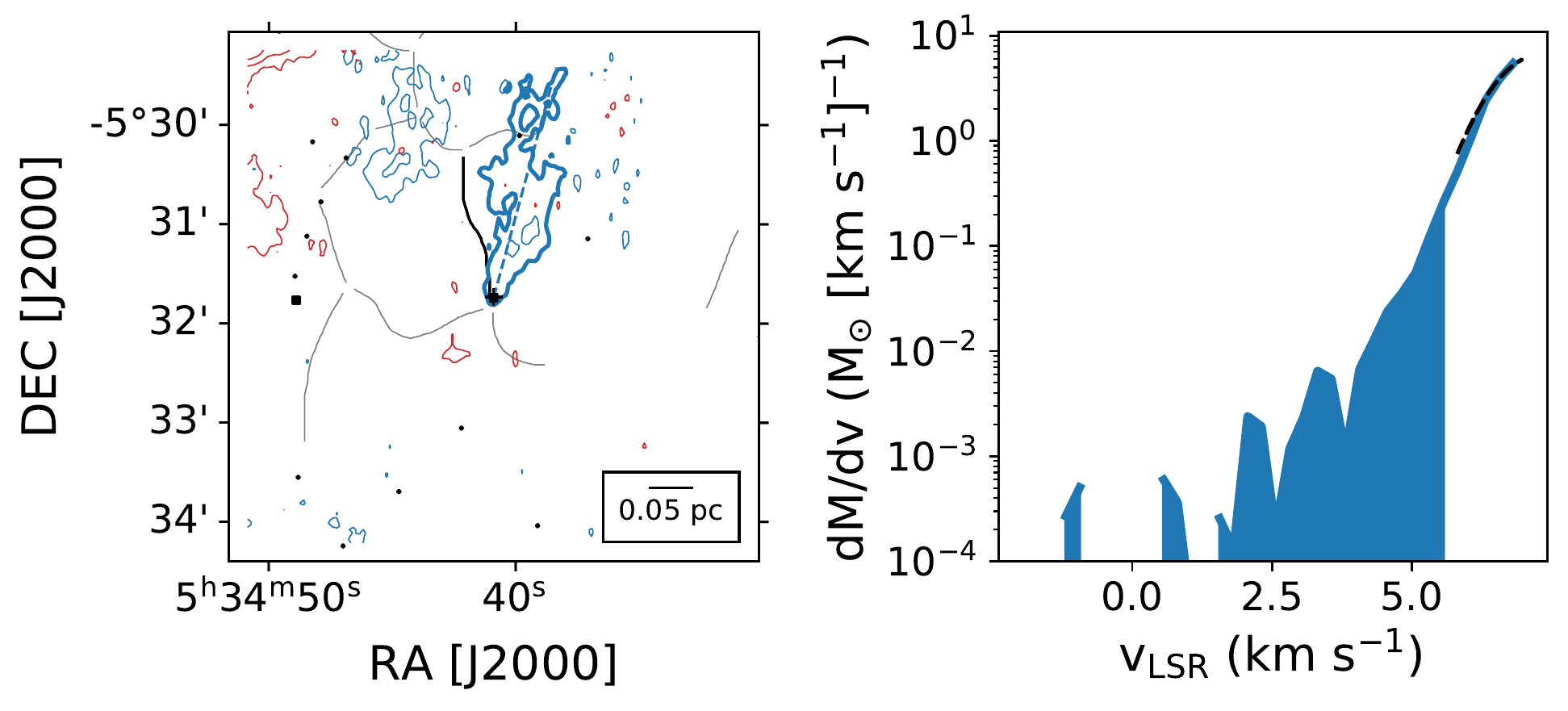}

    \caption{HOPS 50 outflow. The left panel shows the outflow, position angle, nearby sources, and filaments.
    The velocity range of integration is given by v$_{\rm blue}$/v$_{\rm red}$ in Table~\ref{tab:outflows}
    and the contours go from $5$ to $50\sigma$ in steps of 5$\sigma$, where $\sigma$ is the RMS error in the integrated map. Symbols are the same as Figure~\ref{fig:stamp}.
    The right panel shows the mass spectrum with fit, where $\sigma$ is the RMS error in the integrated map. Symbols are the same as Figure~\ref{fig:dmdv}.}
    \end{figure*}
\begin{figure*}[p]
\centering
\includegraphics[width=\textwidth]{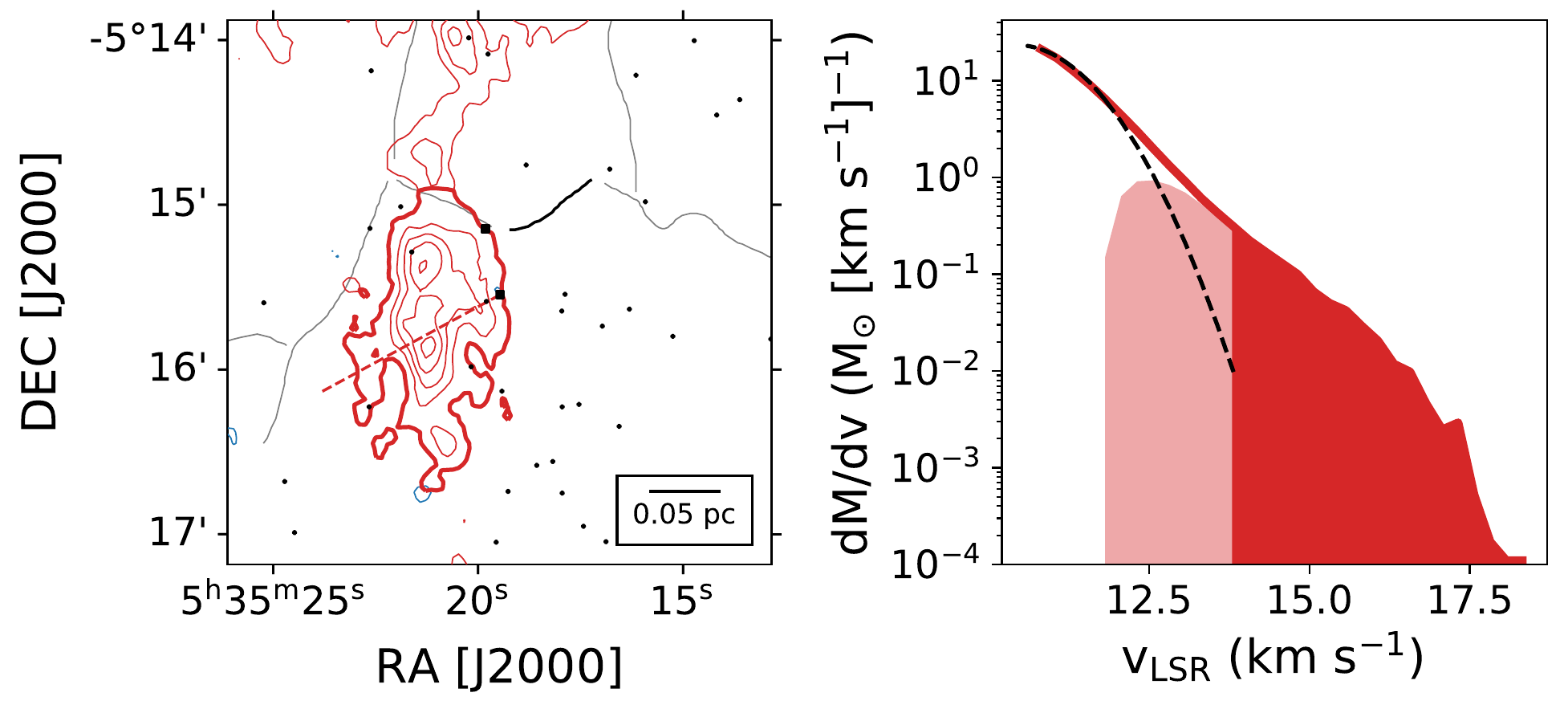}

    \caption{HOPS 56 outflow. The left panel shows the outflow, position angle, nearby sources, and filaments.
    The velocity range of integration is given by v$_{\rm blue}$/v$_{\rm red}$ in Table~\ref{tab:outflows}
    and the contours go from $5$ to $50\sigma$ in steps of 5$\sigma$, where $\sigma$ is the RMS error in the integrated map. Symbols are the same as Figure~\ref{fig:stamp}.
    The right panel shows the mass spectrum with fit, where $\sigma$ is the RMS error in the integrated map. Symbols are the same as Figure~\ref{fig:dmdv}.}
    \end{figure*}
\begin{figure*}[p]
\centering
\includegraphics[width=\textwidth]{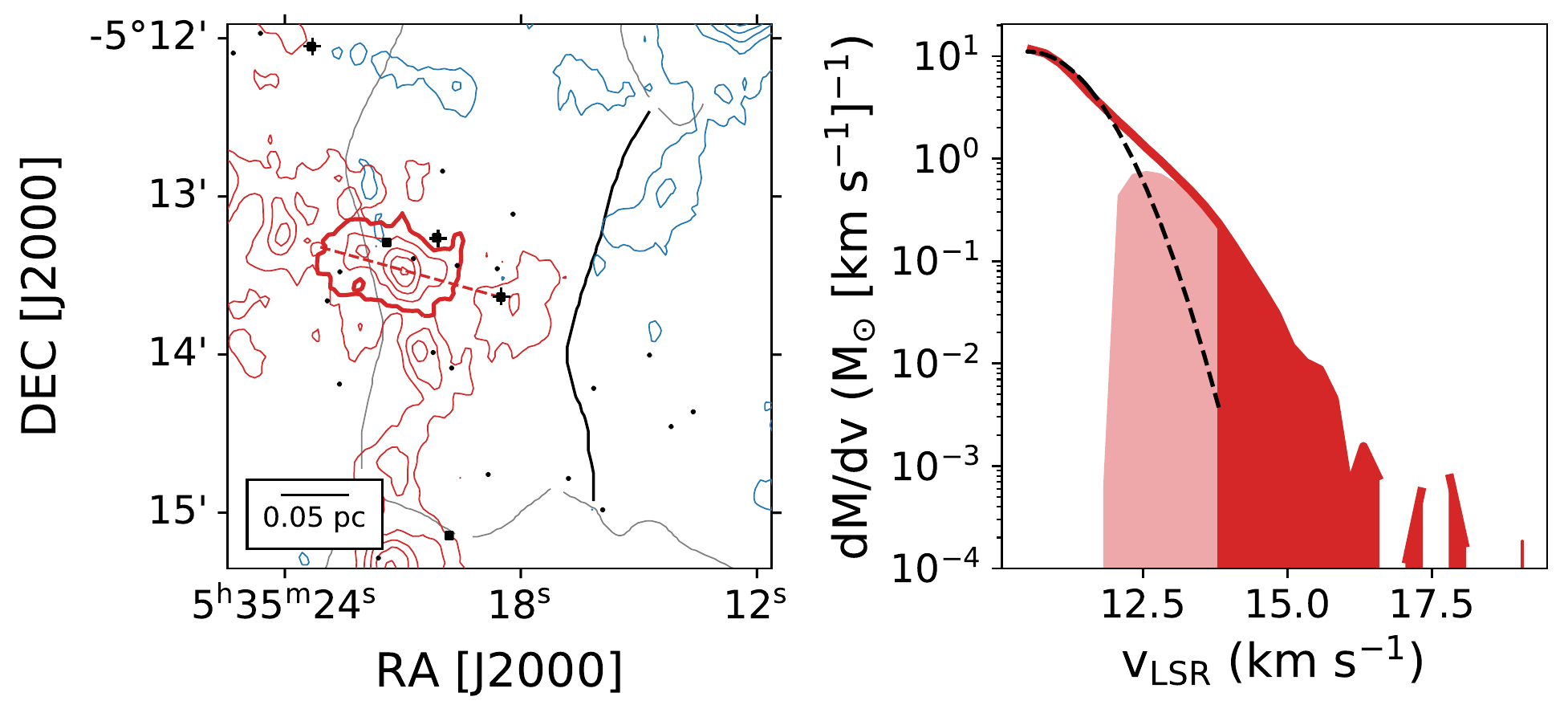}

    \caption{HOPS 58 outflow. The left panel shows the outflow, position angle, nearby sources, and filaments.
    The velocity range of integration is given by v$_{\rm blue}$/v$_{\rm red}$ in Table~\ref{tab:outflows}
    and the contours go from $5$ to $50\sigma$ in steps of 5$\sigma$, where $\sigma$ is the RMS error in the integrated map. Symbols are the same as Figure~\ref{fig:stamp}.
    The right panel shows the mass spectrum with fit, where $\sigma$ is the RMS error in the integrated map. Symbols are the same as Figure~\ref{fig:dmdv}.}
    \end{figure*}
\begin{figure*}[p]
\centering
\includegraphics[width=\textwidth]{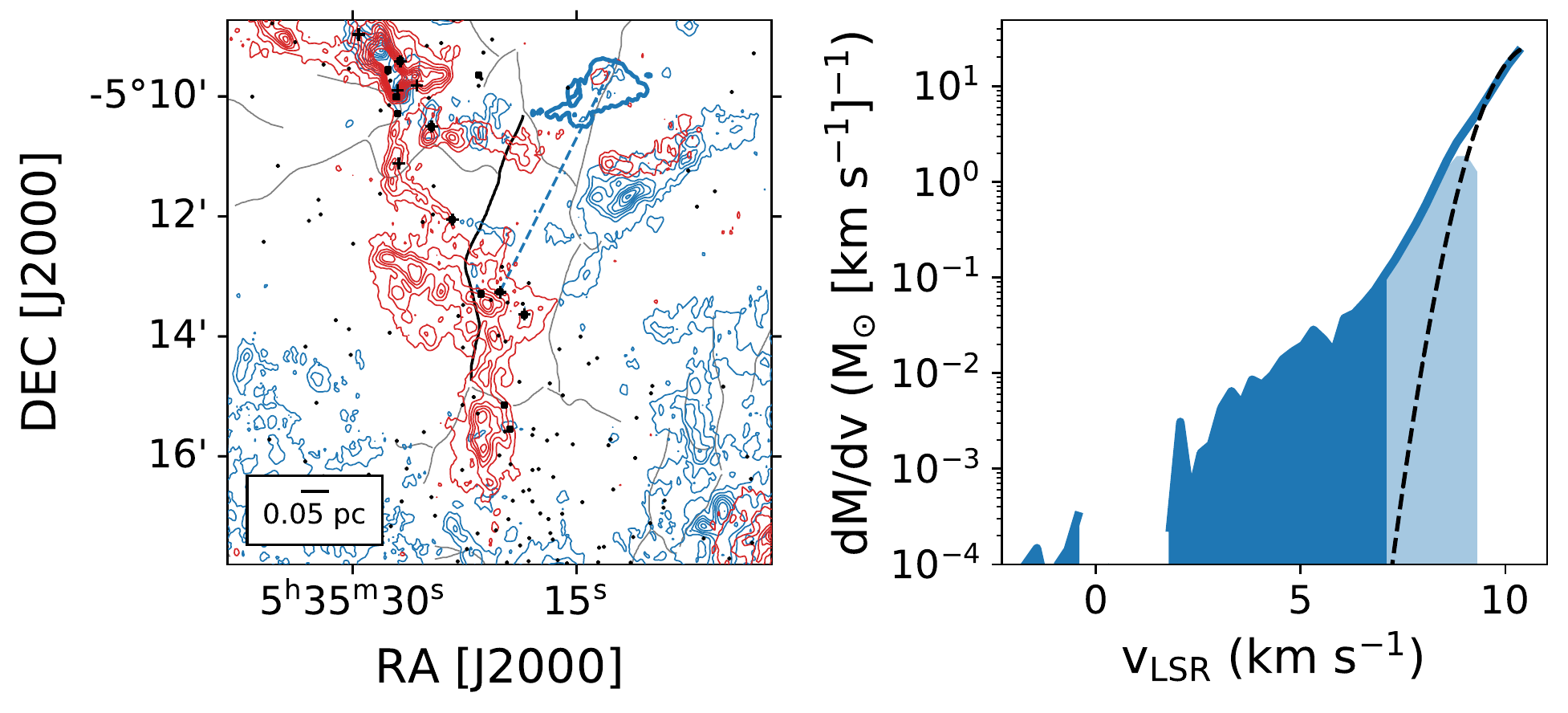}

    \caption{HOPS 59 outflow. The left panel shows the outflow, position angle, nearby sources, and filaments.
    The velocity range of integration is given by v$_{\rm blue}$/v$_{\rm red}$ in Table~\ref{tab:outflows}
    and the contours go from $5$ to $50\sigma$ in steps of 5$\sigma$, where $\sigma$ is the RMS error in the integrated map. Symbols are the same as Figure~\ref{fig:stamp}.
    The right panel shows the mass spectrum with fit, where $\sigma$ is the RMS error in the integrated map. Symbols are the same as Figure~\ref{fig:dmdv}.}
    \end{figure*}
\begin{figure*}[p]
\centering
\includegraphics[width=\textwidth]{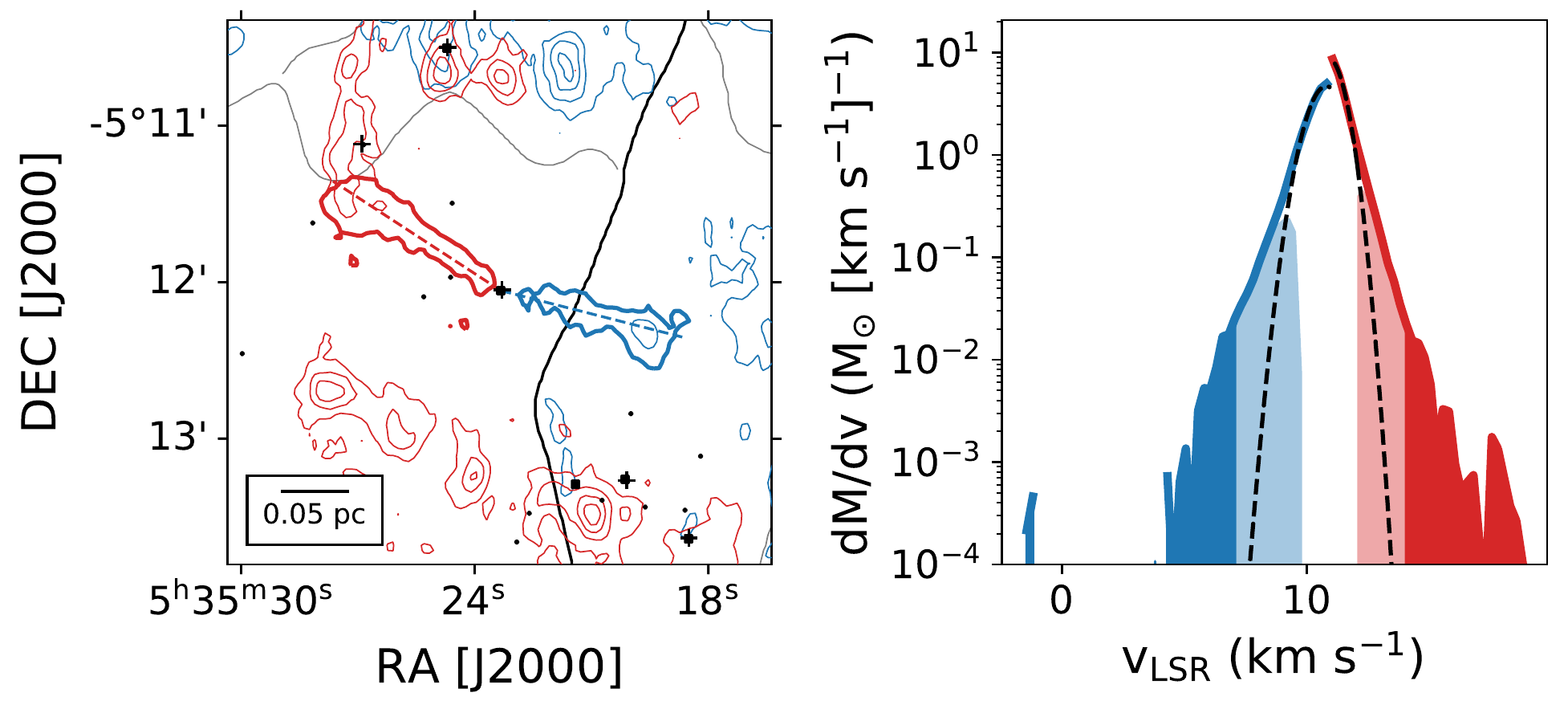}

    \caption{HOPS 60 outflow. The left panel shows the outflow, position angle, nearby sources, and filaments.
    The velocity range of integration is given by v$_{\rm blue}$/v$_{\rm red}$ in Table~\ref{tab:outflows}
    and the contours go from $5$ to $50\sigma$ in steps of 5$\sigma$, where $\sigma$ is the RMS error in the integrated map. Symbols are the same as Figure~\ref{fig:stamp}.
    The right panel shows the mass spectrum with fit, where $\sigma$ is the RMS error in the integrated map. Symbols are the same as Figure~\ref{fig:dmdv}.}
    \end{figure*}
\begin{figure*}[p]
\centering
\includegraphics[width=\textwidth]{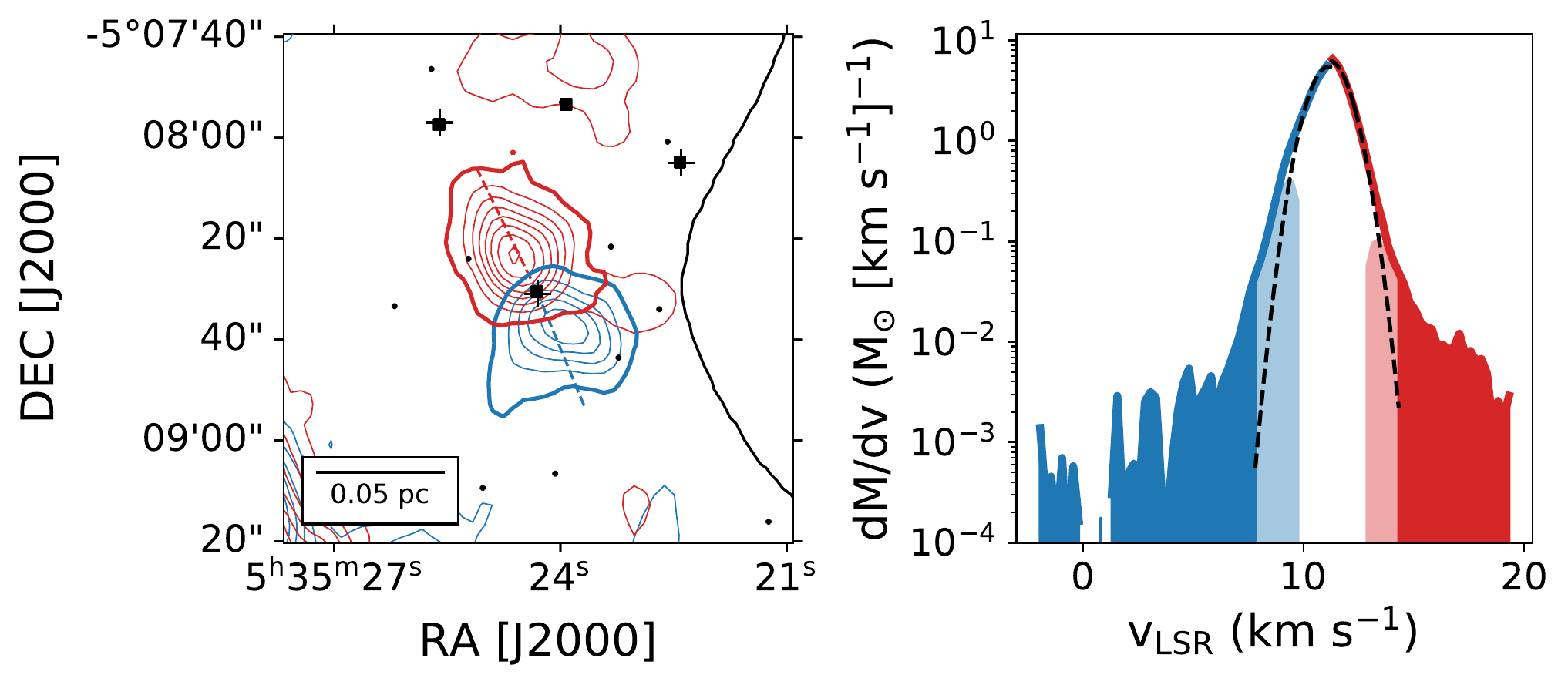}

    \caption{HOPS 68 outflow. The left panel shows the outflow, position angle, nearby sources, and filaments.
    The velocity range of integration is given by v$_{\rm blue}$/v$_{\rm red}$ in Table~\ref{tab:outflows}
    and the contours go from $5$ to $50\sigma$ in steps of 5$\sigma$, where $\sigma$ is the RMS error in the integrated map. Symbols are the same as Figure~\ref{fig:stamp}.
    The right panel shows the mass spectrum with fit, where $\sigma$ is the RMS error in the integrated map. Symbols are the same as Figure~\ref{fig:dmdv}.}
    \end{figure*}
\begin{figure*}[p]
\centering
\includegraphics[width=\textwidth]{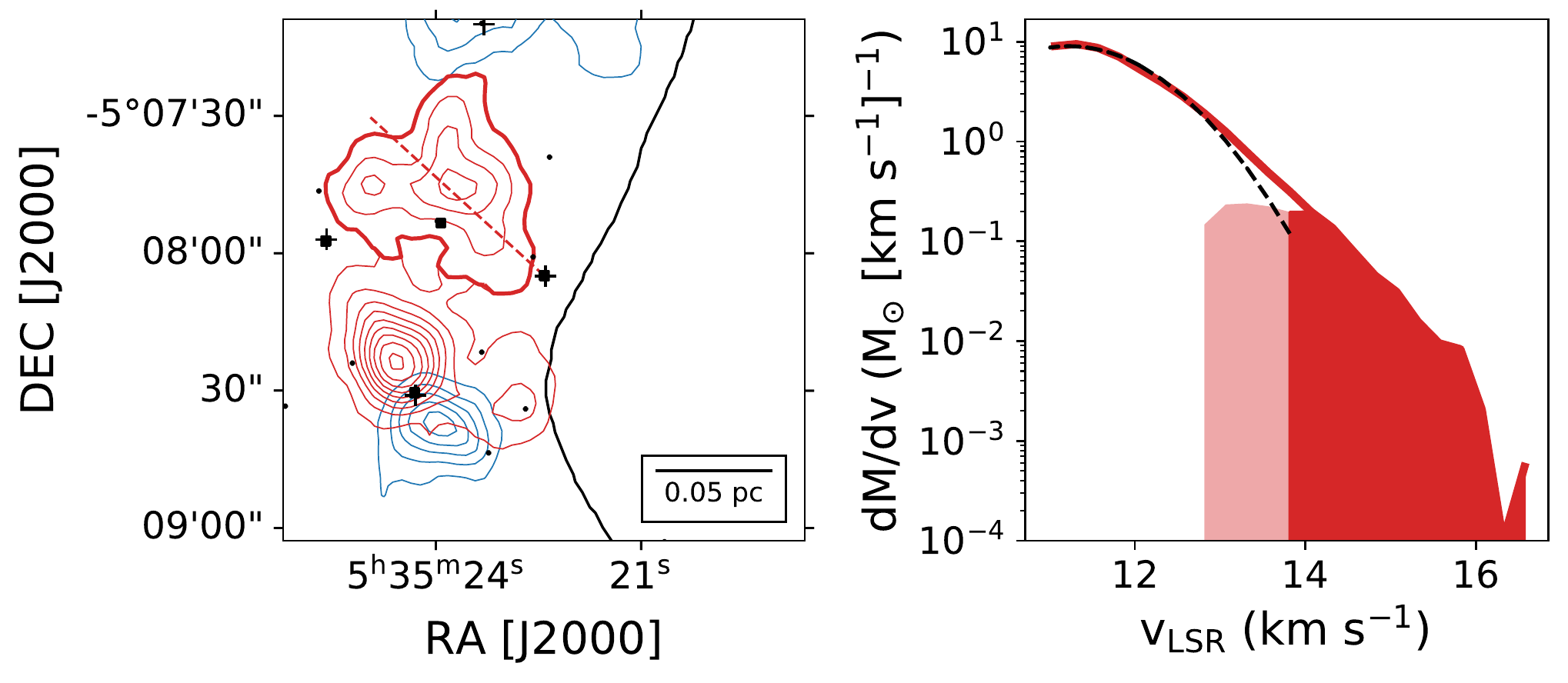}

    \caption{HOPS 70 outflow. The left panel shows the outflow, position angle, nearby sources, and filaments.
    The velocity range of integration is given by v$_{\rm blue}$/v$_{\rm red}$ in Table~\ref{tab:outflows}
    and the contours go from $5$ to $50\sigma$ in steps of 5$\sigma$, where $\sigma$ is the RMS error in the integrated map. Symbols are the same as Figure~\ref{fig:stamp}.
    The right panel shows the mass spectrum with fit, where $\sigma$ is the RMS error in the integrated map. Symbols are the same as Figure~\ref{fig:dmdv}.}
    \end{figure*}
\begin{figure*}[p]
\centering
\includegraphics[width=\textwidth]{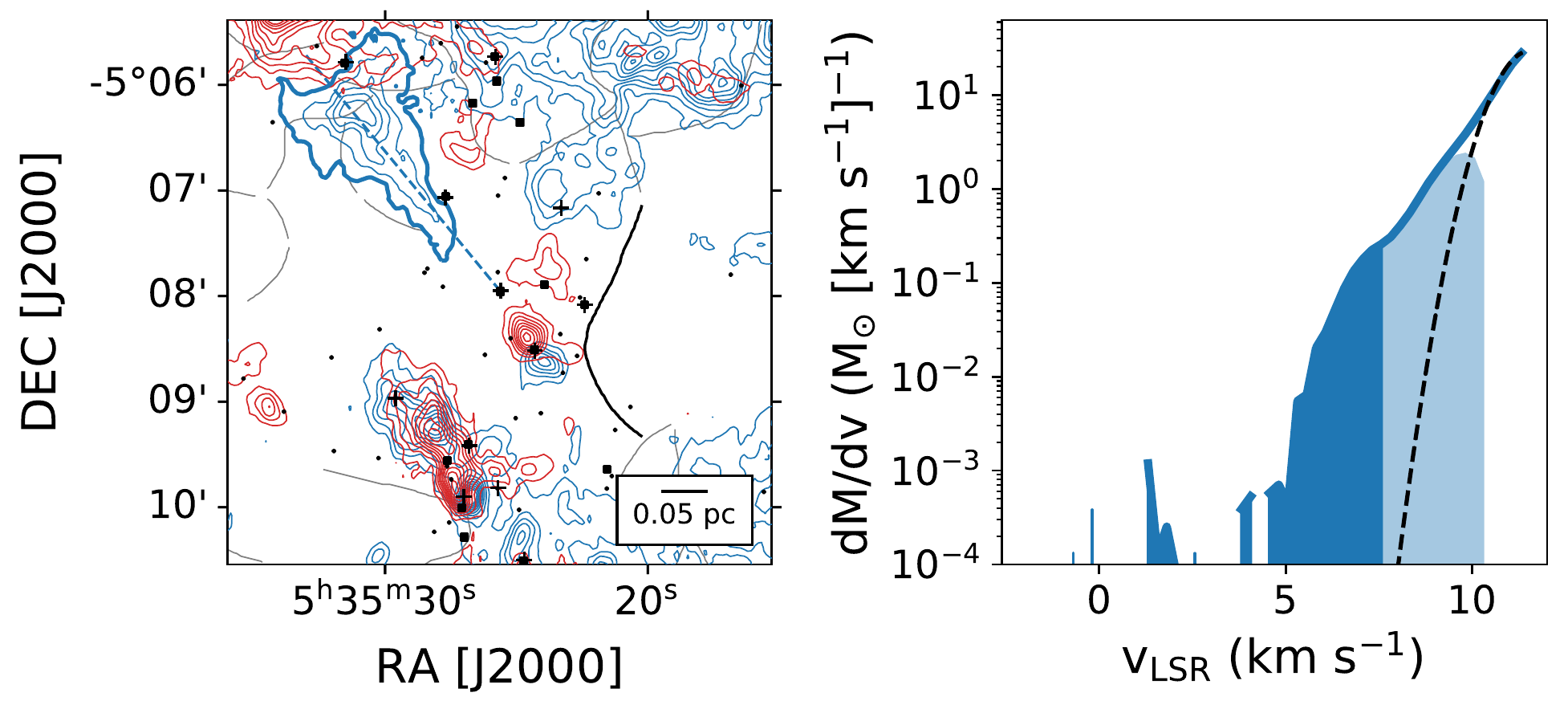}

    \caption{HOPS 71 outflow. The left panel shows the outflow, position angle, nearby sources, and filaments.
    The velocity range of integration is given by v$_{\rm blue}$/v$_{\rm red}$ in Table~\ref{tab:outflows}
    and the contours go from $5$ to $50\sigma$ in steps of 5$\sigma$, where $\sigma$ is the RMS error in the integrated map. Symbols are the same as Figure~\ref{fig:stamp}.
    The right panel shows the mass spectrum with fit, where $\sigma$ is the RMS error in the integrated map. Symbols are the same as Figure~\ref{fig:dmdv}.}
    \end{figure*}
\begin{figure*}[p]
\centering
\includegraphics[width=\textwidth]{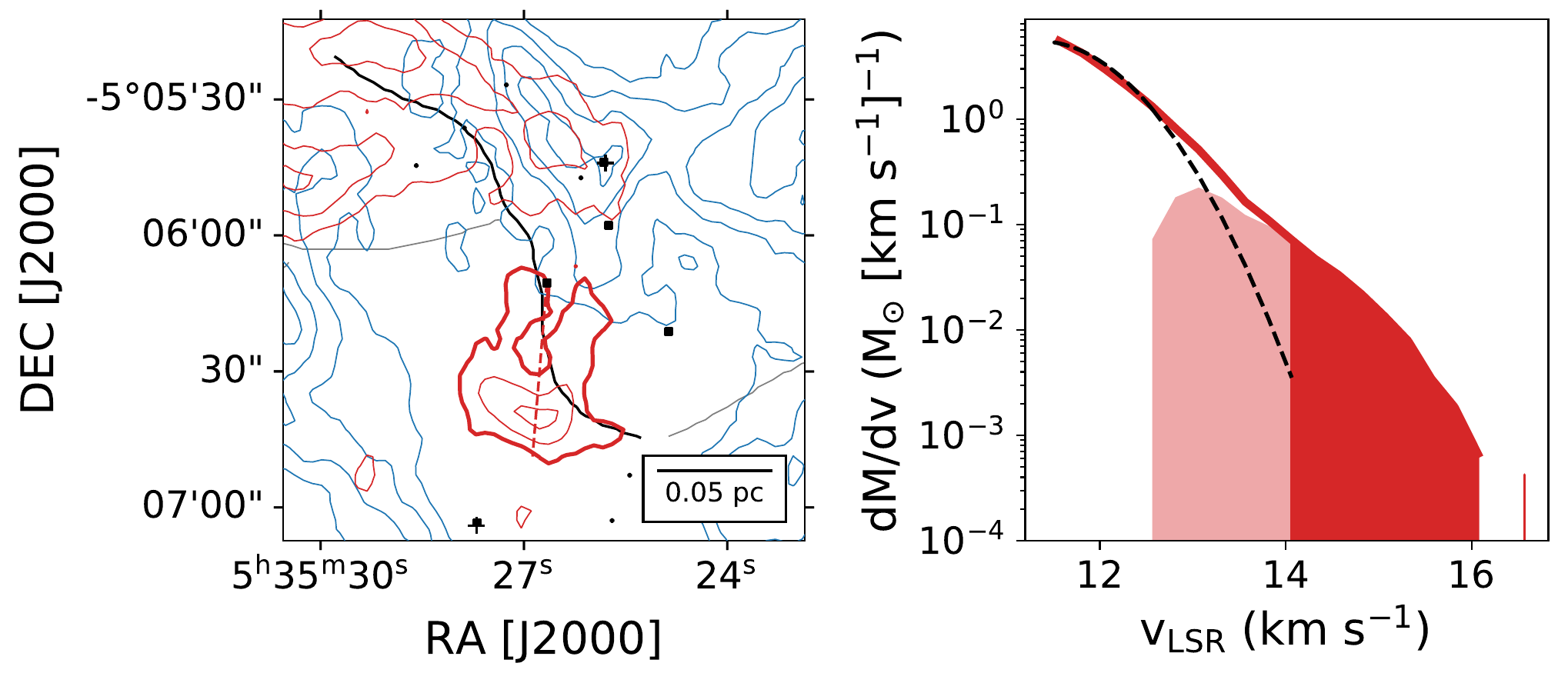}

    \caption{HOPS 75 outflow. The left panel shows the outflow, position angle, nearby sources, and filaments.
    The velocity range of integration is given by v$_{\rm blue}$/v$_{\rm red}$ in Table~\ref{tab:outflows}
    and the contours go from $5$ to $50\sigma$ in steps of 5$\sigma$, where $\sigma$ is the RMS error in the integrated map. Symbols are the same as Figure~\ref{fig:stamp}.
    The right panel shows the mass spectrum with fit, where $\sigma$ is the RMS error in the integrated map. Symbols are the same as Figure~\ref{fig:dmdv}.}
    \end{figure*}
\begin{figure*}[p]
\centering
\includegraphics[width=\textwidth]{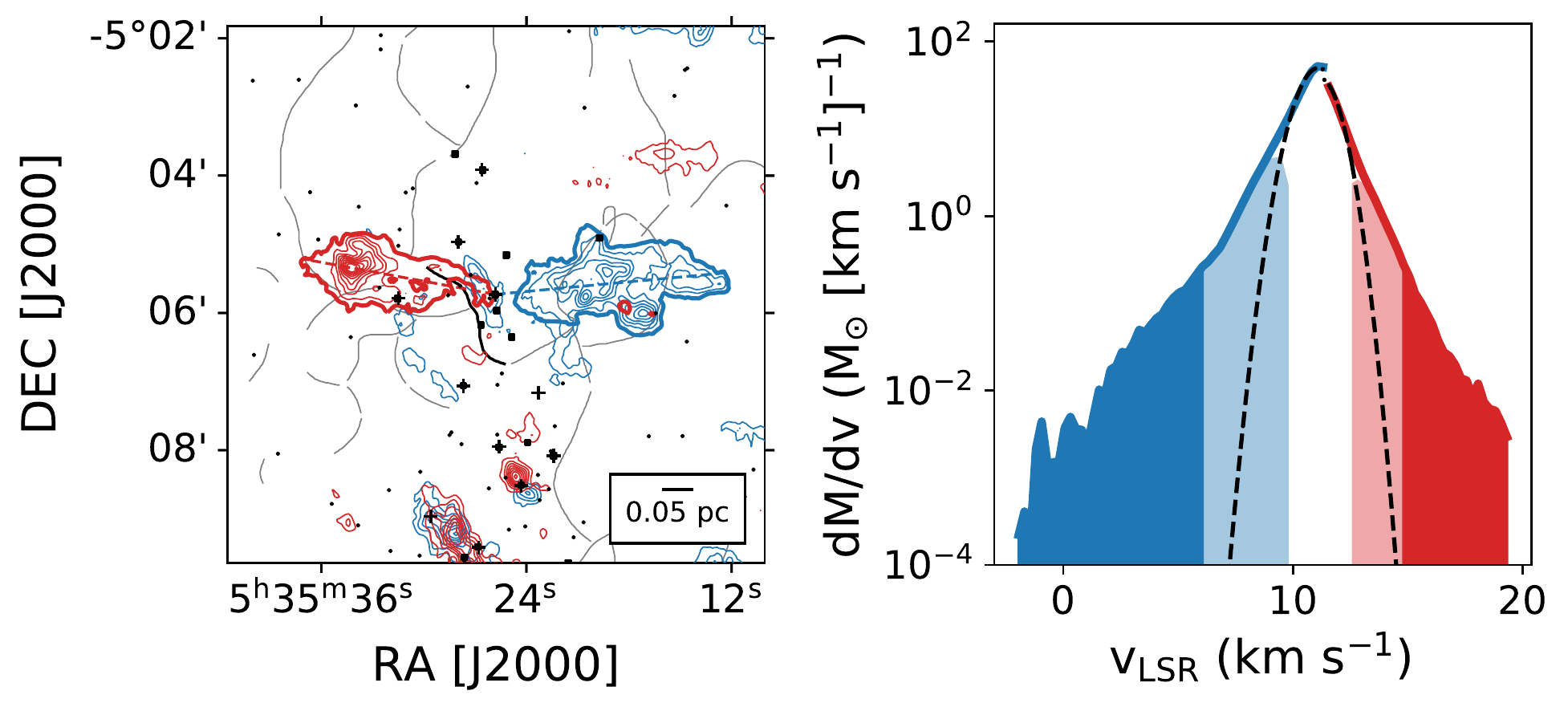}

    \caption{HOPS 78 outflow. The left panel shows the outflow, position angle, nearby sources, and filaments.
    The velocity range of integration is given by v$_{\rm blue}$/v$_{\rm red}$ in Table~\ref{tab:outflows}
    and the contours go from $5$ to $50\sigma$ in steps of 5$\sigma$, where $\sigma$ is the RMS error in the integrated map. Symbols are the same as Figure~\ref{fig:stamp}.
    The right panel shows the mass spectrum with fit, where $\sigma$ is the RMS error in the integrated map. Symbols are the same as Figure~\ref{fig:dmdv}.}
    \end{figure*}
\begin{figure*}[p]
\centering
\includegraphics[width=\textwidth]{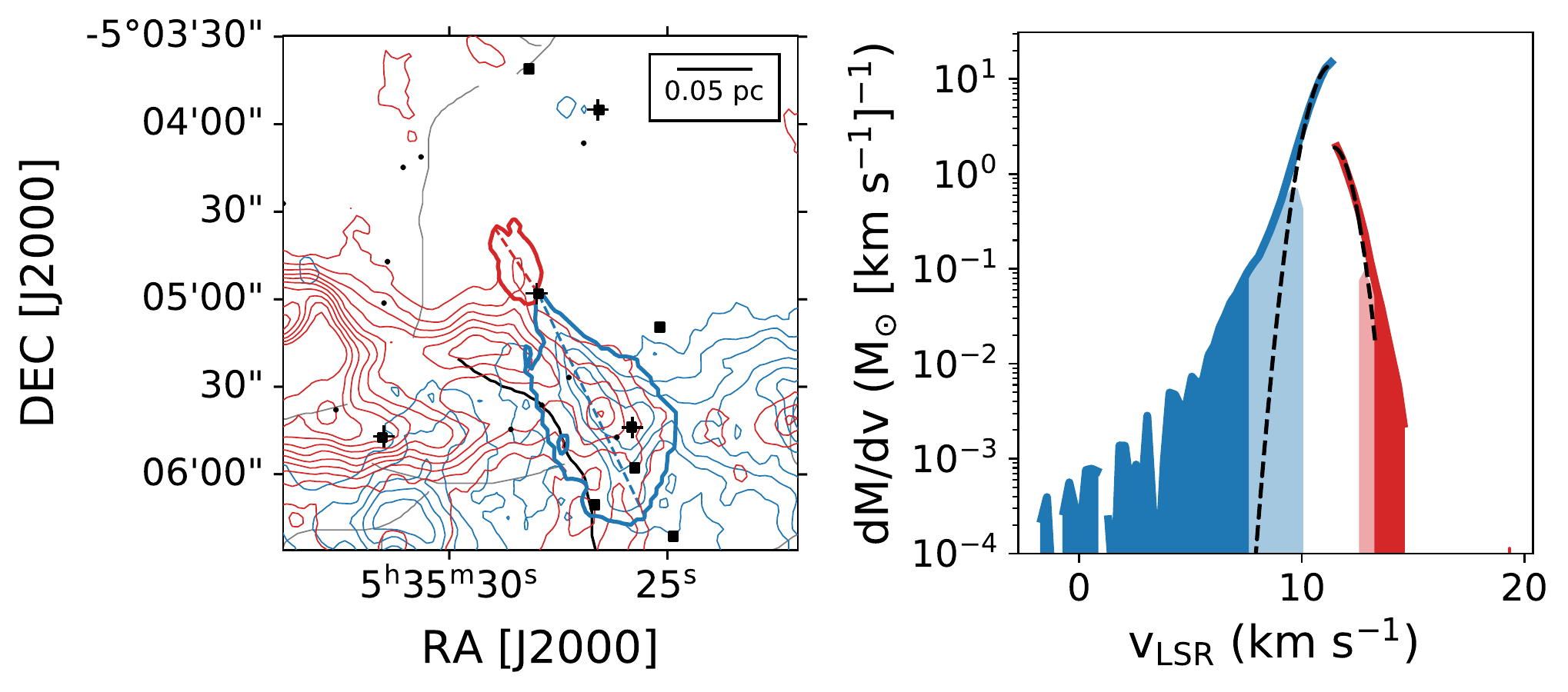}

    \caption{HOPS 81 outflow. The left panel shows the outflow, position angle, nearby sources, and filaments.
    The velocity range of integration is given by v$_{\rm blue}$/v$_{\rm red}$ in Table~\ref{tab:outflows}
    and the contours go from $5$ to $50\sigma$ in steps of 5$\sigma$, where $\sigma$ is the RMS error in the integrated map. Symbols are the same as Figure~\ref{fig:stamp}.
    The right panel shows the mass spectrum with fit, where $\sigma$ is the RMS error in the integrated map. Symbols are the same as Figure~\ref{fig:dmdv}.}
    \end{figure*}
\begin{figure*}[p]
\centering
\includegraphics[width=\textwidth]{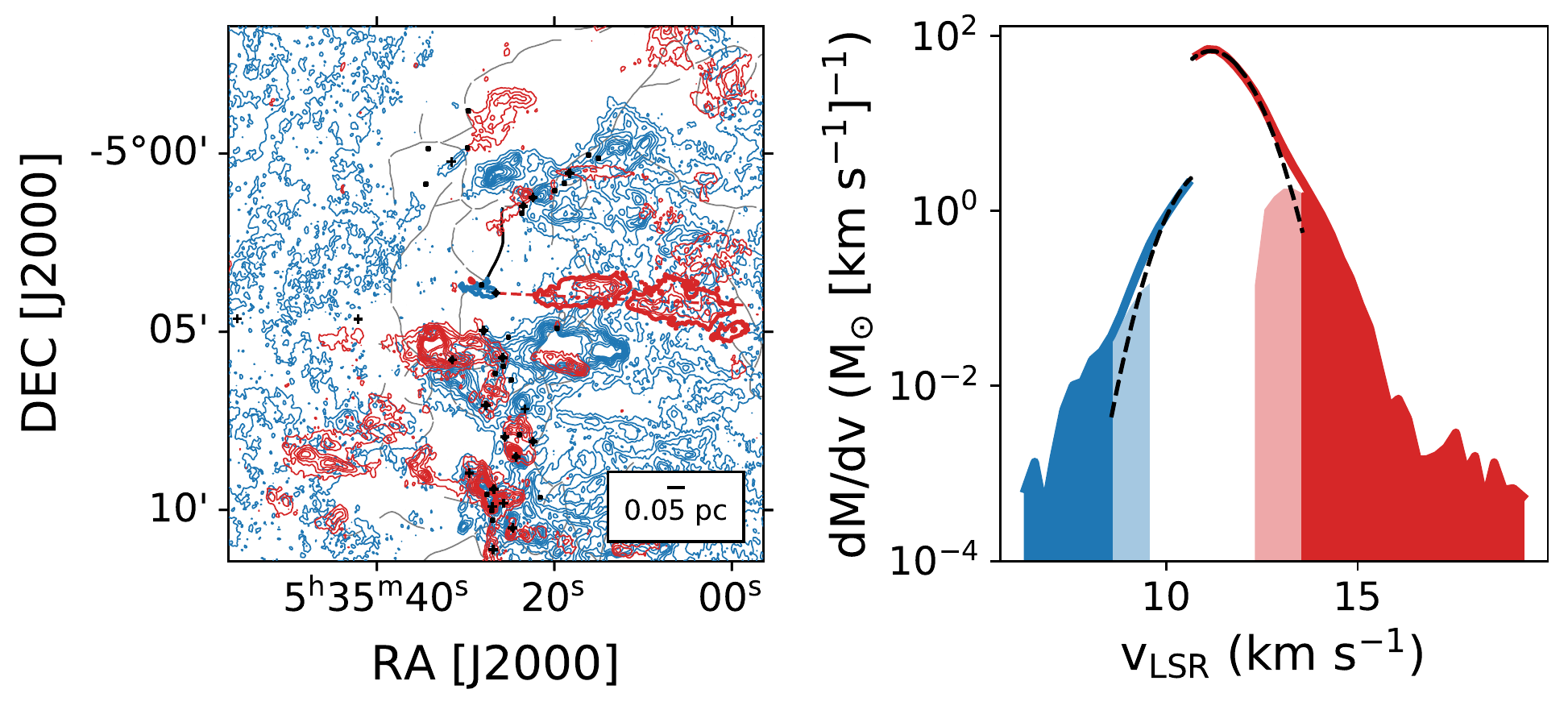}

    \caption{HOPS 84 outflow. The left panel shows the outflow, position angle, nearby sources, and filaments.
    The velocity range of integration is given by v$_{\rm blue}$/v$_{\rm red}$ in Table~\ref{tab:outflows}
    and the contours go from $5$ to $50\sigma$ in steps of 5$\sigma$, where $\sigma$ is the RMS error in the integrated map. Symbols are the same as Figure~\ref{fig:stamp}.
    The right panel shows the mass spectrum with fit, where $\sigma$ is the RMS error in the integrated map. Symbols are the same as Figure~\ref{fig:dmdv}.}
    \end{figure*}
\begin{figure*}[p]
\centering
\includegraphics[width=\textwidth]{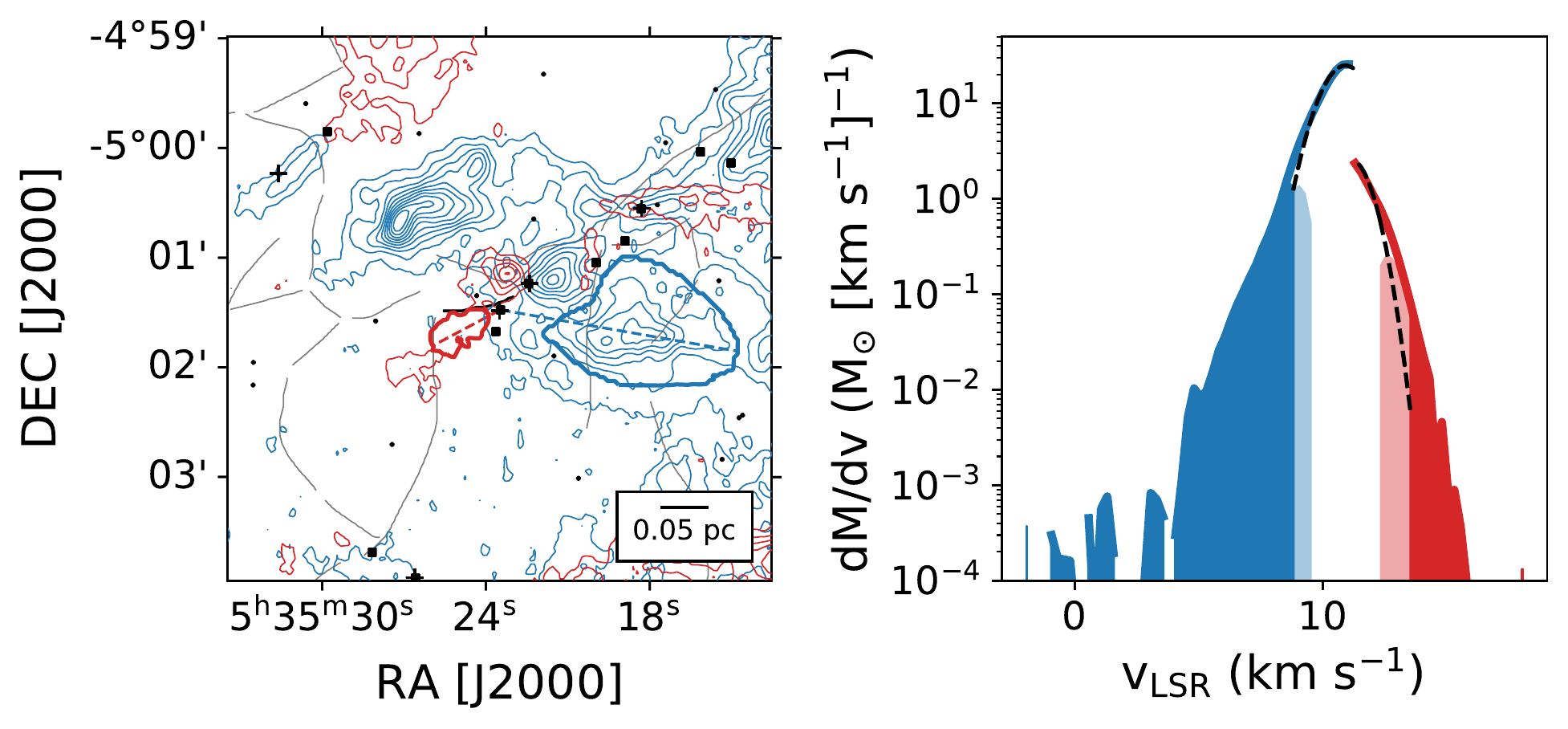}

    \caption{HOPS 87 outflow. The left panel shows the outflow, position angle, nearby sources, and filaments.
    The velocity range of integration is given by v$_{\rm blue}$/v$_{\rm red}$ in Table~\ref{tab:outflows}
    and the contours go from $5$ to $50\sigma$ in steps of 5$\sigma$, where $\sigma$ is the RMS error in the integrated map. Symbols are the same as Figure~\ref{fig:stamp}.
    The right panel shows the mass spectrum with fit, where $\sigma$ is the RMS error in the integrated map. Symbols are the same as Figure~\ref{fig:dmdv}.}
    \end{figure*}
\begin{figure*}[p]
\centering
\includegraphics[width=\textwidth]{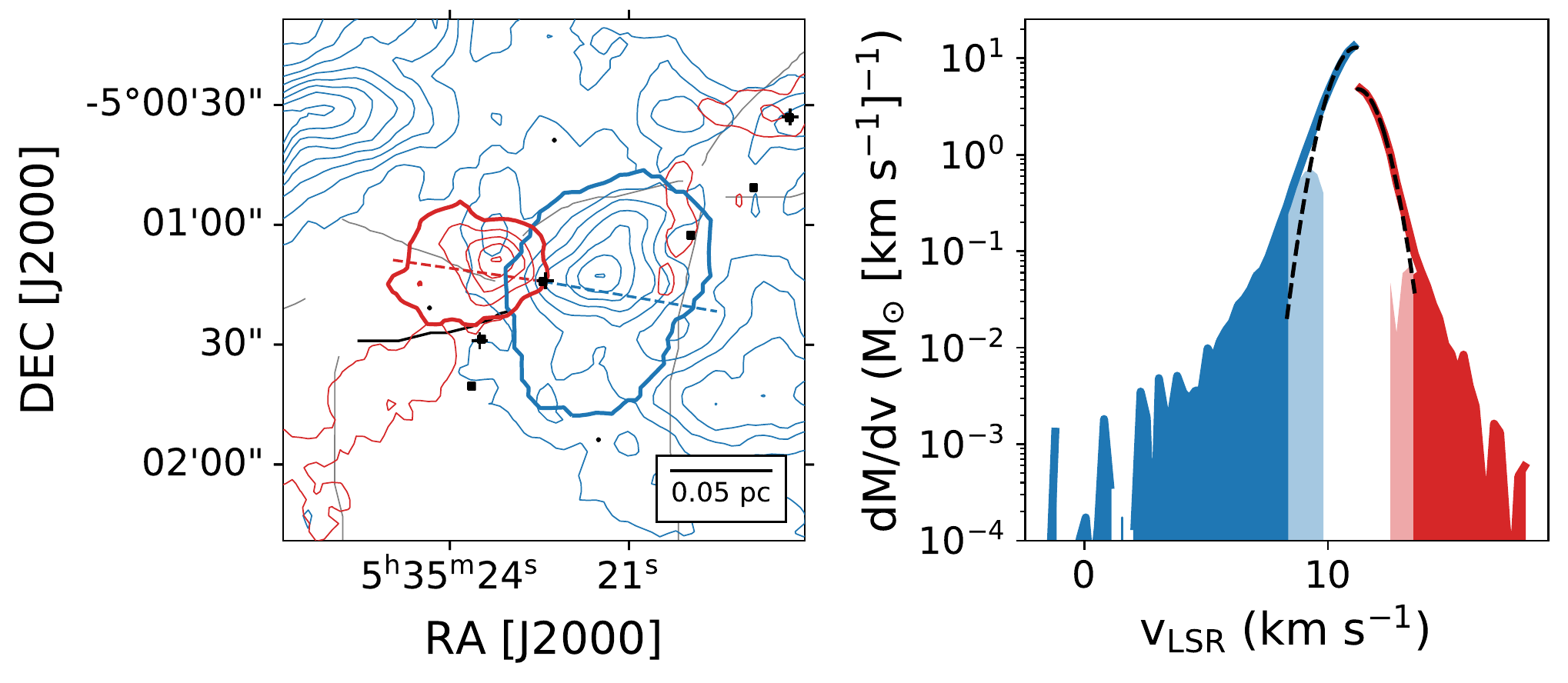}

    \caption{HOPS 88 outflow. The left panel shows the outflow, position angle, nearby sources, and filaments.
    The velocity range of integration is given by v$_{\rm blue}$/v$_{\rm red}$ in Table~\ref{tab:outflows}
    and the contours go from $5$ to $50\sigma$ in steps of 5$\sigma$, where $\sigma$ is the RMS error in the integrated map. Symbols are the same as Figure~\ref{fig:stamp}.
    The right panel shows the mass spectrum with fit, where $\sigma$ is the RMS error in the integrated map. Symbols are the same as Figure~\ref{fig:dmdv}.}
    \end{figure*}
\begin{figure*}[p]
\centering
\includegraphics[width=\textwidth]{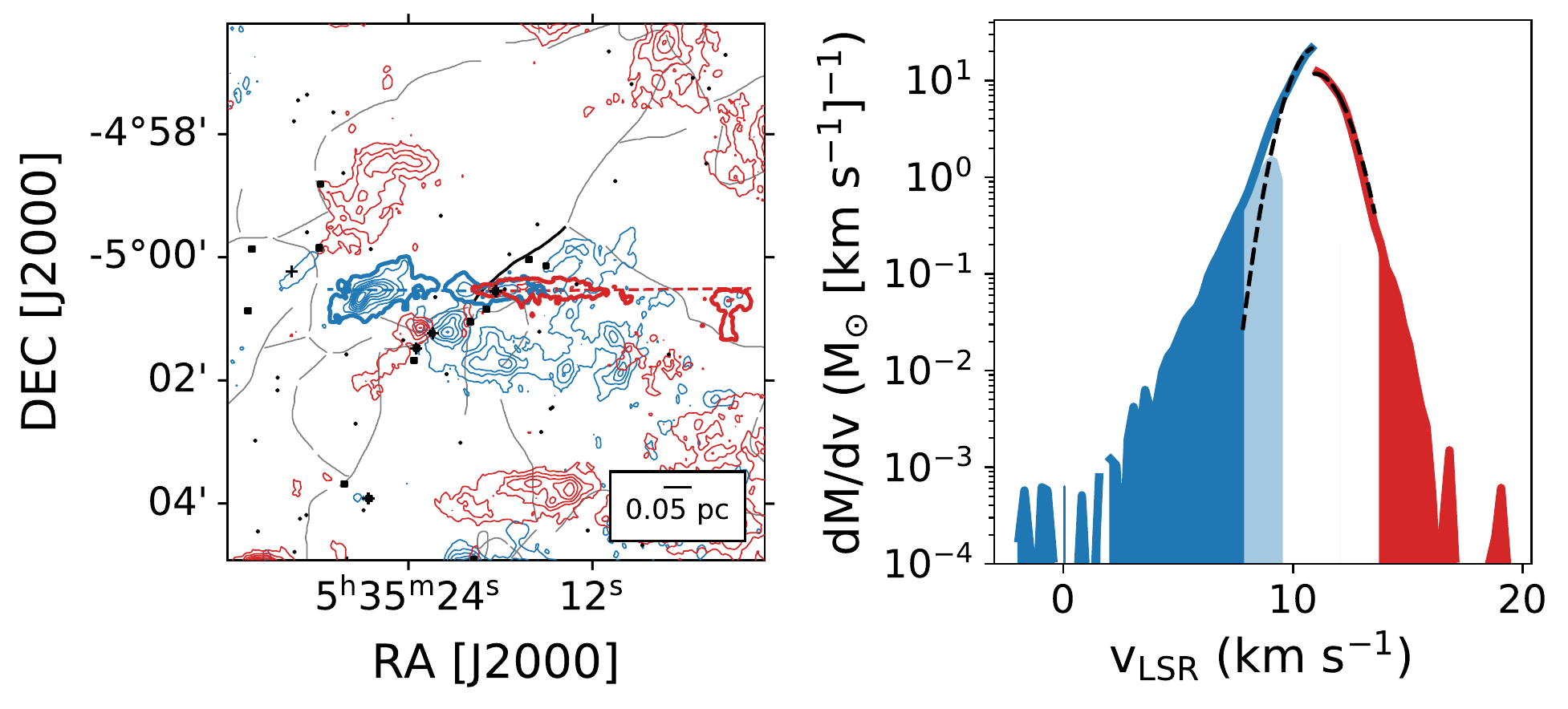}

    \caption{HOPS 92 outflow. The left panel shows the outflow, position angle, nearby sources, and filaments.
    The velocity range of integration is given by v$_{\rm blue}$/v$_{\rm red}$ in Table~\ref{tab:outflows}
    and the contours go from $5$ to $50\sigma$ in steps of 5$\sigma$, where $\sigma$ is the RMS error in the integrated map. Symbols are the same as Figure~\ref{fig:stamp}.
    The right panel shows the mass spectrum with fit, where $\sigma$ is the RMS error in the integrated map. Symbols are the same as Figure~\ref{fig:dmdv}.}
    \end{figure*}
\begin{figure*}[p]
\centering
\includegraphics[width=\textwidth]{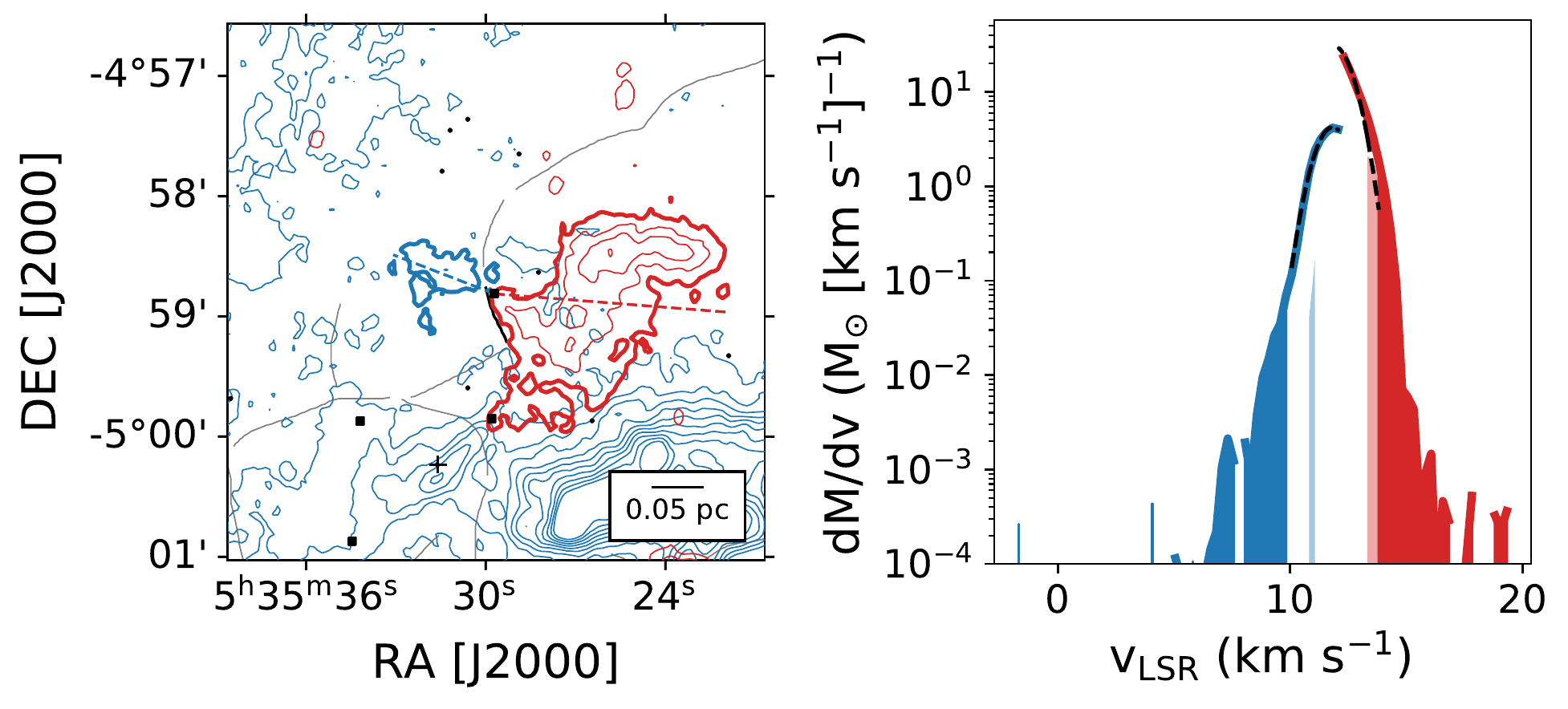}

    \caption{HOPS 96 outflow. The left panel shows the outflow, position angle, nearby sources, and filaments.
    The velocity range of integration is given by v$_{\rm blue}$/v$_{\rm red}$ in Table~\ref{tab:outflows}
    and the contours go from $5$ to $50\sigma$ in steps of 5$\sigma$, where $\sigma$ is the RMS error in the integrated map. Symbols are the same as Figure~\ref{fig:stamp}.
    The right panel shows the mass spectrum with fit, where $\sigma$ is the RMS error in the integrated map. Symbols are the same as Figure~\ref{fig:dmdv}.}
    \end{figure*}
\begin{figure*}[p]
\centering
\includegraphics[width=\textwidth]{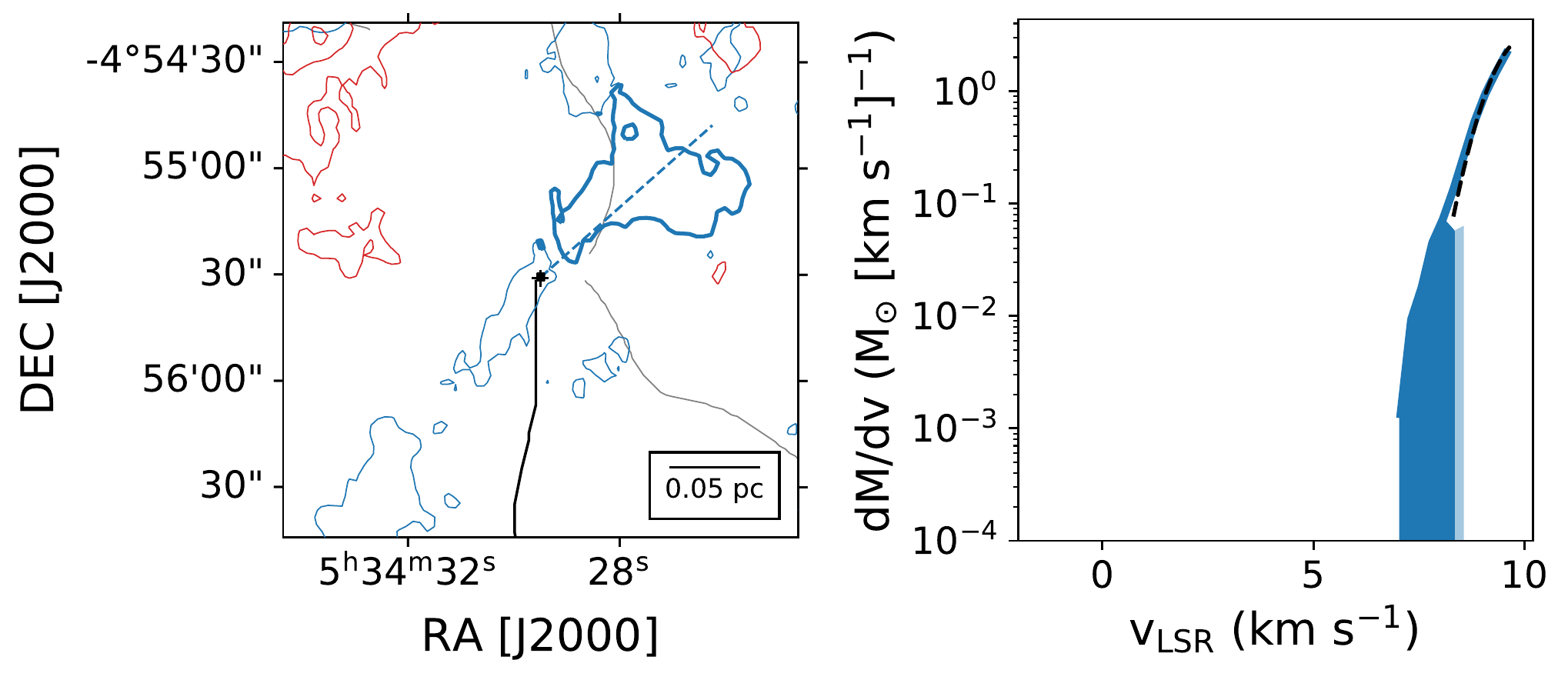}

    \caption{HOPS 99 outflow. The left panel shows the outflow, position angle, nearby sources, and filaments.
    The velocity range of integration is given by v$_{\rm blue}$/v$_{\rm red}$ in Table~\ref{tab:outflows}
    and the contours go from $5$ to $50\sigma$ in steps of 5$\sigma$, where $\sigma$ is the RMS error in the integrated map. Symbols are the same as Figure~\ref{fig:stamp}.
    The right panel shows the mass spectrum with fit, where $\sigma$ is the RMS error in the integrated map. Symbols are the same as Figure~\ref{fig:dmdv}.}
    \end{figure*}
\begin{figure*}[p]
\centering
\includegraphics[width=\textwidth]{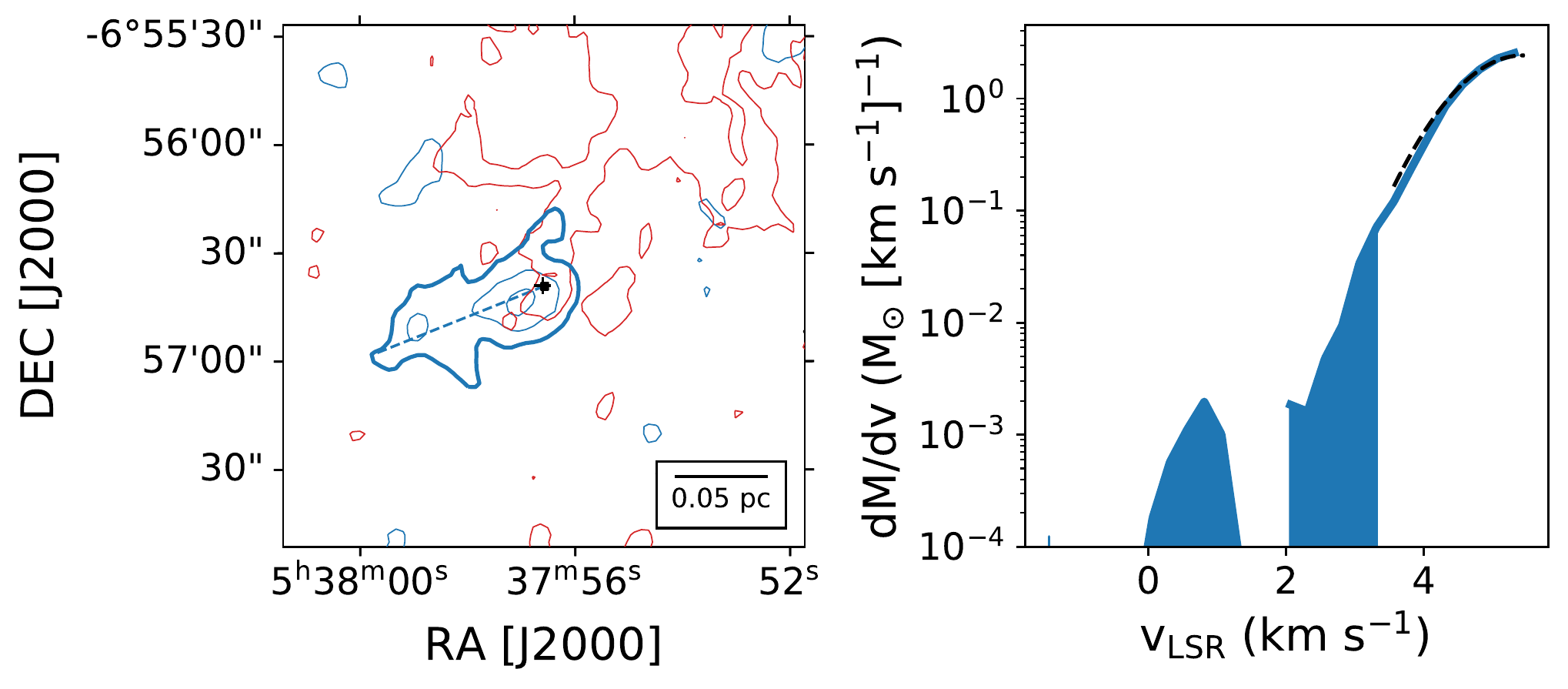}

    \caption{HOPS 157 outflow. The left panel shows the outflow, position angle, nearby sources, and filaments.
    The velocity range of integration is given by v$_{\rm blue}$/v$_{\rm red}$ in Table~\ref{tab:outflows}
    and the contours go from $5$ to $50\sigma$ in steps of 5$\sigma$, where $\sigma$ is the RMS error in the integrated map. Symbols are the same as Figure~\ref{fig:stamp}.
    The right panel shows the mass spectrum with fit, where $\sigma$ is the RMS error in the integrated map. Symbols are the same as Figure~\ref{fig:dmdv}.}
    \end{figure*}
\begin{figure*}[p]
\centering
\includegraphics[width=\textwidth]{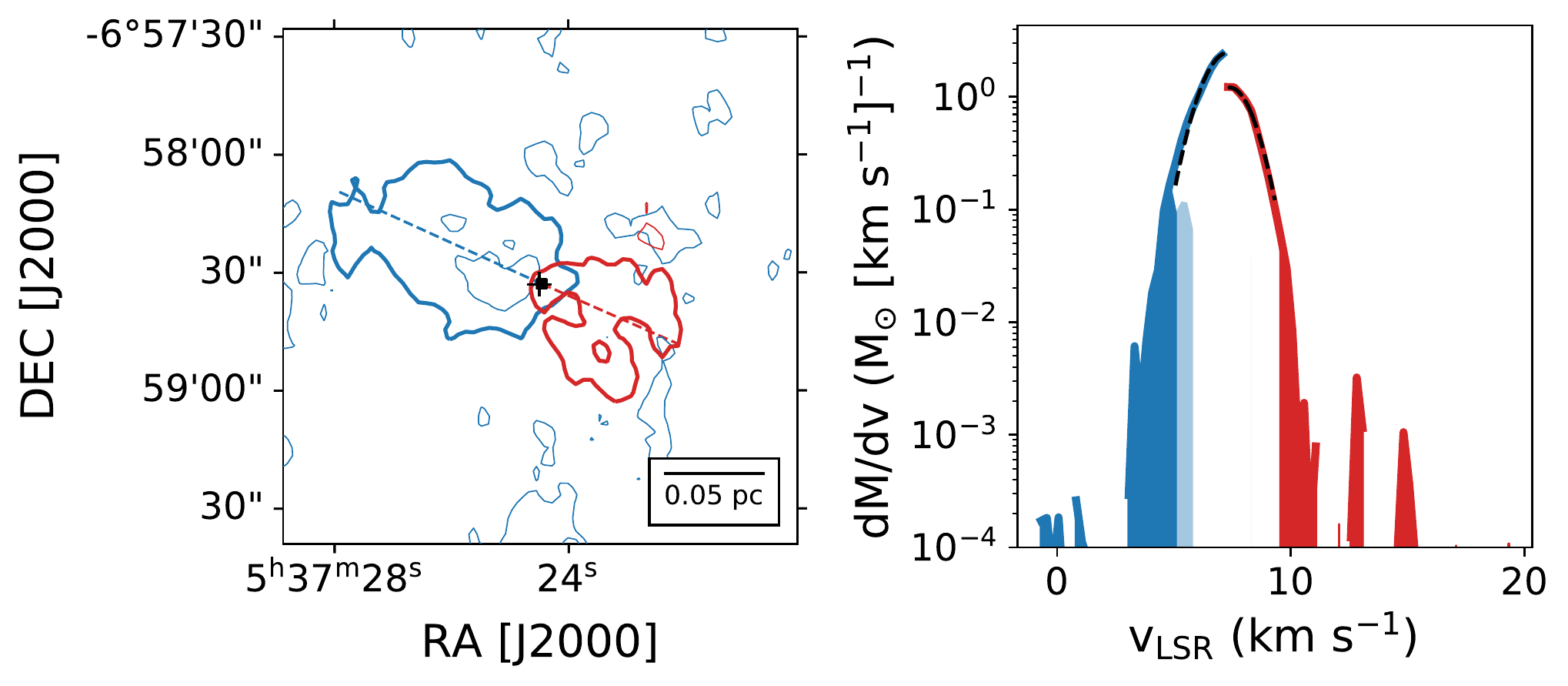}

    \caption{HOPS 158 outflow. The left panel shows the outflow, position angle, nearby sources, and filaments.
    The velocity range of integration is given by v$_{\rm blue}$/v$_{\rm red}$ in Table~\ref{tab:outflows}
    and the contours go from $5$ to $50\sigma$ in steps of 5$\sigma$, where $\sigma$ is the RMS error in the integrated map. Symbols are the same as Figure~\ref{fig:stamp}.
    The right panel shows the mass spectrum with fit, where $\sigma$ is the RMS error in the integrated map. Symbols are the same as Figure~\ref{fig:dmdv}.}
    \end{figure*}
\begin{figure*}[p]
\centering
\includegraphics[width=\textwidth]{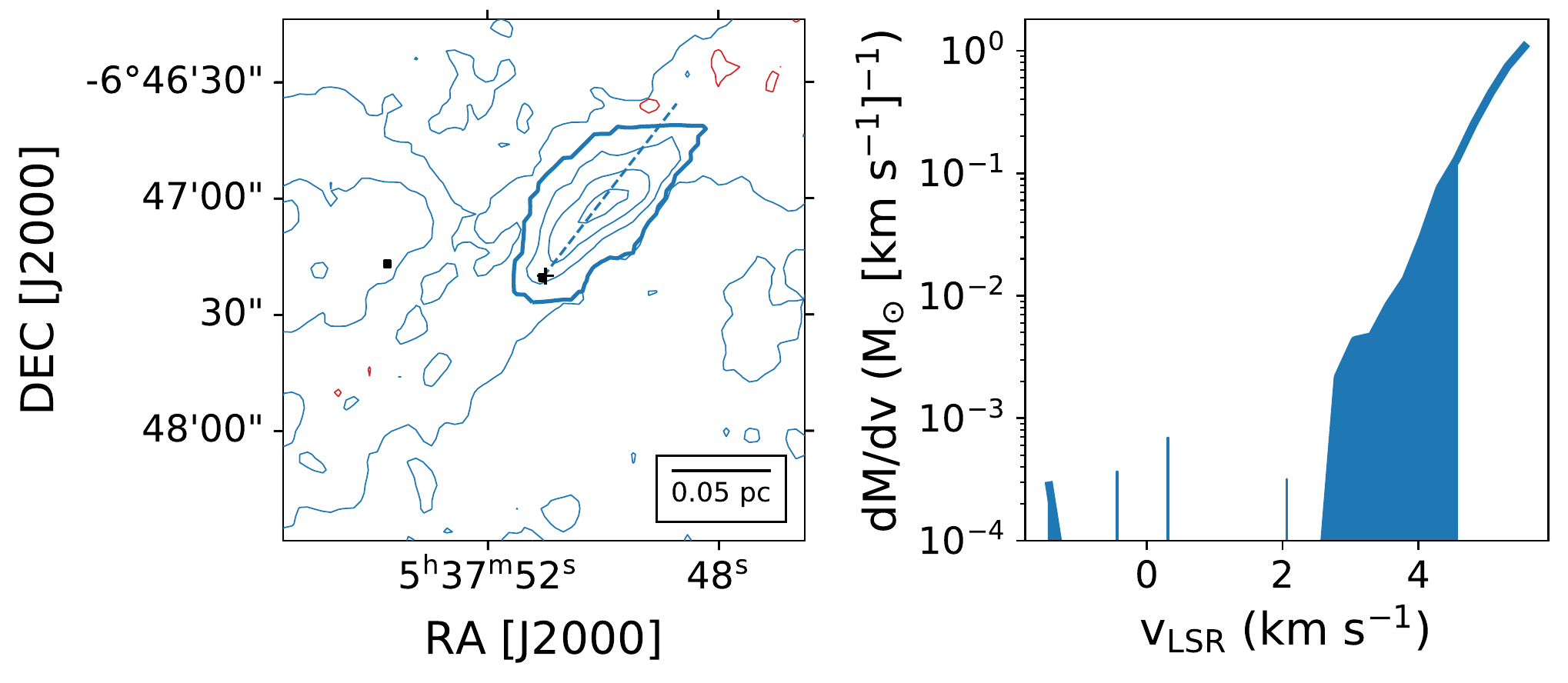}

    \caption{HOPS 160 outflow. The left panel shows the outflow, position angle, nearby sources, and filaments.
    The velocity range of integration is given by v$_{\rm blue}$/v$_{\rm red}$ in Table~\ref{tab:outflows}
    and the contours go from $5$ to $50\sigma$ in steps of 5$\sigma$, where $\sigma$ is the RMS error in the integrated map. Symbols are the same as Figure~\ref{fig:stamp}.
    The right panel shows the mass spectrum with fit, where $\sigma$ is the RMS error in the integrated map. Symbols are the same as Figure~\ref{fig:dmdv}.}
    \end{figure*}
\begin{figure*}[p]
\centering
\includegraphics[width=\textwidth]{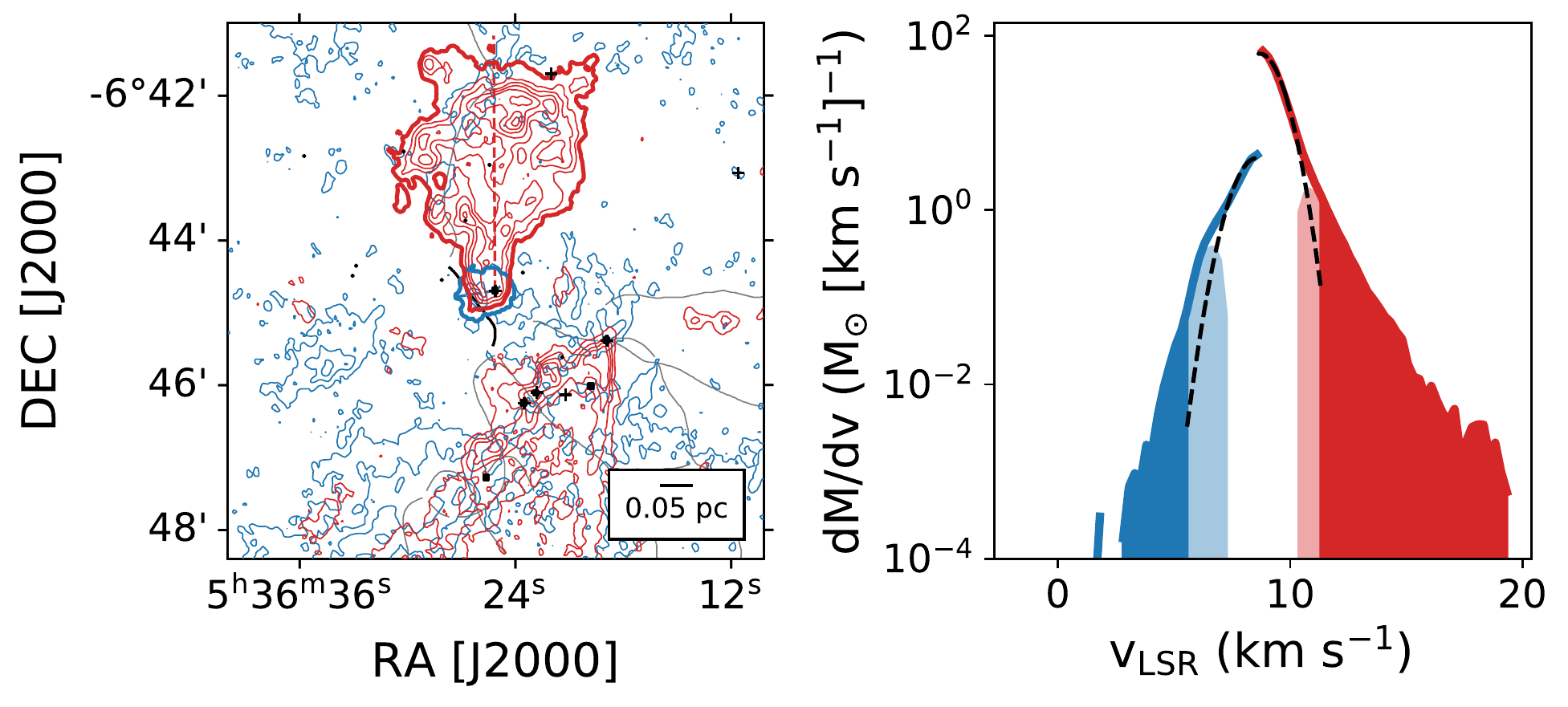}

    \caption{HOPS 166 outflow. The left panel shows the outflow, position angle, nearby sources, and filaments.
    The velocity range of integration is given by v$_{\rm blue}$/v$_{\rm red}$ in Table~\ref{tab:outflows}
    and the contours go from $5$ to $50\sigma$ in steps of 5$\sigma$, where $\sigma$ is the RMS error in the integrated map. Symbols are the same as Figure~\ref{fig:stamp}.
    The right panel shows the mass spectrum with fit, where $\sigma$ is the RMS error in the integrated map. Symbols are the same as Figure~\ref{fig:dmdv}.}
    \end{figure*}
\begin{figure*}[p]
\centering
\includegraphics[width=\textwidth]{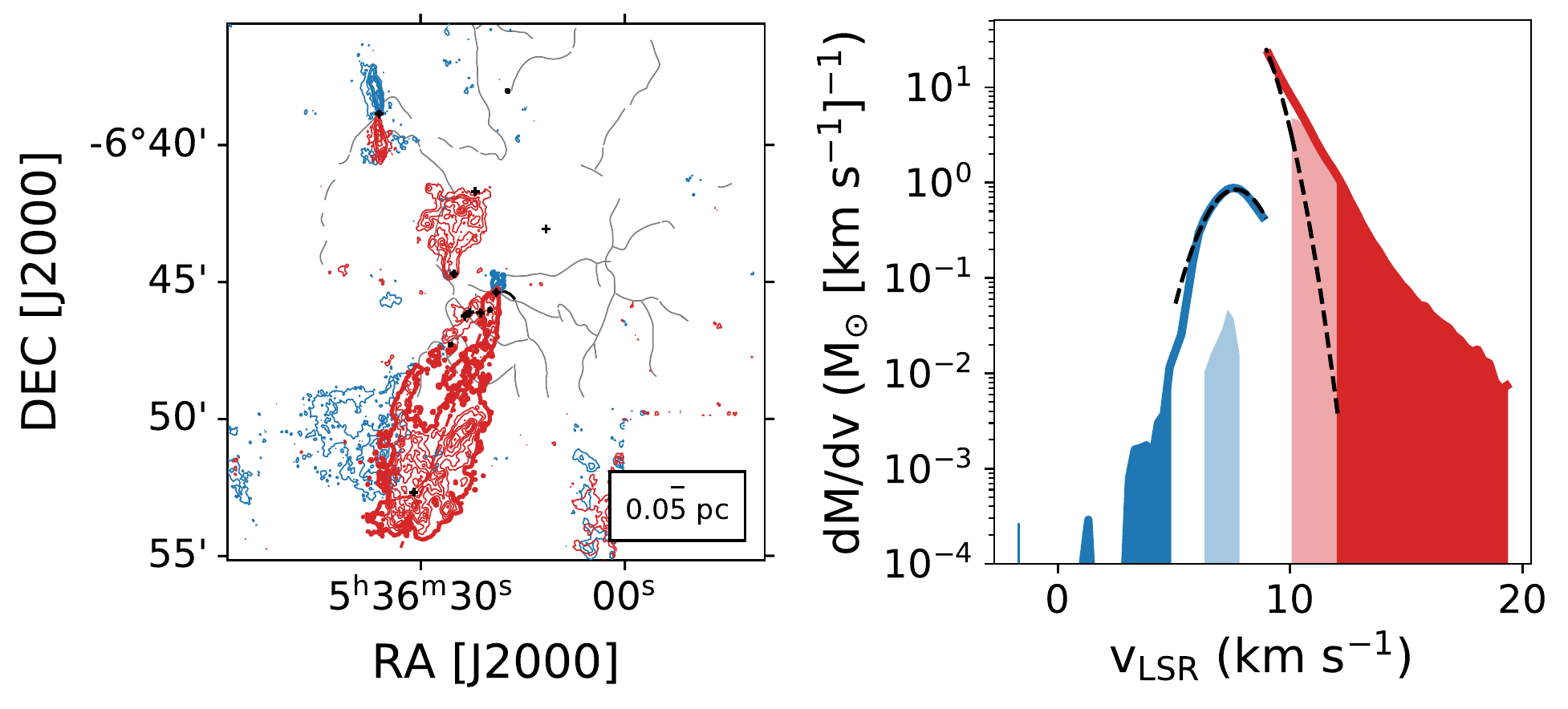}

    \caption{HOPS 168 outflow. The left panel shows the outflow, position angle, nearby sources, and filaments.
    The velocity range of integration is given by v$_{\rm blue}$/v$_{\rm red}$ in Table~\ref{tab:outflows}
    and the contours go from $5$ to $50\sigma$ in steps of 5$\sigma$, where $\sigma$ is the RMS error in the integrated map. Symbols are the same as Figure~\ref{fig:stamp}.
    The right panel shows the mass spectrum with fit, where $\sigma$ is the RMS error in the integrated map. Symbols are the same as Figure~\ref{fig:dmdv}.}
    \end{figure*}
\begin{figure*}[p]
\centering
\includegraphics[width=\textwidth]{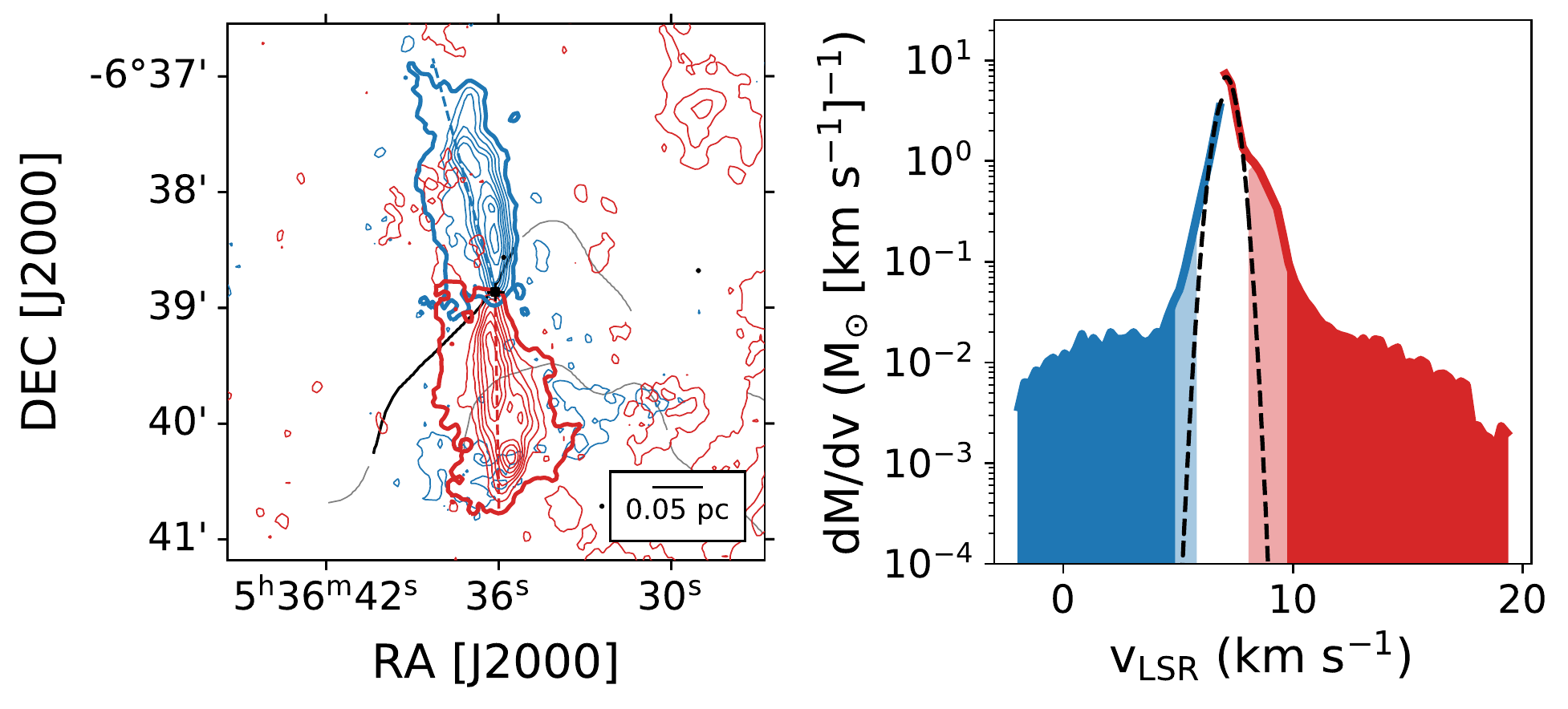}

    \caption{HOPS 169 outflow. The left panel shows the outflow, position angle, nearby sources, and filaments.
    The velocity range of integration is given by v$_{\rm blue}$/v$_{\rm red}$ in Table~\ref{tab:outflows}
    and the contours go from $5$ to $50\sigma$ in steps of 5$\sigma$, where $\sigma$ is the RMS error in the integrated map. Symbols are the same as Figure~\ref{fig:stamp}.
    The right panel shows the mass spectrum with fit, where $\sigma$ is the RMS error in the integrated map. Symbols are the same as Figure~\ref{fig:dmdv}.}
    \end{figure*}
\begin{figure*}[p]
\centering
\includegraphics[width=\textwidth]{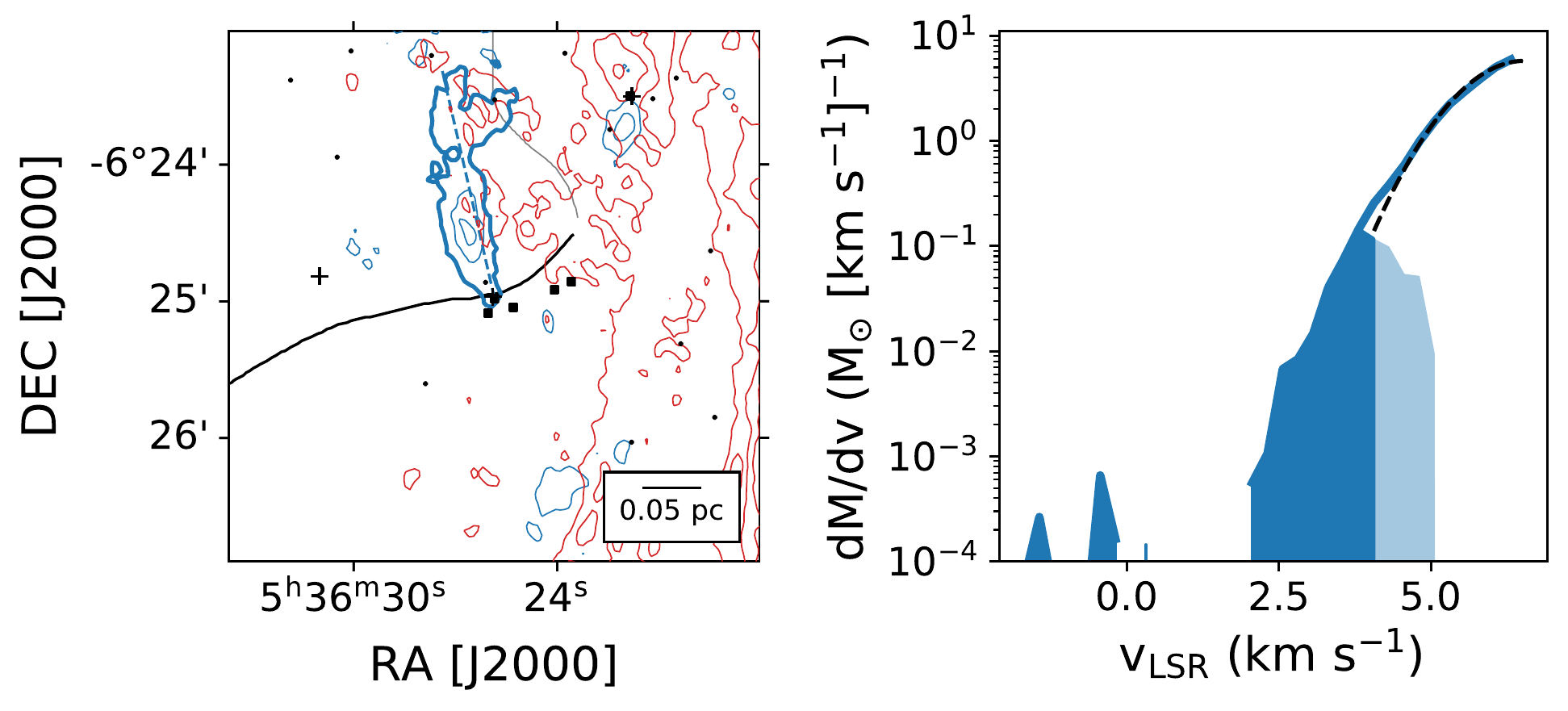}

    \caption{HOPS 174 outflow. The left panel shows the outflow, position angle, nearby sources, and filaments.
    The velocity range of integration is given by v$_{\rm blue}$/v$_{\rm red}$ in Table~\ref{tab:outflows}
    and the contours go from $5$ to $50\sigma$ in steps of 5$\sigma$, where $\sigma$ is the RMS error in the integrated map. Symbols are the same as Figure~\ref{fig:stamp}.
    The right panel shows the mass spectrum with fit, where $\sigma$ is the RMS error in the integrated map. Symbols are the same as Figure~\ref{fig:dmdv}.}
    \end{figure*}
\begin{figure*}[p]
\centering
\includegraphics[width=\textwidth]{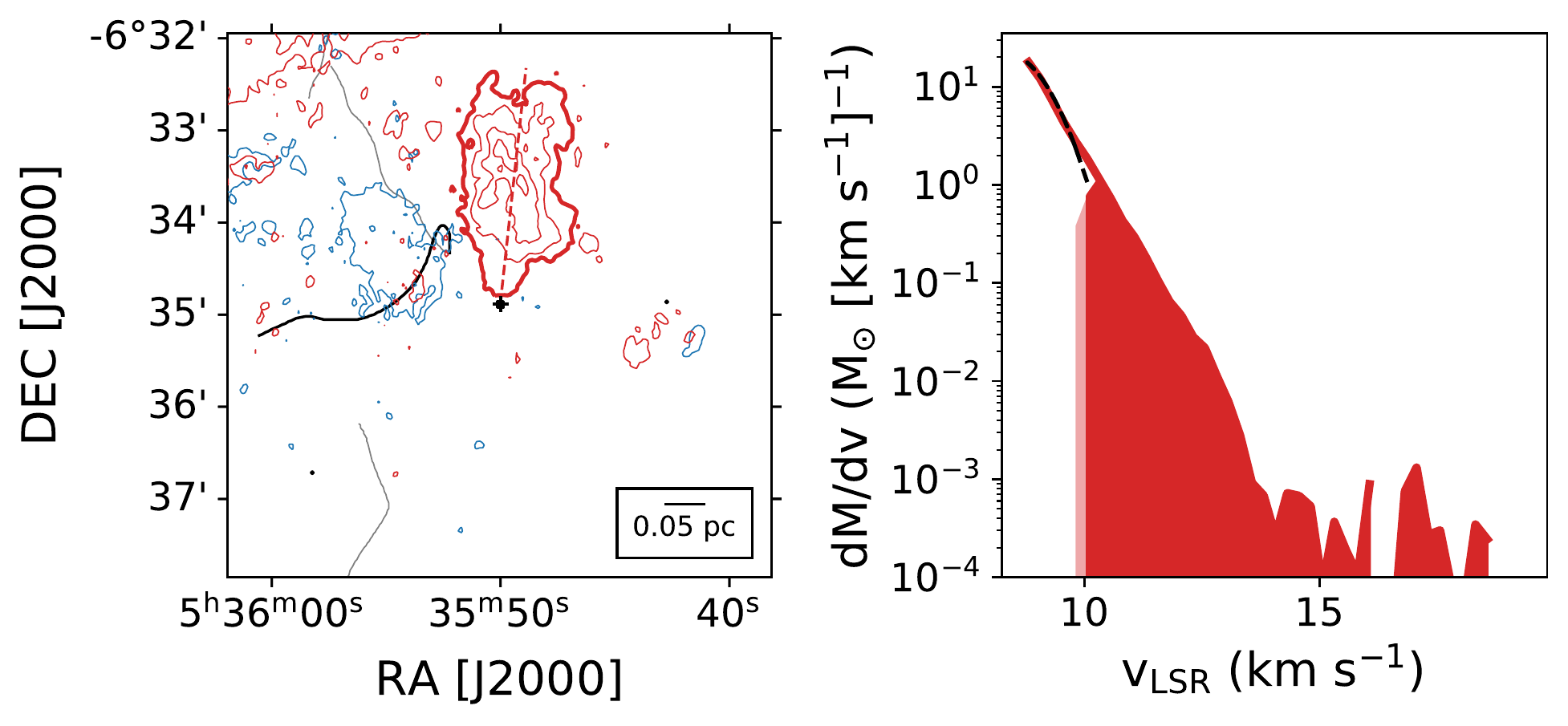}

    \caption{HOPS 177 outflow. The left panel shows the outflow, position angle, nearby sources, and filaments.
    The velocity range of integration is given by v$_{\rm blue}$/v$_{\rm red}$ in Table~\ref{tab:outflows}
    and the contours go from $5$ to $50\sigma$ in steps of 5$\sigma$, where $\sigma$ is the RMS error in the integrated map. Symbols are the same as Figure~\ref{fig:stamp}.
    The right panel shows the mass spectrum with fit, where $\sigma$ is the RMS error in the integrated map. Symbols are the same as Figure~\ref{fig:dmdv}.}
    \end{figure*}
\begin{figure*}[p]
\centering
\includegraphics[width=\textwidth]{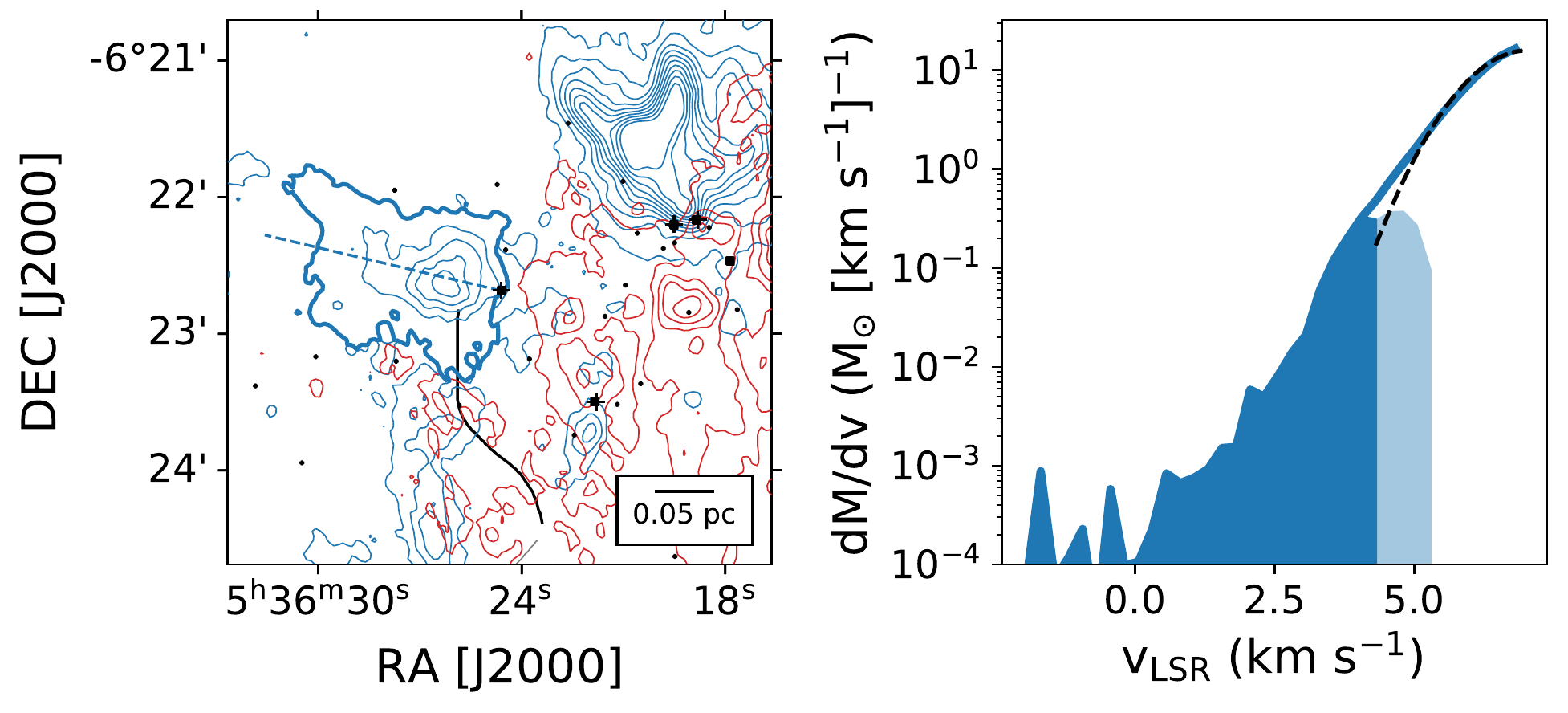}

    \caption{HOPS 178 outflow. The left panel shows the outflow, position angle, nearby sources, and filaments.
    The velocity range of integration is given by v$_{\rm blue}$/v$_{\rm red}$ in Table~\ref{tab:outflows}
    and the contours go from $5$ to $50\sigma$ in steps of 5$\sigma$, where $\sigma$ is the RMS error in the integrated map. Symbols are the same as Figure~\ref{fig:stamp}.
    The right panel shows the mass spectrum with fit, where $\sigma$ is the RMS error in the integrated map. Symbols are the same as Figure~\ref{fig:dmdv}.}
    \end{figure*}
\begin{figure*}[p]
\centering
\includegraphics[width=\textwidth]{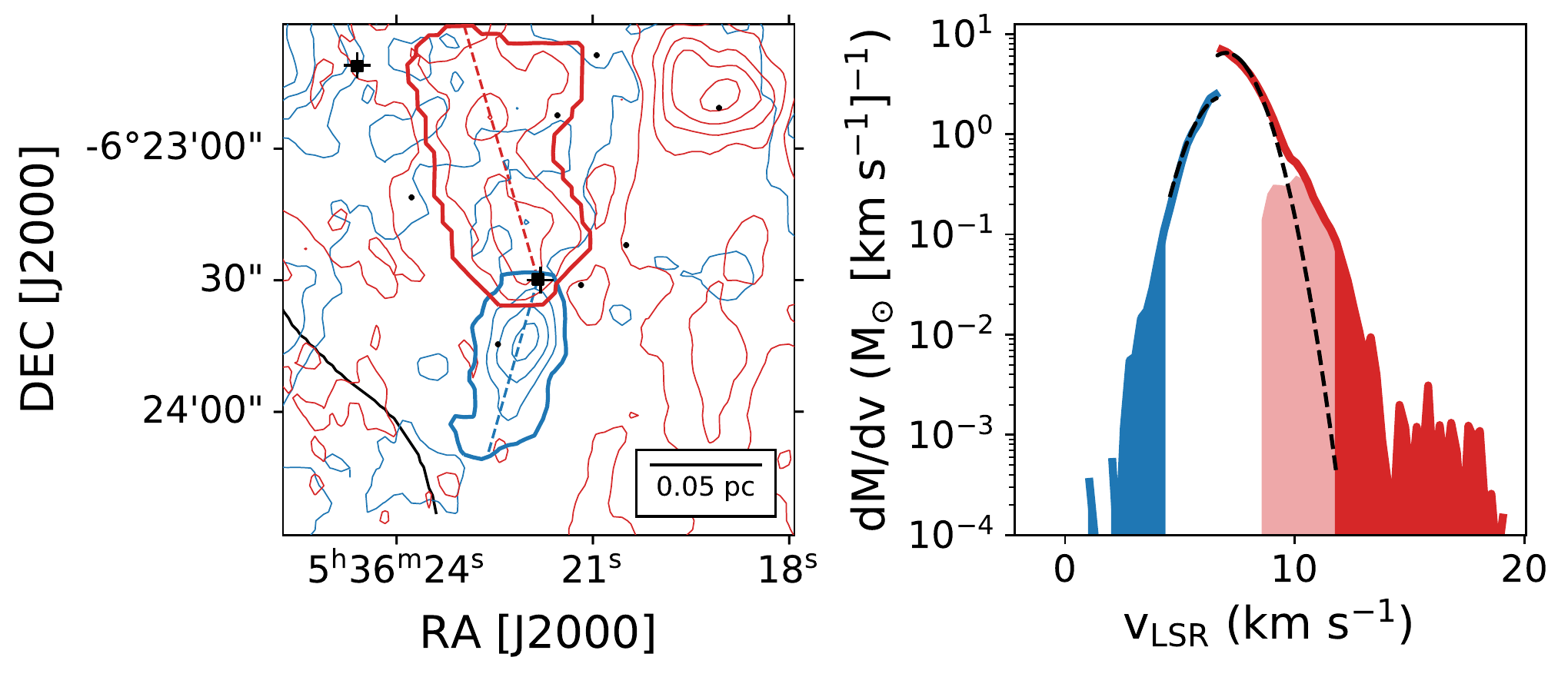}

    \caption{HOPS 179 outflow. The left panel shows the outflow, position angle, nearby sources, and filaments.
    The velocity range of integration is given by v$_{\rm blue}$/v$_{\rm red}$ in Table~\ref{tab:outflows}
    and the contours go from $5$ to $50\sigma$ in steps of 5$\sigma$, where $\sigma$ is the RMS error in the integrated map. Symbols are the same as Figure~\ref{fig:stamp}.
    The right panel shows the mass spectrum with fit, where $\sigma$ is the RMS error in the integrated map. Symbols are the same as Figure~\ref{fig:dmdv}.}
    \end{figure*}
\begin{figure*}[p]
\centering
\includegraphics[width=\textwidth]{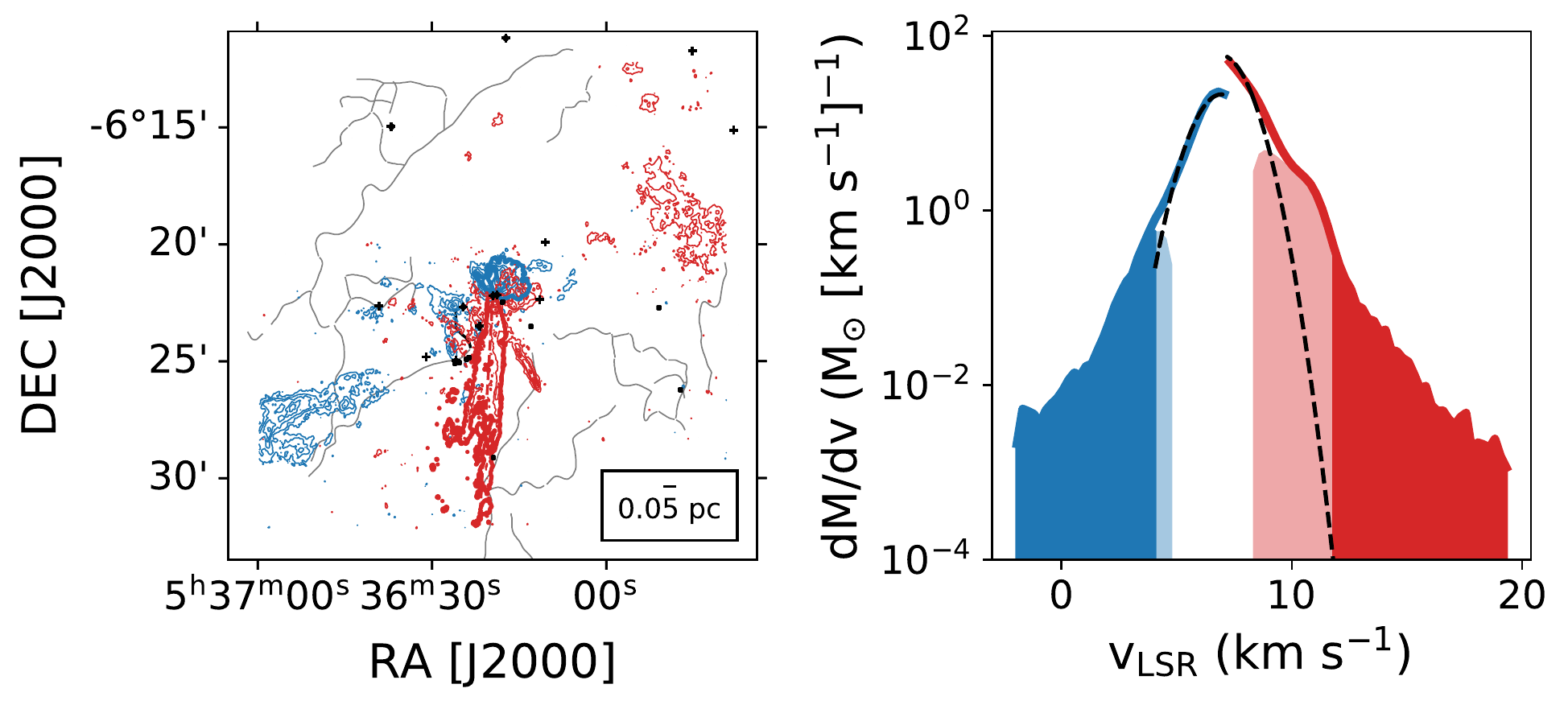}

    \caption{HOPS 181 outflow. The left panel shows the outflow, position angle, nearby sources, and filaments.
    The velocity range of integration is given by v$_{\rm blue}$/v$_{\rm red}$ in Table~\ref{tab:outflows}
    and the contours go from $5$ to $50\sigma$ in steps of 5$\sigma$, where $\sigma$ is the RMS error in the integrated map. Symbols are the same as Figure~\ref{fig:stamp}.
    The right panel shows the mass spectrum with fit, where $\sigma$ is the RMS error in the integrated map. Symbols are the same as Figure~\ref{fig:dmdv}.}
    \end{figure*}
\begin{figure*}[p]
\centering
\includegraphics[width=\textwidth]{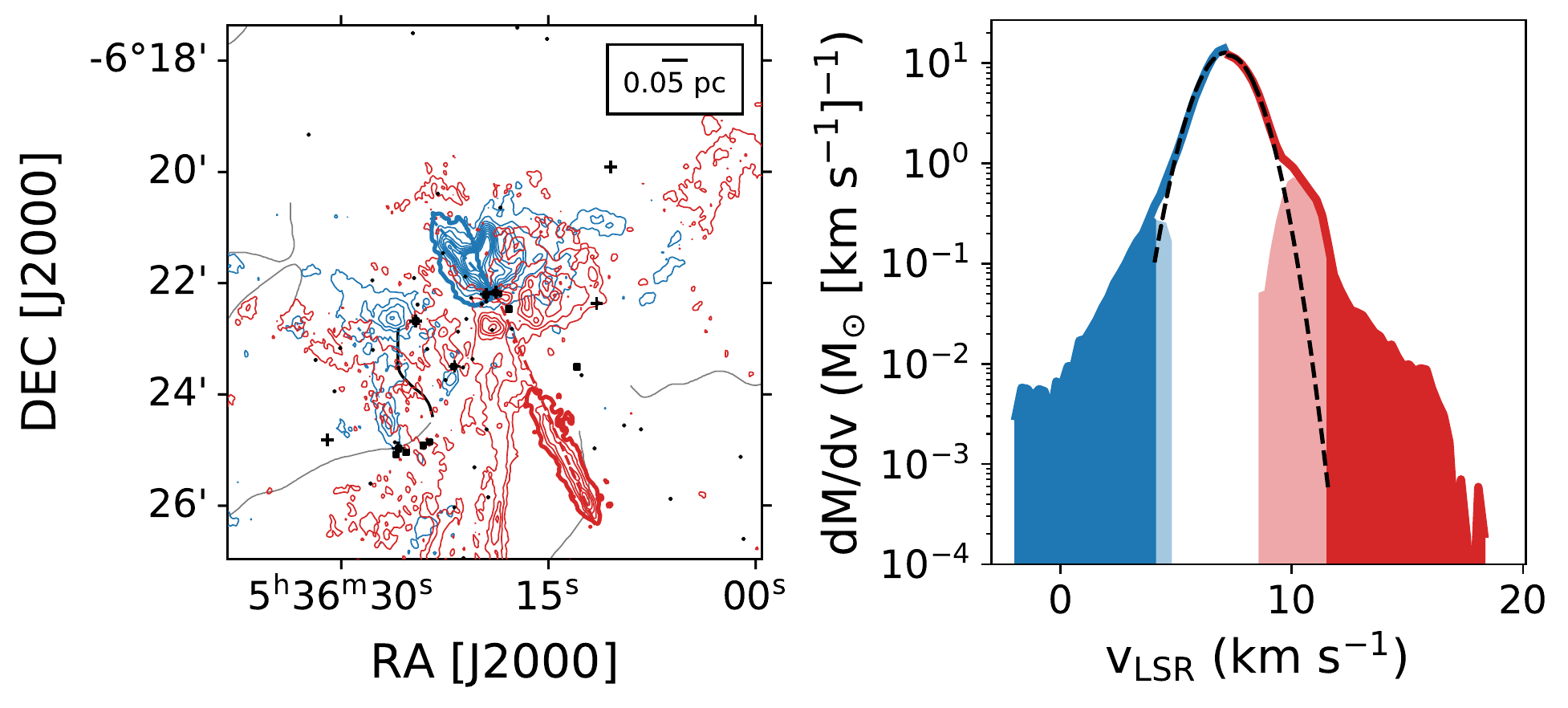}

    \caption{HOPS 182 outflow. The left panel shows the outflow, position angle, nearby sources, and filaments.
    The velocity range of integration is given by v$_{\rm blue}$/v$_{\rm red}$ in Table~\ref{tab:outflows}
    and the contours go from $5$ to $50\sigma$ in steps of 5$\sigma$, where $\sigma$ is the RMS error in the integrated map. Symbols are the same as Figure~\ref{fig:stamp}.
    The right panel shows the mass spectrum with fit, where $\sigma$ is the RMS error in the integrated map. Symbols are the same as Figure~\ref{fig:dmdv}.}
    \end{figure*}
\begin{figure*}[p]
\centering
\includegraphics[width=\textwidth]{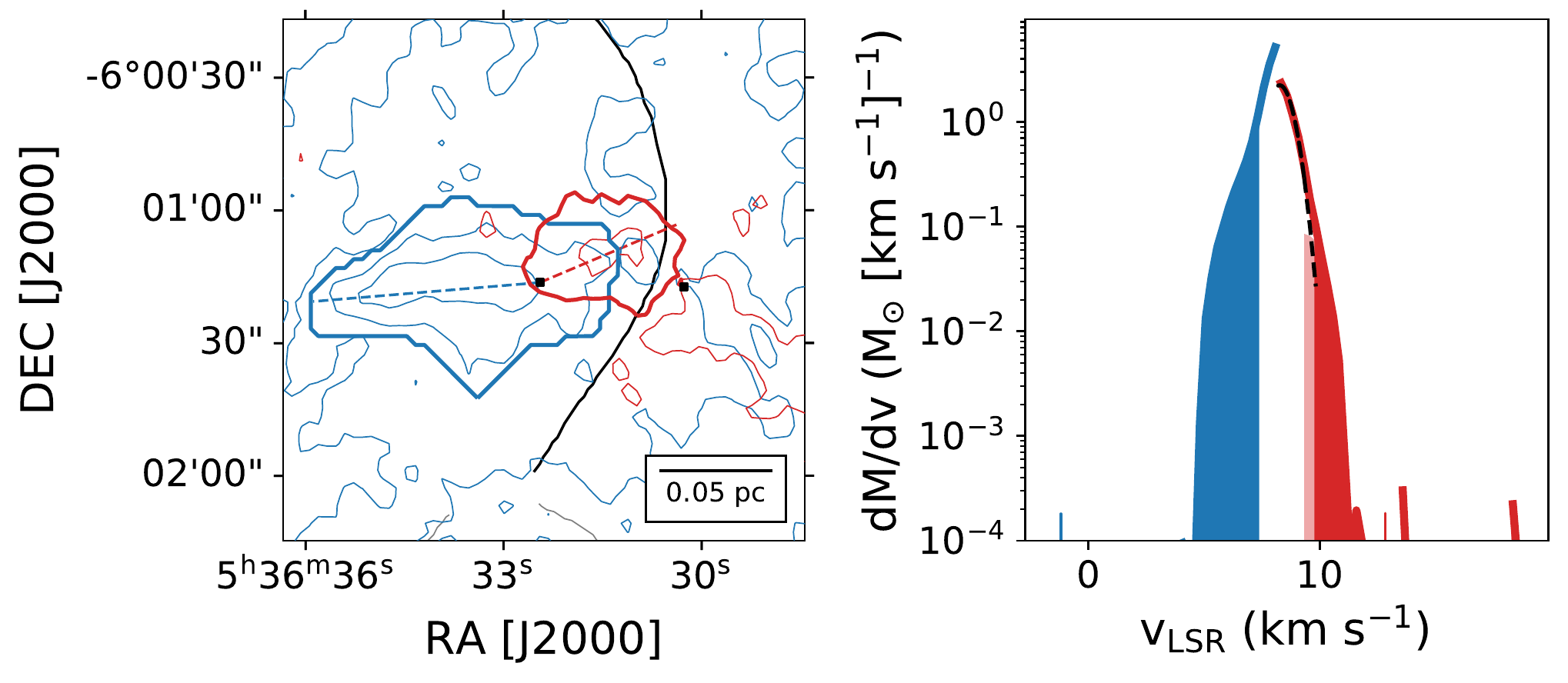}

    \caption{HOPS 192 outflow. The left panel shows the outflow, position angle, nearby sources, and filaments.
    The velocity range of integration is given by v$_{\rm blue}$/v$_{\rm red}$ in Table~\ref{tab:outflows}
    and the contours go from $5$ to $50\sigma$ in steps of 5$\sigma$, where $\sigma$ is the RMS error in the integrated map. Symbols are the same as Figure~\ref{fig:stamp}.
    The right panel shows the mass spectrum with fit, where $\sigma$ is the RMS error in the integrated map. Symbols are the same as Figure~\ref{fig:dmdv}.}
    \end{figure*}
\begin{figure*}[p]
\centering
\includegraphics[width=\textwidth]{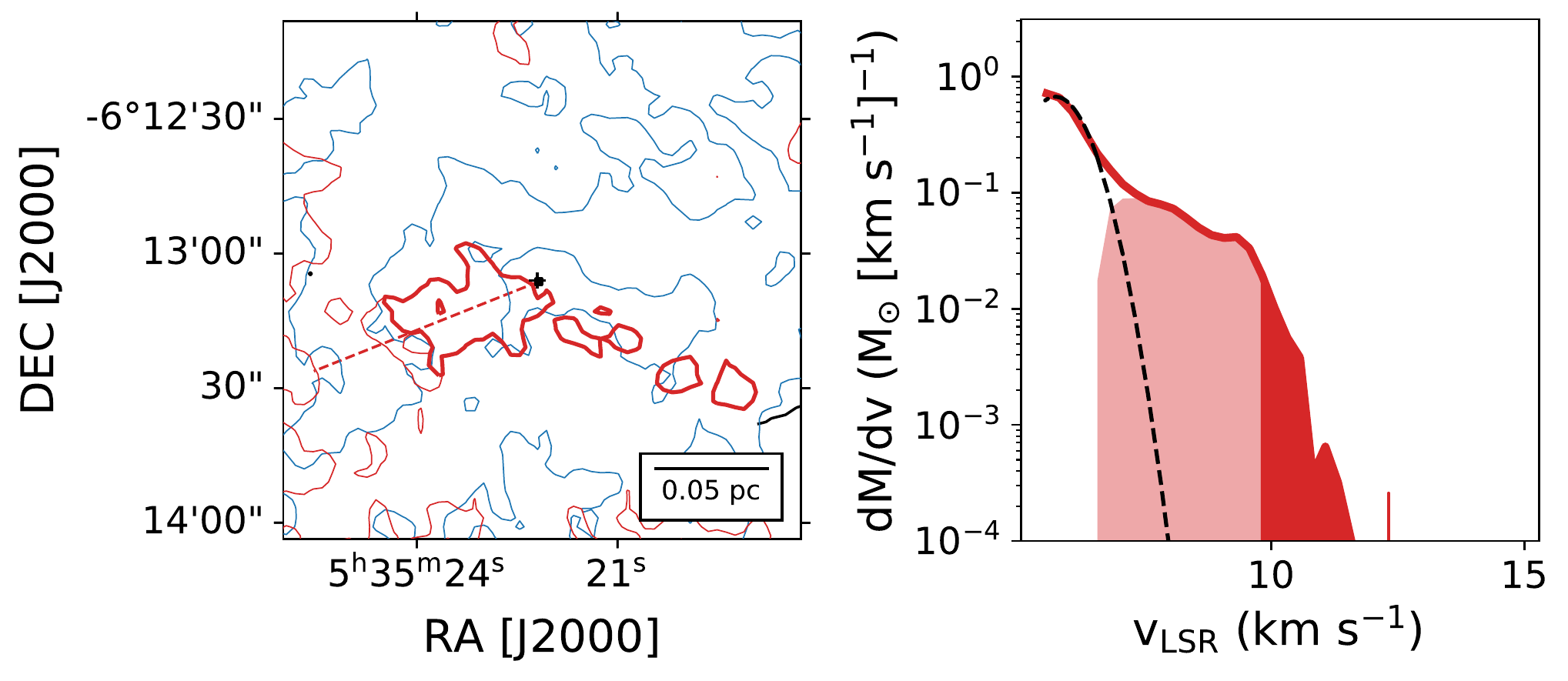}

    \caption{HOPS 198 outflow. The left panel shows the outflow, position angle, nearby sources, and filaments.
    The velocity range of integration is given by v$_{\rm blue}$/v$_{\rm red}$ in Table~\ref{tab:outflows}
    and the contours go from $5$ to $50\sigma$ in steps of 5$\sigma$, where $\sigma$ is the RMS error in the integrated map. Symbols are the same as Figure~\ref{fig:stamp}.
    The right panel shows the mass spectrum with fit, where $\sigma$ is the RMS error in the integrated map. Symbols are the same as Figure~\ref{fig:dmdv}.}
    \end{figure*}
\begin{figure*}[p]
\centering
\includegraphics[width=\textwidth]{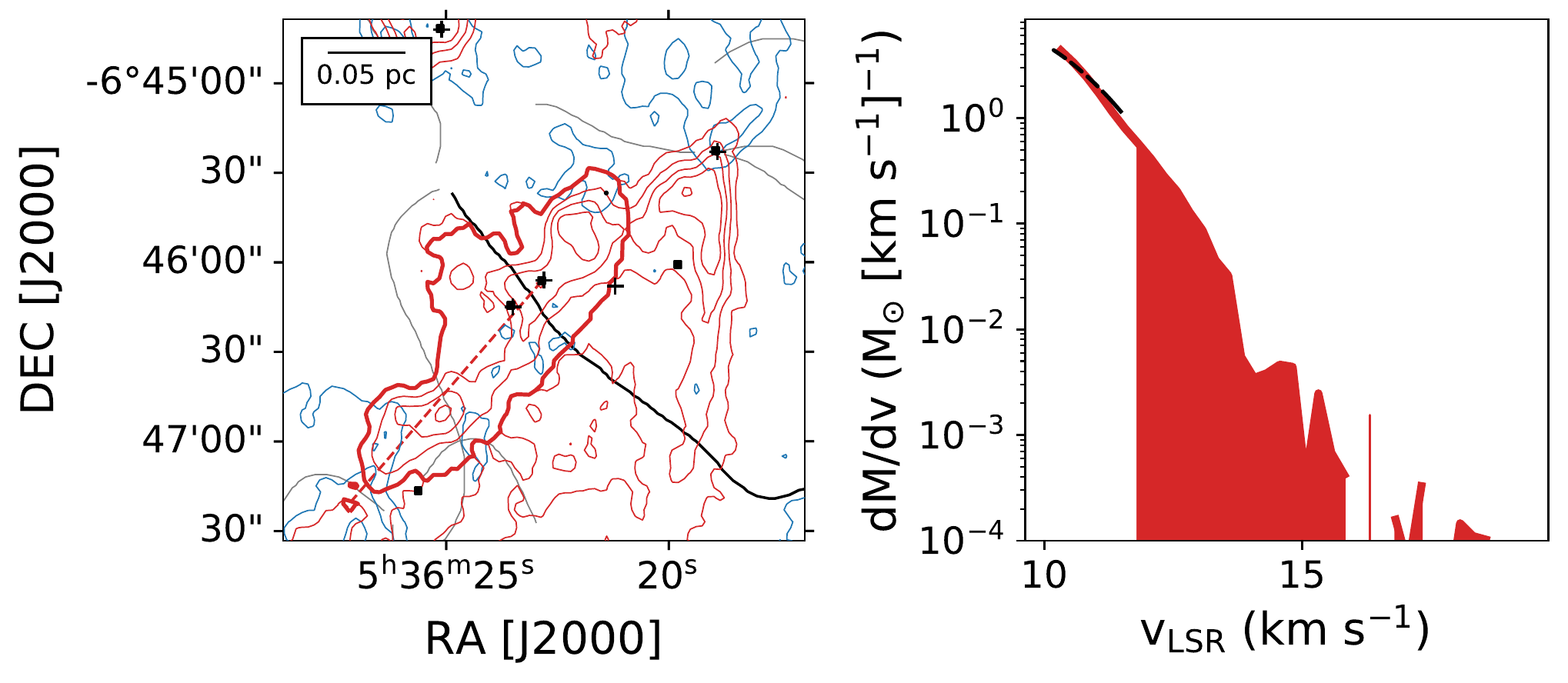}

    \caption{HOPS 203 outflow. The left panel shows the outflow, position angle, nearby sources, and filaments.
    The velocity range of integration is given by v$_{\rm blue}$/v$_{\rm red}$ in Table~\ref{tab:outflows}
    and the contours go from $5$ to $50\sigma$ in steps of 5$\sigma$, where $\sigma$ is the RMS error in the integrated map. Symbols are the same as Figure~\ref{fig:stamp}.
    The right panel shows the mass spectrum with fit, where $\sigma$ is the RMS error in the integrated map. Symbols are the same as Figure~\ref{fig:dmdv}.}
    \end{figure*}
\begin{figure*}[p]
\centering
\includegraphics[width=\textwidth]{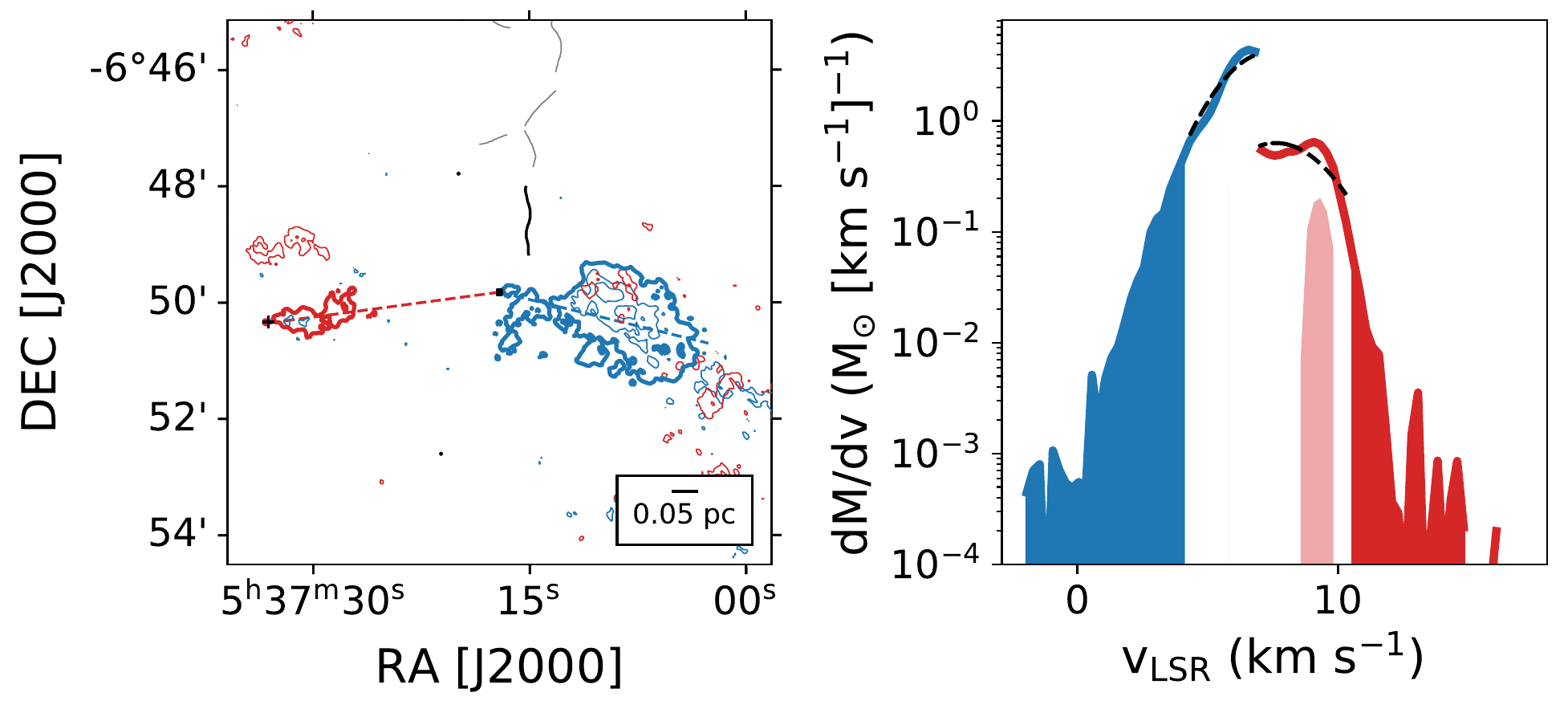}

    \caption{HOPS 355 outflow. The left panel shows the outflow, position angle, nearby sources, and filaments.
    The velocity range of integration is given by v$_{\rm blue}$/v$_{\rm red}$ in Table~\ref{tab:outflows}
    and the contours go from $5$ to $50\sigma$ in steps of 5$\sigma$, where $\sigma$ is the RMS error in the integrated map. Symbols are the same as Figure~\ref{fig:stamp}.
    The right panel shows the mass spectrum with fit, where $\sigma$ is the RMS error in the integrated map. Symbols are the same as Figure~\ref{fig:dmdv}.}
    \end{figure*}
\begin{figure*}[p]
\centering
\includegraphics[width=\textwidth]{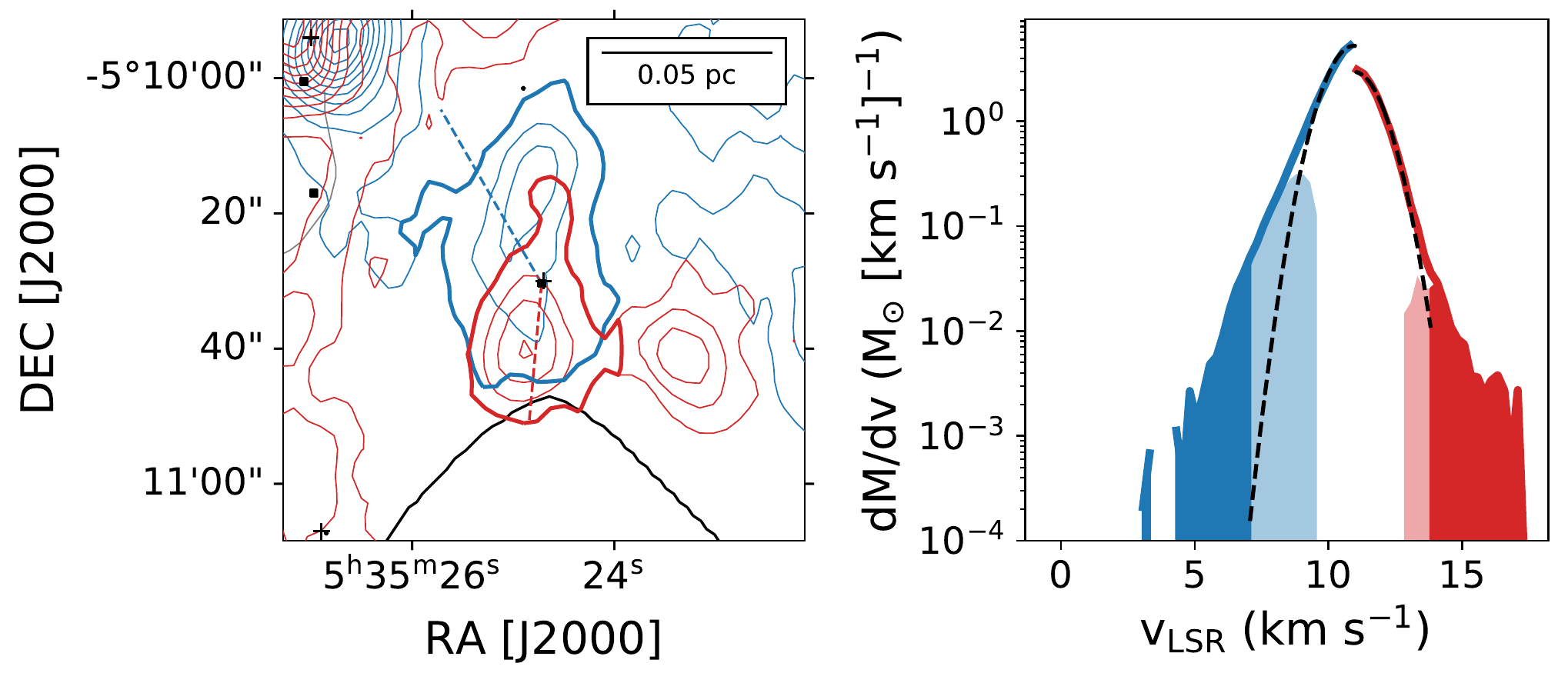}

    \caption{HOPS 368 outflow. The left panel shows the outflow, position angle, nearby sources, and filaments.
    The velocity range of integration is given by v$_{\rm blue}$/v$_{\rm red}$ in Table~\ref{tab:outflows}
    and the contours go from $5$ to $50\sigma$ in steps of 5$\sigma$, where $\sigma$ is the RMS error in the integrated map. Symbols are the same as Figure~\ref{fig:stamp}.
    The right panel shows the mass spectrum with fit, where $\sigma$ is the RMS error in the integrated map. Symbols are the same as Figure~\ref{fig:dmdv}.}
    \end{figure*}
\begin{figure*}[p]
\centering
\includegraphics[width=\textwidth]{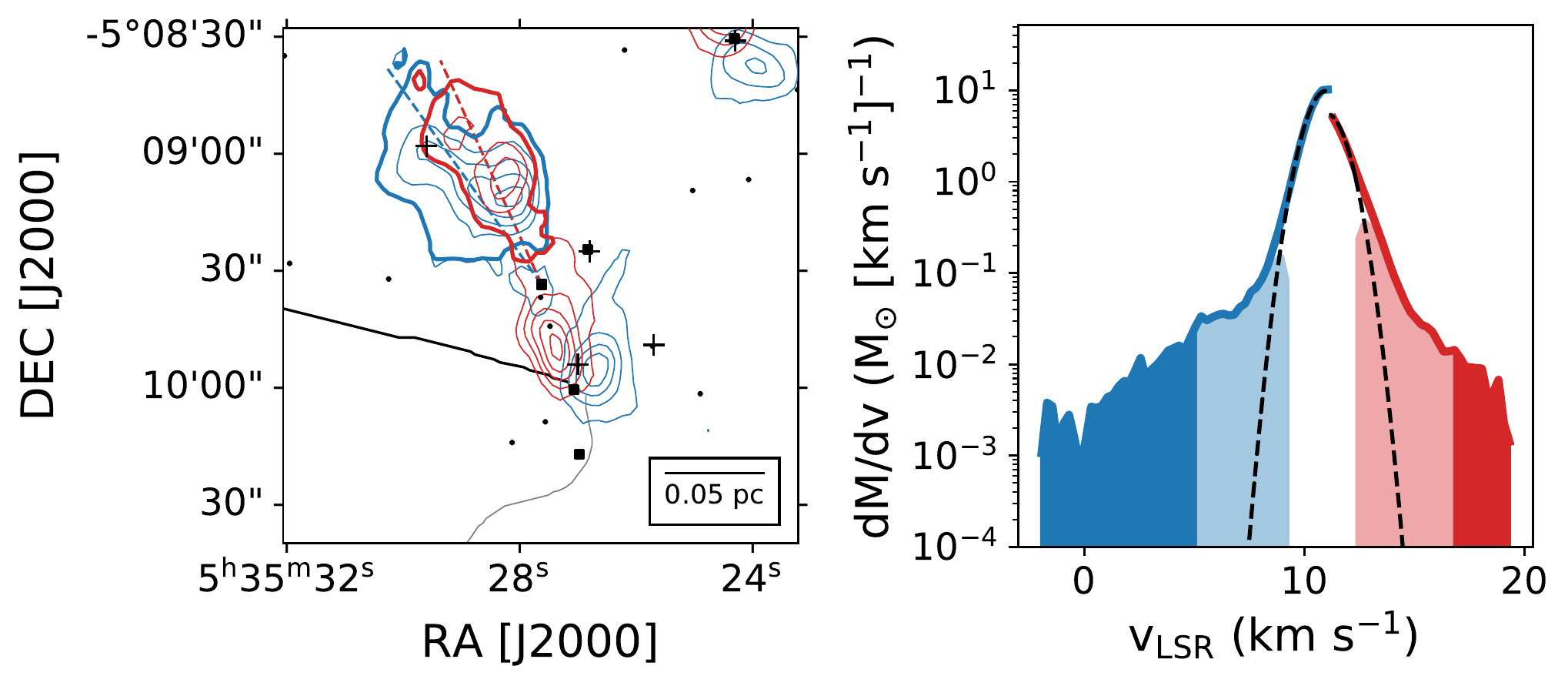}

    \caption{HOPS 370 outflow. The left panel shows the outflow, position angle, nearby sources, and filaments.
    The velocity range of integration is given by v$_{\rm blue}$/v$_{\rm red}$ in Table~\ref{tab:outflows}
    and the contours go from $5$ to $50\sigma$ in steps of 5$\sigma$, where $\sigma$ is the RMS error in the integrated map. Symbols are the same as Figure~\ref{fig:stamp}.
    The right panel shows the mass spectrum with fit, where $\sigma$ is the RMS error in the integrated map. Symbols are the same as Figure~\ref{fig:dmdv}.}
    \end{figure*}
\begin{figure*}[p]
\centering
\includegraphics[width=\textwidth]{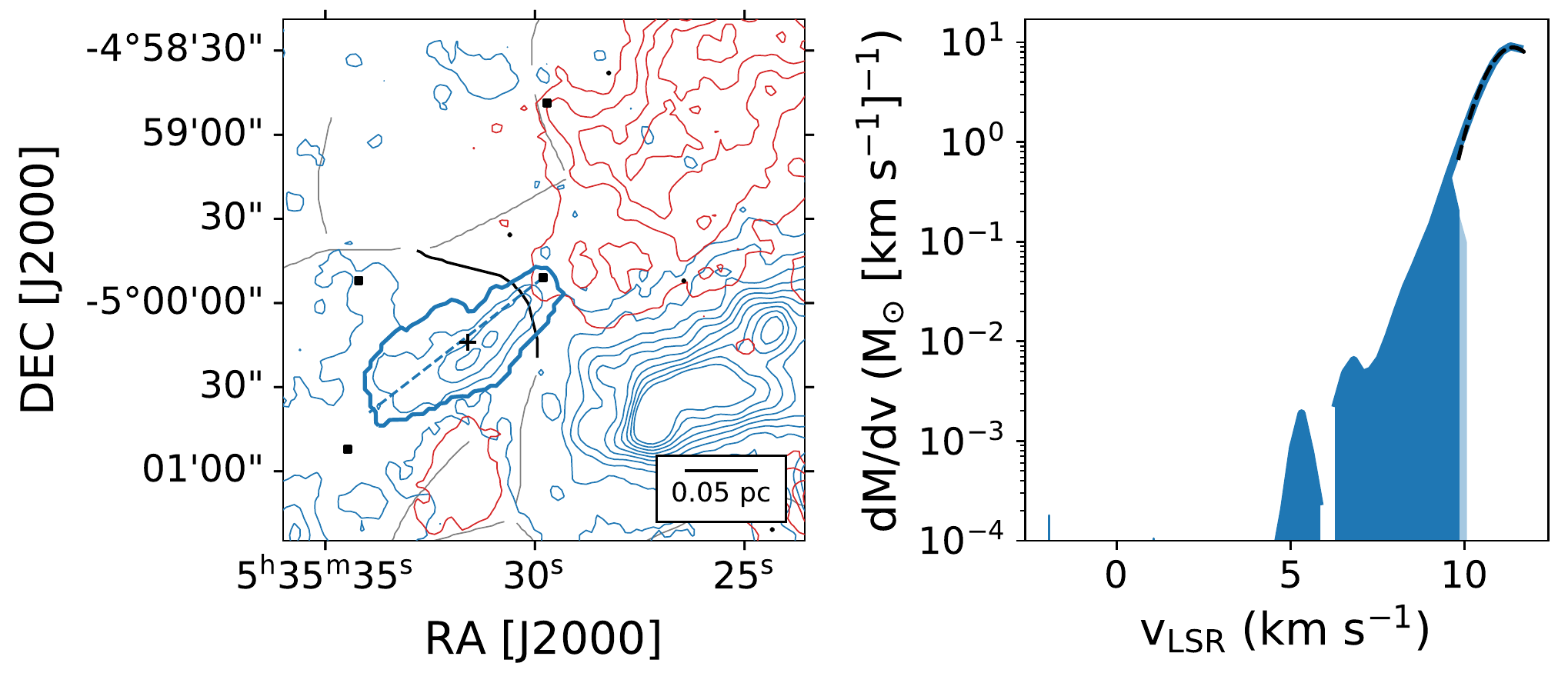}

    \caption{HOPS 383 outflow. The left panel shows the outflow, position angle, nearby sources, and filaments.
    The velocity range of integration is given by v$_{\rm blue}$/v$_{\rm red}$ in Table~\ref{tab:outflows}
    and the contours go from $5$ to $50\sigma$ in steps of 5$\sigma$, where $\sigma$ is the RMS error in the integrated map. Symbols are the same as Figure~\ref{fig:stamp}.
    The right panel shows the mass spectrum with fit, where $\sigma$ is the RMS error in the integrated map. Symbols are the same as Figure~\ref{fig:dmdv}.}
    \end{figure*}

\end{document}